\documentclass[12pt,leqno]{amsart}
\usepackage{graphicx}
\usepackage{indentfirst,csquotes}
\usepackage{url}
\usepackage{amsaddr}
\usepackage{cuted}
\usepackage{anyfontsize}

\usepackage[super,sort&compress,numbers]{natbib}

\usepackage{amssymb,amsthm,amsmath}
\usepackage{xcolor,paralist,hyperref,titlesec,fancyhdr,etoolbox}

\titleformat{\section}[display]{\normalfont\huge\bfseries\centering}{\centering\chaptertitlename\thechapter}{10pt}{\Large}
\titlespacing*{\section}{0pt}{0ex}{0ex}

\hypersetup{colorlinks=true, linkcolor=black, filecolor=black, urlcolor=black }

\usepackage{lipsum}

\begin{document}
\title{Cavity-modified exciton-exciton annihilation in disordered molecular systems} 

\author[]{I. Sokolovskii$^{*{\dagger}}$, B. S. Humphries$^{{\dagger}}$, J. Blumberger$^{{\dagger}}$, \\ and G. Groenhof$^\ddagger$}
\date{\today}
\address{$^{\dagger}$Department of Physics and Astronomy and Thomas Young Centre, University College London, Gower Street, London WC1E 6BT, United Kingdom}
\address{$^{\ddagger}$Nanoscience Center and   Department of Chemistry, University of Jyv\"{a}skyl\"{a}, P.O. Box 35, 40014 Jyv\"{a}skyl\"{a}, Finland}
\email{i.sokolovskii@ucl.ac.uk}

\begin{abstract}
Recent experiments have shown contradictory effects of strong light-matter coupling on exciton-exciton annihilation (EEA) in organic molecular systems. In this work, we perform numerical simulations of polariton dynamics and reveal the role of strong coupling in changing the EEA rate. The results of our simulations suggest that strong coupling allows to partially overcome disorder in systems with poor exciton mobility via delocalisation of excitons owing to the interaction with the common cavity mode. This leads to an enhanced connectivity between excitons and, consequently, to an increase in the EEA rate. Conversely, in systems with high exciton mobility, in which disorder has a much smaller effect on excitation energy transfer, excitons can interact strongly even without coupling to the cavity photons at the exciton densities at which EEA typically occurs. In this case, the EEA rate can be even lower than in bare molecules due to the existence of a competing decay channel associated with photon leakage through the cavity mirrors. We also find that in the weak coupling regime, the EEA rate appears to be suppressed due to this decay channel regardless of the exciton transport properties. Our simulations resolve the experimental controversy on the effect of strong coupling on EEA and provide guidance for minimising the EEA rate towards a more feasible realisation of Bose-Einstein condensation of polaritons. 
\end{abstract} 

\maketitle

\bigskip

\section*{Introduction}\label{sec:intro}

\bigskip

Strong coupling of organic molecules with confined electromagnetic fields in optical resonators opens the potential for improving the efficiency of organic solar cells~\cite{Forrest2022,deJong2024,Peruffo2025}, building logics for classical and quantum computers~\cite{Kavokin2022}, modifying photochemistry~\cite{Ebbesen2021}, and achieving coherent light emission~\cite{Forrest2010}. In particular, polariton lasers represent a low-threshold alternative to conventional lasers, which rely on population inversion and have a high threshold for the injected charge carrier density required to achieve lasing~\cite{Yamamoto1996}. In contrast, polariton lasers rely on Bose-Einstein condensation (BEC), at which a macroscopic number of polaritons occupies the same quantum state~\cite{Bhuyan2023}, usually the bottom of the lower polariton (LP) branch but not necessarily~\cite{Chen2023}. This leads to bosonic stimulation of polaritons, when the transition probability into the common polaritonic state grows linearly with the occupation of this state. Bosonic stimulation is followed by nonlinear amplification, when the emission rate from the occupied state increases non-linearly as the density of polaritons crosses a certain threshold. Owing to a very small effective mass of polaritons, inherited from their partially photonic character, a room-temperature BEC of polaritons is possible, as was first demonstrated by K\'ena-Cohen and Forrest in a single-crystal anthracene microcavity~\cite{Kena-Cohen2010}.

Although organic polariton lasers, which do not require population inversion, allow for lower threshold charge densities compared to conventional photon lasers~\cite{Kena-Cohen2010,Ramezani2017}, an electrically injected laser, which is highly desirable for practical applications, is still beyond reach. This is due to the low mobility of charge carriers in organic materials, singlet/triplet statistics, as well as the effects competing with the population relaxation to the bottom of the LP branch~\cite{Keeling2020}. It is commonly assumed that the relaxation to this state from the reservoir of excitons occurs by the following two mechanisms: radiative pumping of polaritons by uncoupled molecules~\cite{Grant2016,Zaumseil2021} and vibrational-assisted scattering, in which part of the exciton energy is transferred to a specific molecular vibration~\cite{Agranovich2003,Litinskaya2004,Coles2011,Somaschi2011}. The rate of both mechanisms was previously found to scale inversely with the number of strongly coupled molecules~\cite{delPino2015,Eizner2019,Ulusoy2019,Tichauer2022,Davidsson2023,Bhuyan2024,Sokolovskii2024}. Recently, P\'erez-S\'anchez and Yuen-Zhou also proposed to classify polariton relaxation into radiative pumping, polariton-assisted photon recycling, and polariton-assisted Raman scattering~\cite{Perez-Sanchez2025,Schwennicke2025}.

\begin{figure}[!htb]
\centering
\includegraphics[width=0.5\textwidth ]{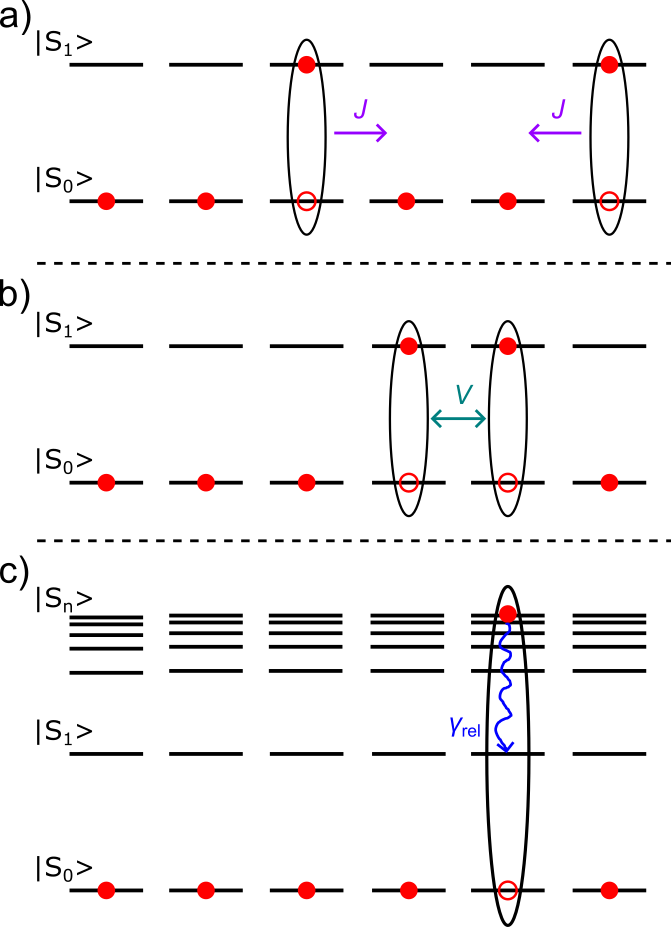}
  \caption{Schematic illustration of the mechanism of exciton-exciton annihilation (EEA). In a chain of molecules, represented as multiple energy levels ($|S_0\rangle,|S_1\rangle,...,|S_n\rangle,...$), a pair of $S_1$-excitons with filled and open circles representing electrons and holes, respectively, is initially created (panel~\textbf{a}). These excitons can hop between adjacent molecules via the dipole-dipole coupling or wave functions overlap, characterised by coupling strength $J$. When the two excitons approach each other, they can interact via coupling term $V$, which describes the excitonic interaction (panel~\textbf{b}). As a result of this interaction, an electron in one of the neighboring molecules can be promoted to the $S_n$-state with energy $E_{S_n}\approx2E_{S_1}$ (panel~\textbf{c}). At the same time, the other molecule returns to the ground state. Subsequently, the first molecule rapidly relaxes into the $S_1$-state, which is characterised by relaxation rate $\gamma_\text{rel}$.     
  }
  \label{fig:figure1}
\end{figure}

Relaxation from the exciton reservoir to the bottom of the LP branch always competes with deactivation processes that deplete the population of polariton states. These deactivation processes include decay of polaritons due to their partially photonic nature and exciton-exciton annihilation (EEA)~\cite{Swenberg1976}. Figure~\ref{fig:figure1} schematically demonstrates the principle of EEA. For a pair of $S_1$-excitons created at a certain distance from each other in a material, each of these excitons can hop independently between adjacent molecules via the F\"{o}rster~\cite{Forster1949} or Dexter~\cite{Dexter1953} mechanism, characterised by small values of coupling strength $J$ (panel~\textbf{a}) compared to reorganization energy, or by transient delocalization for larger values of $J$~\cite{Rao2022,Giannini2022,Stojanovic2024,Ivanovich2025,Ivanovich2026}. Once the excitons approach each other in space (panel~\textbf{b}), the excitonic interaction, $V$, may promote an electron in one of the molecules to the $S_n$-state with energy $E_{S_n}\approx2E_{S_1}$ while simultaneously de-exciting the other molecule. The molecule in the $S_n$-state then rapidly relaxes into the $S_1$-state according to Kasha's rule~\cite{Kasha1950} (panel~\textbf{c}). Therefore, as a result of the interaction, a pair of $S_1$-excitons is converted into a single $S_1$-exciton, for which reason this type of interaction is usually called exciton-exciton annihilation.

If depopulation of the LP branch due to polariton decay and EEA occurs faster than relaxation to the bottom of the LP branch, condensation of polaritons and, consequently, lasing will not be achieved. While the cavity decay rate can be controlled to some extent by the cavity quality factor (Q-factor), the EEA rate is material-specific and is known to increase with charge carrier density~\cite{Mikhnenko2015}. Therefore, EEA can be a serious parasitic effect for polariton lasers, in which high exciton densities are required to achieve condensation~\cite{Akselrod2013}. Furthermore, since polaritons possess a partially excitonic character, the emergence of strong coupling itself can alter the EEA rate. Understanding the mechanism of this effect and whether the rate of EEA can be increased or reduced in the cavity is therefore important for improving organic polariton lasers towards the development of an electrically pumped laser.

The first experimental study of the effect of strong coupling on exciton annihilation was conducted by Akselrod \textit{et al.} for J-aggregates of cyanine dye in a Fabry-P\'erot microcavity~\cite{Akselrod2010}. Photoluminescence measurements revealed an increase in the EEA rate under strong coupling, which was attributed to an enhanced effective exciton diffusion length. Recently, Wu \textit{et al.} demonstrated a reduction in the threshold exciton density for the onset of EEA for photosynthetic light-harvesting 2 (LH2) complexes and Rhodamine 6G (R6G) dye in a planar metallic cavity using pump-probe spectroscopy~\cite{Pullerits2025}. The authors explained this result by improved excitation energy transfer due to strong coupling. In another experiment with R6G, this time in a high Q-factor distributed Bragg reflector microcavity, Qureshi \textit{et al.} employed angle-resolved photoluminescence measurements and obtained the opposite result: the threshold exciton density for the onset of EEA was found to increase in the cavity~\cite{Qureshi2025}. The suppression of the EEA was attributed to \textit{i)} a more efficient depopulation of the exciton reservoir through cavity decay, \textit{ii)} an increase in the propagation velocity of excitons, which facilitates their escape from neighbouring excitons, and \textit{iii)} a more efficient coupling of excitons with the cavity mode than with each other. Furthermore, suppression of the EEA rate was recently demonstrated in a two-dimensional perovskite (PEA)$_2$PbI$_4$ in a Fabry-P\'erot microcavity~\cite{Lagemaat2024} and in a monolayer of WS$_2$, in which excitons were intermediately coupled to Mie resonances of a dielectric nanoantenna~\cite{Maie2023}. These results were attributed to the hybrid nature of polaritons and an increase in the spontaneous emission rate of the material through the Purcell effect. Finally, no modification of EEA was reported in a recent study of polycrystalline zinc phthalocyanine strongly coupled to surface-plasmon polaritons~\cite{Buttner2025}.

To reconcile these conflicting experimental results and study the effect of the cavity-modified EEA, we performed numerical simulations of polariton dynamics in the second excitation subspace, with the inclusion of interaction terms responsible for exciton-exciton annihilation. The results of the simulations indicate a competition between two mechanisms: \textit{cavity-enhanced exciton delocalisation}, 
which leads to better connectivity between excitons in the presence of disorder and, therefore, enhances the EEA rate, and \textit{polariton decay}, which depletes the excited-state population, reducing exciton collisions; therefore, this effect decreases the EEA rate. Thus, the overall positive or negative influence of a cavity on the EEA rate is determined by the relationship between the magnitudes of disorder strength, of exciton coupling, and of cavity decay rate in each specific molecule-cavity system. This implies that a careful selection of the strongly coupled material and the cavity structure is required to achieve a balance between the rates of EEA and of polariton decay in order to minimise the threshold for BEC of polaritons.

\bigskip

\section*{Theoretical model}\label{sec:model}

\bigskip

\subsection*{System's description and model Hamiltonian}\label{section:model_a}
Following previous theoretical studies of exciton-exciton annihilation~\cite{Malyshev1999,Knoester2001}, we consider a closed chain of $N$ three-level molecules (TLMs) with the following electronic states: the ground state with energy $E_{S_0}$, the first excited state with energy $E_{S_1}$, and the $n^\text{th}$ excited state with energy $E_{S_n}=2E_{S_0}$. In such a system, two types of excitons can be created in each molecule, namely $S_1$-excitons and $S_n$-excitons. Furthermore, $S_1$-excitons can hop between adjacent molecules due to exciton coupling, characterised by strength $J$, and a pair of such excitons can interact through the coupling term $V$, resulting in EEA. The $S_1$-excitons are also coupled to the fundamental cavity mode with energy $E_c=E_{S_1}$, and the coupling is characterised by strength $g$. Since the oscillator strength of the $S_n$-state is often significantly smaller compared to the $S_1$-state, as is also the case for R6G molecule used as a model system in this work (Section~3 in the Supporting Information, SI), the $S_n$-state is assumed to interact with the second cavity mode of energy $E_{c2}=E_{S_n}$ very weakly. Therefore, we do not include the corresponding coupling term in the model. Based on the same reasoning about oscillator strength, we also assume that the dipole-dipole interaction between $S_n$-states is very weak and therefore neglect intermolecular transfer of the $S_n$-excitons.

The molecule-cavity system is described by the following Hamiltonian:
\begin{equation}
\begin{array}{ccl}
{\hat{H}} &=&\sum_i^{N} E_{S_1,i}\hat{\sigma}^+_{S_1,i}\hat{\sigma}^-_{S_1,i}+E_{\text{c}}\hat{a}^\dagger\hat{a}-\sum_i^Ng_i\left(\hat{\sigma}_{S_1,i}^++\hat{\sigma}_{S_1,i}^-\right)\left(\hat{a}^\dagger+\hat{a}\right) + \\
\\
&&\sum_{i,k}^NJ_{i,k}\left(\hat{\sigma}_{S_1,i}^{+}\hat{\sigma}_{S_1,k}^{-}+\text{h.c.}\right) + \\
\\
&&\sum_j^NE_{S_n,j}\hat{\sigma}^+_{S_n,j}\hat{\sigma}^-_{S_n,j} + \sum_{i,k}^NV_{i,k}\left[\left(\hat{\sigma}^+_{S_n,i}+\hat{\sigma}^+_{S_n,k}\right)\hat{\sigma}^-_{S_1,i}\hat{\sigma}^-_{S_1,k} +\text{h.c.}\right].
\label{eq:Ham_Rabi}
\end{array}
\end{equation}
Here, the first line corresponds to the standard Dicke Hamiltonian that describes the interaction between several molecules and one cavity mode~\cite{Kirton2019}. Operators $\hat{\sigma}^+_{S_1,i}$ and $\hat{\sigma}^-_{S_1,i}$ create and destroy a $S_1$-exciton in molecule $i$, whereas operators $\hat{a}^\dagger$ and $\hat{a}$ create and annihilate a cavity photon. The second line in Equation~\ref{eq:Ham_Rabi} describes the interaction between TLMs $i$ and $k$, which allows $S_1$-excitons to jump between the corresponding molecules~\cite{Sokolovskii2025}. In the last line of Equation~\ref{eq:Ham_Rabi}, the first term contains the energies of $S_n$-excitons in each TLM, and the second term describes EEA in TLMs $i$ and $k$, which results in the destruction of $S_1$-excitons at both TLMs (operators $\hat{\sigma}^-_{S_1,i}$ and $\hat{\sigma}^-_{S_1,k}$) and the simultaneous formation of a $S_n$-exciton at either TLM $i$ or TLM $k$ (operators $\hat{\sigma}^+_{S_n,i}$ and $\hat{\sigma}^+_{S_n,k}$)~\cite{Knoester2001,Renger2001,Fan2023}.
Furthermore, the coupling elements $J_{i,k}$ and $V_{i,k}$ responsible for the transfer of $S_1$-excitons and EEA, respectively, are assumed to decay with the distance as $J_{i,k}=J/|i-k|^3$ and $V_{i,k}=V/|i-k|^3$, according to the point dipole approximation~\cite{Beljonne2025}.

Because EEA involves interaction of two excitons, the Hamiltonian matrix is constructed up to the second excitation subspace (Section~1 in the SI). Within the Rotating-wave approximation (RWA), which is valid below ultrastrong coupling, \textit{i.e.} when $g\sqrt{N}<0.1E_{S_1}$~\cite{Solano2019}, and which we employ in this work, the counter-rotating terms, $\hat{\sigma}^-_{S_1,i}\hat{a}$ and $\hat{\sigma}^+_{S_1,i}\hat{a}^\dagger$, are neglected, and hence the zeroth 
excitation subspace is no longer directly coupled to the second excitation subspace. Therefore, for the purpose of this study, we may only consider the part of the Hamiltonian that corresponds to the second excitation subspace. In this subspace, four types of product states are distinguished: \textit{i)} $|\phi_{S_n,0}^i\rangle\equiv|S_0^1,S_0^2,...,S_n^i,...,S_0^N\rangle\otimes|0\rangle$, which describes the situation in which TLM $i$ is in the $S_n$-state, while the other TLMs are in the ground state and no cavity photon is present; \textit{ii)} $|\phi_{2S_1,0}^i\rangle\equiv|S_0^1,...,S_1^i,...,S_1^j,...,S_0^N\rangle\otimes|0\rangle$, which is the state with simultaneously excited TLMs $i$ and $j$ into the $S_1$-state, while the other TLMs are in the ground state and no cavity photon is present; \textit{iii)} $|\phi_{S_1,1}^i\rangle\equiv|S_0^1,S_0^2,...,S_1^i,...,S_0^N\rangle\otimes|1\rangle$, which describes the situation in which TLM $i$ is in the $S_1$-state and a single photon is present in the cavity; and \textit{iv)} $|\phi_{S_0,2}\rangle\equiv|S_0^1,S_0^2,...,S_0^N\rangle\otimes|2\rangle$, which corresponds to two photons present in the cavity, while all TLMs are in the ground state. In total, there are $N_\text{st}=2N+\frac{N(N-1)}{2}+1=\frac{(N+1)(N+2)}{2}$ states, $N$ of which are $|\phi_{S_n,0}\rangle$ states, $N(N-1)/{2}$ states are $|\phi_{2S_1,0}\rangle$ states, another $N$ states are $|\phi_{S_1,1}\rangle$ states, and the remaining state is the $|\phi_{S_0,2}\rangle$ state. 

Following Ryzhov \textit{et al.}, we assume that EEA leads to the promotion of an electron in one of the two interacting TLMs to an excited vibrational state of the $n^\text{th}$ electronic state, \textit{i.e.} $|S_n,v>0\rangle$~\cite{Knoester2001}. From there, efficient phonon-assisted relaxation to the vibrational ground state, $|S_n,v=0\rangle$, occurs, followed by relaxation to the $S_1$-state in accordance with Kasha's rule~\cite{Kasha1950}. Subsequently, the TLM can also decay to the ground state via radiative or nonradiative pathways. Modelling internal conversion to the $S_1$-state requires the use of advanced methods of quantum or semiclassical dynamics, which is intractable for the system under consideration. Therefore, we instead approximate the three relaxation processes (phonon-assisted relaxation within the $S_n$-state, internal conversion to the $S_1$-state, and decay to the ground state) as a single process by introducing the deactivation terms, $-i\hbar\gamma_v$, from the Lindblad operator to the diagonal elements of the system's Hamiltonian (Equation~\ref{eq:Ham_Rabi}) that correspond to the $|\phi_{S_n,0}\rangle$ states~\cite{Manzano2020,Sokolovskii2024b}. In other words, we assume that after internal conversion to the $S_1$-state and hence entering the first excitation subspace, a TLM becomes decoupled from the second excitation subspace and can only decay to the ground state. Future work will aim at removing this approximation and simulating exciton annihilation including not only the second but also the zeroth and first excitation subspaces explicitly.

With the deactivation terms and under the RWA, the matrix form of the Hamiltonian in the second excitation subspace reads as follows:

\begin{equation}
\text{\makebox[\textwidth][c]{$
{\fontsize{8pt}{10pt}\selectfont
\setlength{\arraycolsep}{0pt}
\textbf{H} = 
 \begin{pmatrix}\begin{array}{cccccccccc} 
   E_{S_n,1}-i\hbar\gamma_v & 0 & 0 & V & V & 0 & 0 & 0 & 0 & 0 \\
   0 & E_{S_n,2}-i\hbar\gamma_v & 0 & V & 0 & V & 0 & 0 & 0 & 0 \\
  0 & 0 & E_{S_n,3}-i\hbar\gamma_v & 0 & V & V & 0 & 0 & 0 & 0 \\
  V & V & 0 & E_{S_1,1}+E_{S_1,2} & J & J & g & g & 0 & 0 \\
  V & 0 & V & J & E_{S_1,1}+E_{S_1,3} & J & g & 0 & g & 0 \\
  0 & V & V & J & J &  E_{S_1,2}+E_{S_1,3} & 0 & g & g & 0\\
  0 & 0 & 0 & g & g & 0 & E_{S_1,1}+E_c-i\hbar\gamma_c & J & J & \sqrt{2}g \\
  0 & 0 & 0 & g & 0 & g & J & E_{S_1,2}+E_c-i\hbar\gamma_c & J &  \sqrt{2}g \\
  0 & 0 & 0 & 0 & g & g & J & J & E_{S_1,3}+E_c-i\hbar\gamma_c & \sqrt{2}g \\
  0 & 0 & 0 & 0 & 0 & 0 & \sqrt{2}g & \sqrt{2}g & \sqrt{2}g & 2E_c-2i\hbar\gamma_c
\end{array}
\end{pmatrix}
\label{eq:Hamiltonian_matrix}}$}}
\end{equation}

where, to save space, the number of TLMs is selected to be $N=3$. Additionally, deactivation terms $-i\hbar\gamma_c$ and $-2i\hbar\gamma_c$ with $\gamma_c$ being the cavity decay rate, are added to the diagonal elements corresponding to states $|\phi_{S_1,1}\rangle$ and $|\phi_{S_0,2}\rangle$, respectively, to account for photon decay due to the imperfections of the cavity mirrors. We also note that the light-matter coupling terms corresponding to the exchange of a photon between states $|\phi_{S_1,1}\rangle$ and $|\phi_{S_0,2}\rangle$ (\textit{i.e.} the terms in the last row and last column of the Hamiltonian matrix in Equation~\ref{eq:Hamiltonian_matrix}) are multiplied by the square root of two since $\langle2|\langle S_0|\hat{\sigma}\hat{a}^\dagger|S_1\rangle|1\rangle=\sqrt{2}$.

Diagonalisation of the Hamiltonian in Equation~\ref{eq:Hamiltonian_matrix} yields adiabatic light-matter states, which can be divided into three types~\cite{Yuen-Zhou2022,Kowalewski2025}: multi-polariton states with a significant photonic content, dark polariton states with a small but nonzero photonic content, and dark states without cavity contribution (Figure~S1 in the SI). In the absence of energetic disorder, decay channels ($\gamma_v=0$ and $\gamma_c=0$), exciton and EEA couplings ($J=0$ and $V=0$), the splitting between the multi-polariton states is given by $\Omega_\text{MP}=2 g\sqrt{4N-2}\approx4g\sqrt{N}$, \textit{i.e.} approximately twice as high as the Rabi splitting in the first excitation subspace, $\Omega_\text{R}=2g\sqrt{N}$ (Section~2 in the SI). At the same time, the splitting between dark polariton states is slightly smaller than the Rabi splitting and is given by $\Omega_\text{DP}=2g\sqrt{N-2}$.

To simulate the dynamics of polaritons, the total polariton wave function is constructed as a linear combination of the product states:
\begin{equation}
\begin{array}{ccl}
\vert\Psi\left(t\right)\rangle &=&\sum_{i=1}^Nd_i(t)\vert\phi^i_{S_n,0}\rangle + \sum_{i=N+1}^{N(N-1)/2}d_i(t)\vert\phi^i_{2S_1,0}\rangle + \\
\\
&&\sum_{i=N(N-1)/2+1}^{N_\text{st}-1}d_i(t)\vert\phi^i_{S_1,1}\rangle + d_{N_\text{st}}(t)\vert\phi^{N_\text{st}}_{S_0,2}\rangle.
\label{eq:total_wf}
\end{array}
\end{equation}

The expansion coefficients, $d_i(t)$, are propagated by numerical integration of the Schr\"{o}dinger equation over discrete time intervals, $\Delta t$, 
\begin{equation}
d_i(t+\Delta t)=d_i(t)\exp\left(-i\textbf{H}\Delta t/\hbar\right).
\label{eq:SE_diabatic}
\end{equation}

Because of the deactivation terms, $-i\hbar\gamma_v$, $-i\hbar\gamma_c$ and $-i2\hbar\gamma_c$, the norm of the total wave function, $|\Psi(t)|^2=\sum_i^{N_\text{st}}|d_i(t)|^2$ (Equation~\ref{eq:total_wf}), is not conserved and decreases with time. Therefore, population of the ground state (GS), $|\phi_{S_0,0}\rangle\equiv|S_0^1,S_0^2,...,S_0^N\rangle\otimes|0\rangle$, is determined as $P_\text{GS}(t)=1-|\Psi(t)|^2$ at each time step. Furthermore, the total populations of $|\phi_{S_n,0}\rangle$, $|\phi_{2S_1,0}\rangle$, $|\phi_{S_1,1}\rangle$, and $|\phi_{S_0,2}\rangle$ states are defined as $P_{S_n,0}(t)=\sum_{i=1}^N|d_i(t)|^2$, $P_{2S_1,0}(t)=\sum_{i=N+1}^{N(N-1)/2}|d_i(t)|^2$, $P_{S_1,1}(t)=\sum_{i=N(N-1)/2+1}^{N_\text{st}-1}|d_i(t)|^2$, and $P_{S_0,2}(t)=|d_{N_\text{st}}(t)|^2$, respectively.

Equation~\ref{eq:SE_diabatic} describes the evolution of a closed quantum system, in which the decay to the ground state is modelled implicitly with imaginary deactivation terms. To model this process explicitly, as well as to take a weak Markovian interaction with the environment at $T=0K$ into account, the Lindblad master equation can be employed~\cite{Manzano2020,Scala2007}:
\begin{equation}
    \dot{\hat{\rho}} = -\frac{i}{\hbar}\left[\hat{H},\hat{\rho}\right] + \sum_n\left(\hat{L}_n\hat{\rho}\hat{L}_n^\dagger - \frac{1}{2}\{\hat{L}_n^\dagger\hat{L}_n,\hat{\rho}\}\right),
\label{eq:Lindblad_standard}
\end{equation}
where $\hat{\rho}=\sum_j^{N_\text{st}}|\phi_j\rangle\langle\phi_j|$ is the system's density matrix with $|\phi_j\rangle\in\{|\phi_{S_n,0}\rangle\}$, $\{|\phi_{2S_1,0}\rangle\}$, $\{|\phi_{S_1,1}\rangle\}$, $\{|\phi_{S_0,2}\rangle\}$. Operators $\hat{L}_n$ and $\hat{L}_n^\dagger$ are the so-called Lindblad jump operators, which describe the dissipation of energy to the environment and the loss of coherence. Here, we consider the latter effect by including the pure dephasing operator $\hat{L}_\text{deph}=\sqrt{\gamma_\text{deph}}\hat{\sigma}_z$ acting on the states with one TLM excited, \textit{i.e.} $|\phi_{S_n,0}\rangle$ and $|\phi_{S_1,1}\rangle$, and operator $\hat{L}_\text{deph}^\prime=\sqrt{2\gamma_\text{deph}}\hat{\sigma}_z$ acting on the states with two TLMs excited, \textit{i.e.} $|\phi_{2S_1,0}\rangle$, in addition to cavity decay, described by operators $\hat{L}_c=\sqrt{\gamma_c}\hat{a}$ and $\hat{L}_c^\prime=\sqrt{2\gamma_c}\hat{a}$ acting on the states $|\phi_{S_1,1}\rangle$ and $|\phi_{S_0,2}\rangle$, respectively.

\subsection*{Simulation details}\label{section:model_b}

To study how strong coupling modifies the EEA rate in systems with different exciton mobilities and concentrations, we simulated polariton dynamics over a broad range of numbers of molecules, from $N=10$ to $N=100$, and exciton coupling strengths, from $\langle J\rangle=5$~meV to $\langle J\rangle=150$~meV. The maximum value of the exciton coupling corresponded to the time-averaged value obtained from molecular dynamics (MD) simulations of the H-dimer of R6G~\cite{Doveiko2025}, the chromophore molecule used in the experiments of Wu \textit{et al.}~\cite{Pullerits2025} and Qureshi \textit{et al.}~\cite{Qureshi2025} on cavity-modified EEA. This value can be thought of as an extreme case of high molecular concentration leading to the formation of aggregates. The details of the MD simulations are discussed in Section~3 of the SI.

In realistic systems, molecules never have exactly the same energies. Therefore, static disorder was introduced by randomly sampling excitation energies of TLMs from the Gaussian distribution:
\begin{equation}
  p(E)=\frac{1}{\sqrt{2\pi}\sigma_E}\exp\left[-\frac{\left(E-\langle E\rangle\right)^2}{2\sigma_E^2}\right],
\label{eq:guass_distrE}
\end{equation} 
where, for the $S_1$-state, $\langle E\rangle=2.3$~eV was the mean value corresponding to the absorption maximum of R6G in aqueous solutions~\cite{Selwyn1972}, and $\sigma_E$ was the disorder strength, previously estimated to be around $100$~meV from quantum mechanics/molecular mechanics (QM/MM) simulations of Rhodamine molecules in water~\cite{Sokolovskii2023}, and we used this value in all simulations unless otherwise stated. The energy of the $S_n$-state of each TLM was also randomly taken from Equation~\ref{eq:guass_distrE} and then multiplied by two. 

The coupling terms responsible for exciton hops and EEA were also randomly drawn from the Gaussian distribution with the corresponding means $\langle J\rangle$ and $\langle V\rangle$ and standard deviations $\sigma_J$ and $\sigma_V$, respectively. All results presented in the main text were obtained with the following values: $\langle V\rangle=20$~meV, $\sigma_J=10$~meV, $\sigma_V=10$~meV. We note that because there are several high-energy states nearly in resonance with the double energy of the $S_1$-state, $2E_{S_1}$, all of these states can contribute to EEA, effectively increasing the EEA rate compared to a single $S_0\rightarrow S_n$ transition. Furthermore, for some of these states, the transition dipole moments (TDMs) can undergo significant fluctuations during the dynamics, leading to large fluctuations in the corresponding coupling terms (Figure~S2). Therefore, we conducted additional simulations of polariton dynamics with elevated EEA coupling strength and its mean, namely $\langle V\rangle=50$~meV and $\sigma_V=50$~meV, and confirmed that the conclusions drawn in this study are independent of the choice of $\langle V\rangle$ and $\sigma_V$ (Figure~S3). 

The time propagation of the total wave function according to Equation~\ref{eq:SE_diabatic} was run for $t=2$~ps with a time step of $\Delta t=1$~fs. The latter is smaller than the shortest time scale in the dynamics of the light-matter system, namely the period of Rabi oscillations, and therefore is small enough to capture all the relevant physical effects. Indeed, performing simulations with the chosen time step gives the same populations of product states as in simulations with a shorter time step of $\Delta t=0.1$~fs (Figure~S4). In all simulations, the dynamics were initialised in a two-exciton product state $|\phi_{2S_1,0}\rangle$ with the expansion coefficient $d_i(0)=1$, where $N<i\le N(N-1)/2$, and the starting positions of the two excitons were selected either randomly or to correspond to the greatest distance between TLMs in the chain. The energy of the cavity photon was chosen to be in resonance with the average value of the $S_1$-state, \textit{i.e.} $E_c=\langle E\rangle=2.3$~eV. In simulations without cavity decay ($\gamma_c=0$), the collective light-matter coupling strength was set to $g\sqrt{N}=175$~meV, which corresponded to the Rabi splittings achieved for R6G in a metallic and DBR microcavity in the experiments on cavity-modified EEA~\cite{Pullerits2025,Qureshi2025}. In simulations with cavity decay ($\gamma_c\ne0$), the collective coupling strength was reduced to $g\sqrt{N}=100$~meV in order to investigate whether the suppression/enhancement of the EEA rate can be achieved not only in the strong coupling regime but also in the weak coupling regime, \textit{i.e.} when $\gamma_c>\Omega_\text{R}=2g\sqrt{N}=200$~meV $-$ or, equivalently, $\tau_c<1/\Omega_\text{R}=21$~fs, with $\tau_c=1/\gamma_c$ being the cavity lifetime. Accordingly, the excitation energy disorder strength was reduced to $\sigma_E=57.14$~meV to retain the same ratio between the collective light-matter coupling strength and the disorder strength, $g\sqrt{N}/\sigma=1.75$, as in simulations without cavity decay. For the deactivation rate of the $S_n$-state, we chose a value of $\gamma_v=41.36$~meV, corresponding to a lifetime of $\tau_v=100$~
fs, which is within the sub-picosecond timescale typical for internal conversion of organic dyes to the $S_1$-state~\cite{Birks1973}. All simulations were repeated $1000$ times with different realisations of the disordered Hamiltonian ($\sigma_E$, $\sigma_J$ and $\sigma_V$) to ensure convergence of the populations of the product states and of the ground state. 

Simulations of molecular systems with dephasing in the Lindblad formalism were performed with $N=10$ and $20$ TLMs for $t=2$~ps with a time step of $\Delta t=50$~as and averaged over $100$ disorder realisations. Because the size of the density matrix in Equation~\ref{eq:Lindblad_standard} scales as $N_\text{st}^2$ with the number of product states, modelling molecular ensembles with $N>20$ becomes computationally prohibitive (see Section~5.2 in the SI for discussion). To avoid this problem and perform simulations of larger molecular ensembles, instead of solving the Lindblad master equation, the total wave function can be propagated by numerical integration of the Schr\"{o}dinger equation (Equation~\ref{eq:SE_diabatic}) with stochastic Markovian jumps associated with the Lindblad operators~\cite{Dalibard1992,Molmer1993,Herrera2022,Tremblay2022,Davidsson2023}. This so-called Monte Carlo wave function (MCWF) method allowed us to run the dynamics of $N=50$ TLMs with the inclusion of dephasing. Because the method requires averaging not only over disorder realisations but also over stochastic quantum trajectories, the number of runs was increased to 10000. In Section~5 of the SI, we provide a detailed description of simulations with both the Lindblad master equation and MCWF, as well as validate the MCWF approach against a direct solution of the Lindblad master equation for a system with $N=10$.

Simulations of polariton dynamics were conducted with our home-built Python script. The geometry optimisation of the H-dimer of R6G was performed in ORCA 6.0~\cite{ORCA}, and the MD simulations were performed in NAMD 3.0~\cite{NAMD}. The coupling terms $J$ and $V$ were calculated using the transition electrostatic potential charges, which were in turn computed with Multiwf~\cite{Tian2012,Tian2024}, and the atomic coordinates were extracted from MD simulations with MDAnalysis~\cite{Michaud2011,Gowers2016}.

\bigskip

\section*{Results and discussion}

\bigskip

\begin{figure*}[!htb]
\centering
\includegraphics[width=1\textwidth ]{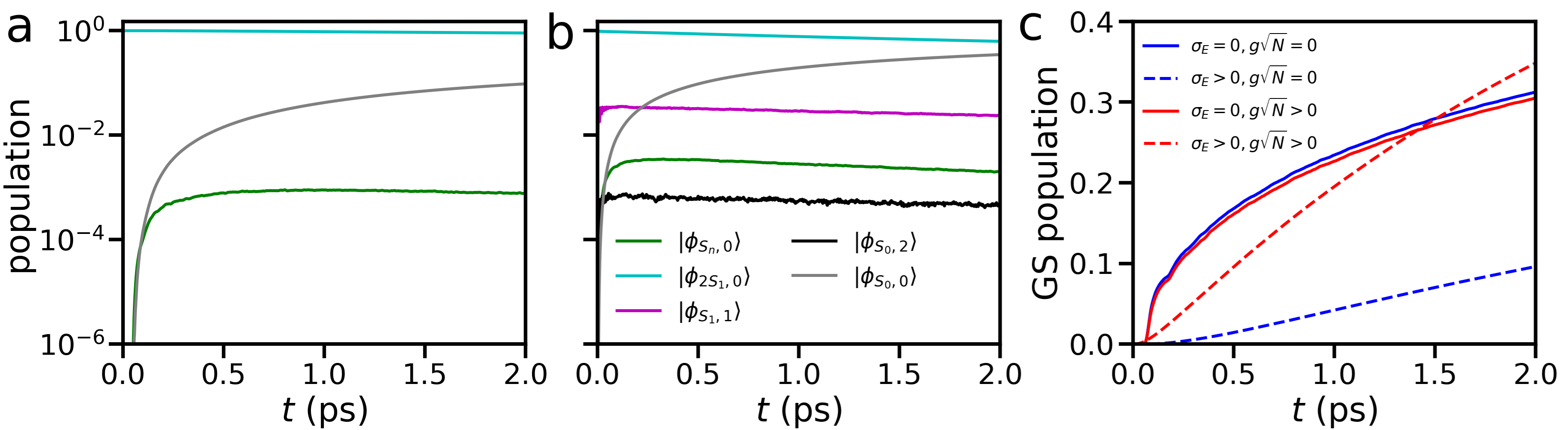}
  \caption{Panels~\textbf{a}~and~\textbf{b}: total populations of all $\vert\phi_{S_n,0}\rangle$ states (green), $\vert\phi_{2S_1,0}\rangle$ states (cyan), $\vert\phi_{S_1,1}\rangle$ states (purple), as well as of the $\vert\phi_{S_0,2}\rangle$ state (black) and the ground state, $\vert\phi_{S_0,0}\rangle$ (grey), in simulations of polariton dynamics in a system of $N=50$ molecules with an exciton coupling strength of $\langle J\rangle=70$~meV outside ($g\sqrt{N}=0$~meV, \textbf{a}) and inside the cavity ($g\sqrt{N}=175$~meV, \textbf{b}). Panel~\textbf{c}: Comparison between the ground state populations in simulations of ordered ($\sigma_E=0$, solid lines) and disordered ($\sigma_E>0$, dashed lines) molecular ensembles inside the cavity ($g\sqrt{N}>0$, red) and outside the cavity ($g\sqrt{N}=0$, blue). The data corresponding to the molecules with disorder (dashed lines) were replotted from panels~\textbf{a}~and~\textbf{b} (grey lines). All lines are averages over $1000$ realisations of the disordered Hamiltonian (Equation~\ref{eq:Hamiltonian_matrix}). Note that the populations in panels~\textbf{a}~and~\textbf{b} are plotted in logarithmic scale.
  }
  \label{fig:figure2}
\end{figure*}

We first explore the influence of strong coupling on the EEA rate by performing simulations of polariton dynamics in a system of $N=50$ TLMs with the exciton coupling strength $\langle J\rangle=70$~meV in a cavity without decay ($\gamma_c=0$). In this case, the only deactivation channel is decay of $S_n$-excitons, and the GS population thus serves as a measure of the efficiency of EEA. Figure~\ref{fig:figure2}\textbf{a} shows the total populations of the two-$S_1$-exciton states ($\vert\phi_{2S_1,0}\rangle$, cyan), the one-$S_n$-exciton states ($\vert\phi_{S_n,0}\rangle$, green), and the ground state ($\vert\phi_{S_0,0}\rangle$, grey) as a function of time in simulations outside the cavity (\textit{i.e.} $g\sqrt{N}=0$). After initial excitation into the $\vert\phi_{2S_1,0}\rangle$ state with one exciton created in molecule $i=13$ and the other exciton created in molecule $j=37$, \textit{i.e.} at the maximum possible distance from each other, the total population of the states with two $S_1$-excitons slowly decreases with the concomitant build-up of the states with one $S_n$-exciton. The latter allows the population to decay to the ground state via deactivation terms $-i\hbar\gamma_v$, resulting in a growth of the population of this state. Noteworthy, the build-up of the $\vert\phi_{S_n,0}\rangle$ and $\vert\phi_{S_0,0}\rangle$ populations does not follow the excitation immediately but starts after a delay of around 50~fs. This delay reflects the time the excitons need to travel before getting sufficiently near for the EEA to become efficient (Figure~S5).

Placing the TLMs inside the cavity (\textit{i.e.} $g\sqrt{N}=175$~meV) opens up relaxation channels involving the hybrid light-matter states, resulting in a non-zero population of the $\vert\phi_{S_1,1}\rangle$ states (Figure~\ref{fig:figure2}\textbf{b}, purple) and the $\vert\phi_{S_0,2}\rangle$ state (black). Importantly, the GS population increases faster than outside the cavity (dashed lines in Figure~\ref{fig:figure2}\textbf{c}), indicating an enhancement of the EEA rate due to strong coupling. In contrast to the EEA process outside the cavity, where disorder disrupts the excitation energy transfer and hence protects excitons against EEA, there is no such protection in the cavity. Through polariton formation, the cavity provides an additional transfer channel that is robust against disorder. Because a faster increase in the GS population inside the cavity is not observed in a system without excitation energy disorder (i.e. when $\sigma_E=0$, solid lines in Figure~\ref{fig:figure2}\textbf{c}), we attribute cavity-enhanced EEA in a disordered system to the ability of the cavity to provide additional connectivity between excitons via interaction with the common electromagnetic mode, which allows the excitons to partially overcome disorder.

We would like to emphasize that the enhancement of the EEA rate cannot be explained simply by Rabi oscillations arising from exciton-cavity coupling, which can occasionally populate the states with two neighbouring excitons. Indeed, as shown in Figure~S6a for a simulation with suppressed exciton coupling, $J=0$, in which EEA can only arise from the interaction of excitons with the common cavity mode, Rabi oscillations make only a minor contribution to the GS population. Moreover, this contribution quickly decreases with the number of molecules (Figure~S6b,c). This is because the number of pairs of neighbouring excitons grows linearly with $N$, whereas the total number of available two-$S_1$-exciton states grows as $N(N-1)/2\propto N^2$. Thus, the probability of targeting a state with two neighbouring excitons within one Rabi cycle is $\propto1/N$, which falls off rapidly with increasing $N$.

It is important to note that in the ideal case with no disorder, when $\sigma_E=\sigma_J=\sigma_V=0$~meV, the GS population in the cavity is actually slightly smaller than in the case of bare molecules (solid lines in Figure~\ref{fig:figure2}\textbf{c}), which is in line with the theoretical model of Siltanen \textit{et al.}, in which an idealised molecule-cavity system was considered~\cite{Siltanen2025}. This slight suppression of EEA is because part of the $\vert\phi_{2S_1,0}\rangle$ population that is transferred to the $\vert\phi_{S_n,0}\rangle$ states in the absence of strong coupling is transferred to the $\vert\phi_{S_1,1}\rangle$ and $\vert\phi_{S_0,2}\rangle$ states in the presence of strong coupling instead. Only in a disordered system does the enhancement of the EEA rate become possible.

Another curious observation that can be made from Figure~\ref{fig:figure2}\textbf{c} is the difference in the character of the GS population build-up. In the absence of disorder (solid lines), the build-up is uneven and slows down after a rapid initial increase, whilst in the presence of disorder (dashed lines), nearly linear growth of the GS population is observed. The latter is because in addition to the disorder-induced localisation of the exciton population at the position of the two initially excited TLMs, an extended population uniformly distributed over the whole molecular chain is formed when TLMs are strongly coupled to the cavity (Figure~S7a,b). This results in a nonzero exciton coupling between any pair of neighbouring TLMs, which allows for EEA to continuously occur. Due to the linear increase in the disordered TLM-cavity system in contrast to the sub-linear increase in the system without disorder, the GS population in the former case eventually overtakes the GS population in the case of no disorder ($\approx1.4$~ps for $\langle J\rangle=70$~meV, Figure~\ref{fig:figure2}\textbf{c}). Interestingly, this occurs earlier in the strong light-matter coupling regime as the exciton coupling strength $J$ increases (Figure~S7c). This can be explained by the fact that, while the GS population slightly increases with $\langle J\rangle$ in a disordered system, as we show below, the GS population decreases with $\langle J\rangle$ in the system without disorder. The latter is because an increase in exciton mobility reduces the time that excitons spend near each other, which leads to a decrease in the probability of exciton interaction. As a consequence, the GS population grows slower with increasing $\langle J\rangle$, in contrast to the disordered system (Figure~S7c).

\begin{figure*}[!htb]
\centering
\includegraphics[width=1\textwidth ]{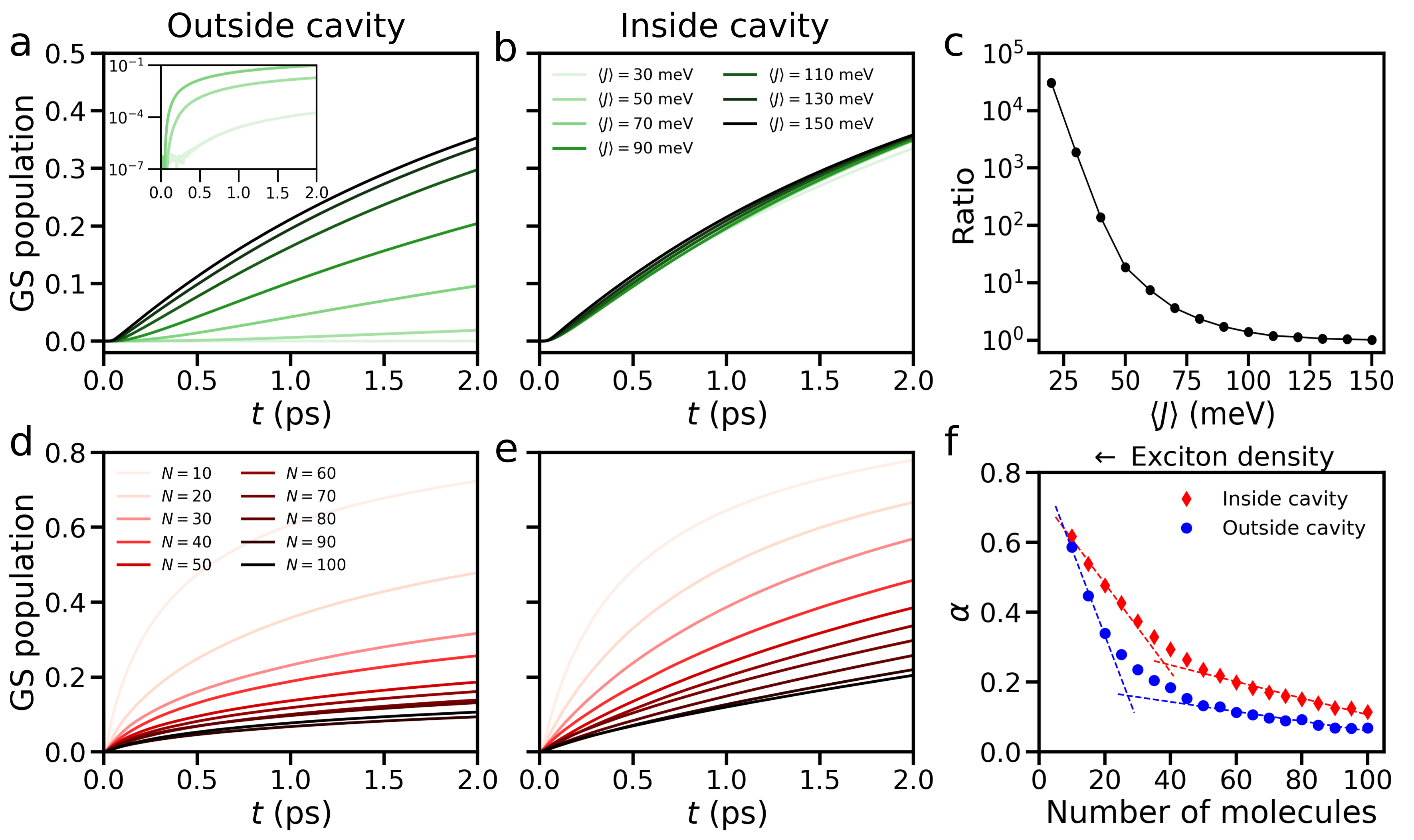}
  \caption{Panels~a~and~b: Population of the ground state (GS), $\vert\phi_{S_0,0}\rangle$, in simulations with variable coupling strength, $\langle J\rangle$, and fixed number of molecules, $N=50$, outside (a) and inside the cavity (b). The inset in panel~a depicts the logarithmic scale plots of the GS population in simulations with $\langle J\rangle=30$~meV, $50$~meV, and $70$~meV. Panel~c: Ratio between the GS populations inside and outside the cavity at the end of the simulation, \textit{i.e.} $R=P^\text{inside}_\text{GS}(t^\prime)/P^\text{outside}_\text{GS}(t^\prime)$ with $t^\prime=2$~ps. Panels~d~and~e: the GS population in simulations with variable number of molecules and fixed exciton coupling strength, $\langle J\rangle=50$~meV, outside (d) and inside the cavity (e). Panel~f: Dependence of $\alpha$ on the number of molecules, $N$, inside (red diamonds) vs. outside the cavity (blue circles), with $\alpha$ extracted from fitting the GS population at different values of $N$ to the function $P_\text{GS}(t)=\alpha t^\beta$. The dashed lines are fits to a linear function at small and large numbers of molecules. The intersection of the lines indicates the onset of EEA, which occurs at a larger $N$ (smaller exciton density) when molecules are strongly coupled to the cavity. All lines in panels a-b and d-e are averages over $1000$ realisations of the disordered Hamiltonian (Equation~\ref{eq:Hamiltonian_matrix}).
  }
  \label{fig:figure3}
\end{figure*}

To further elaborate on the interplay between disorder, exciton mobility and polariton formation, we performed simulations at different exciton coupling strengths between $\langle J\rangle=20$~meV and $\langle J\rangle=150$~meV in a system of $N=50$ TLMs, to gradually transition between the regimes of poor exciton mobility ($\langle J\rangle\ll\sigma_E$ with $\sigma_E=100$~meV), moderate exciton mobility ($\langle J\rangle\leq\sigma_E$), and high exciton mobility ($\langle J\rangle>\sigma_E$). The standard deviation of the exciton coupling distribution was kept $\sigma_J=10$~meV for all values of $\langle J\rangle$. Figure~\ref{fig:figure3}\textbf{a} demonstrates the time-dependence of the GS population for different values of $\langle J\rangle$ in the absence of strong coupling. When exciton coupling is much smaller than the disorder strength, $\langle J\rangle\ll\sigma_E$, exciton propagation is suppressed due to Anderson localisation (Figure~S8a), resulting in negligible annihilation and therefore the GS population remains low (light line in the inset of Figure~\ref{fig:figure3}\textbf{a}). Increasing $\langle J\rangle$ improves exciton transport (Figure~S8b,c) and increases the probability of exciton interaction. As a result, the GS population also increases with $\langle J\rangle$, reaching the value of $\approx0.35$ after two picoseconds in the case of the highest coupling strength of $\langle J\rangle=150$~meV.

In contrast, the GS population experiences very little dependence on the exciton coupling strength under strong light-matter coupling and in all cases is similar to the highest GS population outside the cavity at $\langle J\rangle=150$~meV (Figure~\ref{fig:figure3}\textbf{b}). There, some of the probability amplitude of the total wave function (Equation~\ref{eq:total_wf}) associated with the exciton population, delocalises over the molecular chain (Figure~S7a,b), in which case the highest EEA rate can be expected due to continuous interaction via the coupling term $V$ (Equation~\ref{eq:Hamiltonian_matrix}). Therefore, the role of the cavity in enhancing the EEA rate is in expanding the exciton population over many molecules through their interaction with the common electromagnetic mode, which provides better connectivity between excitons and hence more efficient interaction. We note that with two excitons in a chain of $N=50$ molecules considered in these simulations, the exciton density was $4\%$, which was close to an exciton density of $1\%$ achieved in a study of EEA in J-aggregates of a cyanine dye~\cite{Akselrod2010}. Therefore, we expect the picture described above to hold in realistic molecule-cavity systems.

By expanding the exciton population over many molecules, the cavity allows for efficient EEA even in the case of the smallest exciton coupling, in contrast to uncoupled TLMs, in which the exciton population is localised on the initially excited molecules due to disorder (Figure~S7b). To quantify the extent of exciton expansion, we calculated the escape probability of the excitonic wave packet as~\cite{Aroeira2024}
\begin{equation}
\chi(t) = 1 - \frac{\sum_i\sum_{j=i-2}^{i+2}P^\text{exc}_j}{\sum_{k=1}^N P^\text{exc}_k},
\label{eq:esc_prob}
\end{equation}
where $P^\text{exc}_j$ is the exciton probability density in TLM $j$ (see Section~4.2 in the SI for details), and the terms in the numerator on the right-hand side of Equation~\ref{eq:esc_prob} are summed over a narrow range around the initially excited TLMs, which we chose to be the TLMs with indices $i=13$ and $i=37$. An escape probability of zero corresponds to the situation where the entire population is in the initially excited TLMs, whereas $\chi=1$ indicates that no population remains in these TLMs. That is, the escape probability estimates the likelihood that the exciton wave packet will propagate beyond the initially excited spot. We found that the time-averaged escape probability increased gradually from $\overline{\chi}=0.09$ to $\overline{\chi}=0.69$ in the exciton coupling range from $\langle J\rangle=20$~meV to $\langle J\rangle=150$~meV in the absence of light-matter coupling (Figure~S9). In contrast, the escape probability increases only slightly between $\overline{\chi}=0.63$ and $\overline{\chi}=0.71$ within the same range of $\langle J\rangle$ in the strong coupling regime. The different character of the escape probability between $g\sqrt{N}=0$ and $g\sqrt{N}>0$ is consistent with the difference in the GS population observed in Figure~\ref{fig:figure3}\textbf{a},\textbf{b} and further supports our claim that strong coupling provides delocalisation of excitons, which allows for efficient EEA even in systems with weak exciton interaction compared to the disorder strength.

In Figure~\ref{fig:figure3}\textbf{c} we summarise the results presented in Figure~\ref{fig:figure3}\textbf{a},\textbf{b} by plotting the ratio between the GS populations in the presence and absence of strong light-matter coupling, $R=P^\text{inside}_\text{GS}(t^\prime)/P^\text{outside}_\text{GS}(t^\prime)$, after $t^\prime=2$~ps of simulation. At high exciton mobilities, $\langle J\rangle>\sigma_E$, excitons are already well connected, and the cavity has no effect on the EEA rate, resulting in $R\approx1$. As $\langle J\rangle$ decreases, exciton propagation outside the cavity becomes increasingly suppressed due to excitation energy disorder, and Anderson localisation of excitons takes place when $\langle J\rangle\ll\sigma_E$. On the contrary, the suppression of exciton spreading is mitigated inside the cavity due to strong coupling, resulting in enhanced EEA rates and substantial increase in $R$ (\textit{e.g.}, $R>10^4$ at $\langle J\rangle=20$~meV and $\sigma_E=100$~meV), indicating that strong coupling can significantly boost EEA in systems with low exciton mobility. 

It is known that exciton annihilation manifests itself when the density of excitons reaches a certain threshold~\cite{Mikhnenko2015}. Because the EEA rate is enhanced in a perfect cavity without decay, we can anticipate the onset of EEA to occur at a lower exciton density than in bare molecules. To test this, we performed simulations with a variable number of TLMs, ranging from $N=10$ to $N=100$, with a fixed exciton coupling strength of $\langle J\rangle=50$~meV. This range of the number of TLMs corresponds to an exciton density of $2\%-20\%$. This is higher than a critical density of $0.1\%-1\%$, typical for organic films with a critical exciton concentration around $10^{18}-10^{19}$~cm$^{-3}$ and a molecular concentration of $10^{21}$~cm$^{-3}$~\cite{Pope1999,Brutting2005,Plint2025}. Nevertheless, as we demonstrate below, it is still possible to qualitatively compare the threshold exciton densities in molecules inside and outside the cavity, even if the exact numbers may be overestimated. Because the length of the molecular chain was different in the simulations with variable $N$, the initial state was chosen to correspond to two $S_1$-excitons located at random positions in each realisation of the disordered Hamiltonian (Equation~\ref{eq:Hamiltonian_matrix}). Figure~\ref{fig:figure3}\textbf{d},\textbf{e} shows the time-dependence of the GS population at different values of $N$ outside and inside the cavity. In both cases, the GS population builds up faster when the number of TLMs is smaller $-$ or, equivalently, exciton density is higher $-$ since excitons become more likely to meet and interact.

The onset of annihilation is experimentally identified as a breakdown of the linear relationship between excitation fluence, which is proportional to the exciton density, and a measured observable, which is proportional to the excited-state population~\cite{Qureshi2025}. Here, we imitate this approach by fitting the GS populations in Figure~\ref{fig:figure3}\textbf{d},\textbf{e} to the power-law function, $P_\text{GS}(t)=\alpha t^\beta$, and plotting the extracted coefficient $\alpha$, which quantifies the speed with which the GS population grows, as a function of the number of molecules in Figure~\ref{fig:figure3}\textbf{f}. In the plot, we can distinguish two regimes, characterised by a linear increase in $\alpha$ with different slopes, shown as dashed lines. The threshold exciton density was then estimated from the intersection of these lines and found to be reduced under strong coupling with a lossless cavity ($\gamma_c=0$). Specifically, the threshold exciton density~$\rho$ (number of molecules) was found to be $\rho=5.6\%$ ($N_\text{th}=36$) in the cavity and $\rho=7.7\%$ ($N_\text{th}=26$) in the absence of strong coupling. This result provides an additional argument to the statement that EEA can be enhanced in an ideal cavity. A qualitatively similar result was also found by plotting the $\beta$ coefficient with respect to the number of molecules and extracting the threshold from the intersection of linear fits at small and large $N$ (Figure~S10).

\begin{figure*}[!htb]
\centering
\includegraphics[width=1\textwidth ]{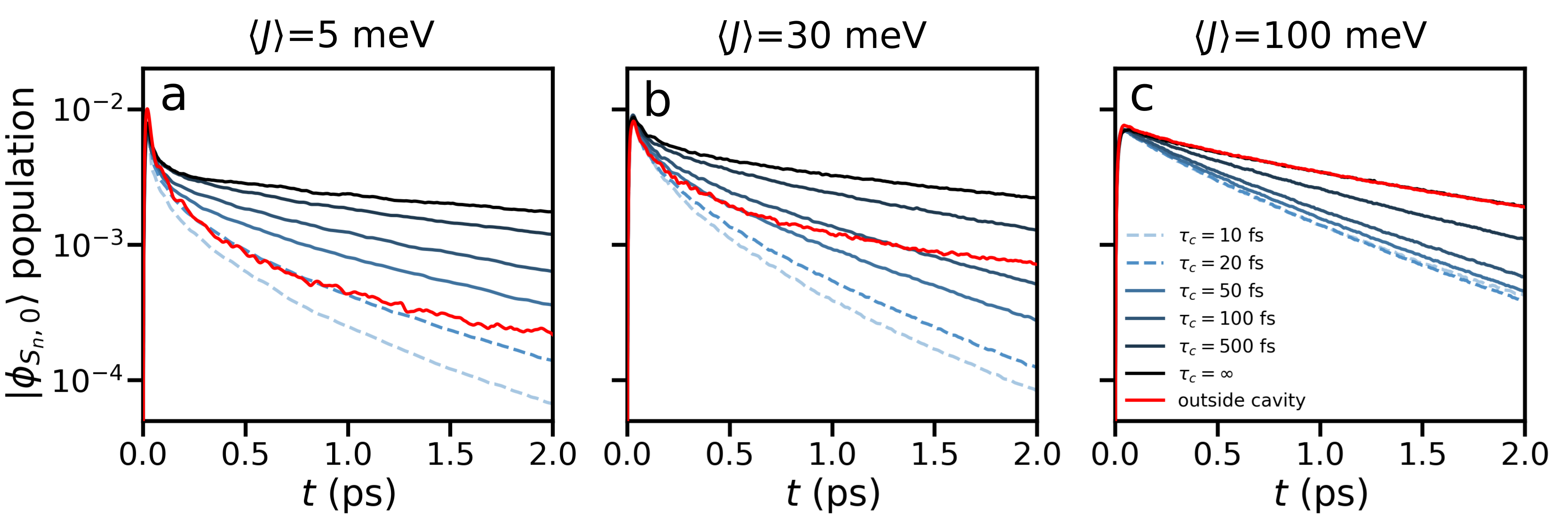}
  \caption{Total population of the $\vert\phi_{S_n,0}\rangle$ states in simulations of polariton dynamics in a system of $N=50$ three-level molecules with an exciton coupling strength of $\langle J\rangle=5$~meV (a), $\langle J\rangle=30$~meV (b), and $\langle J\rangle=100$~meV (c), and the cavity lifetime ranging between $\tau_c=10$~fs and $\tau_c=500$~fs, as well as in simulations in the ideal cavity with no decay ($\gamma_c=1/\tau_c=0$) and outside the cavity. All lines are averages over $1000$ realisations of the disordered Hamiltonian (Equation~\ref{eq:Hamiltonian_matrix}).}
  \label{fig:figure4}
\end{figure*}

So far, we have considered the TLMs in a perfect cavity without decay, \textit{i.e.} $\gamma_c=0$. However, real cavities have a limited lifetime, $\tau_c$, for the electromagnetic modes they support. To explore how the EEA rate changes with the quality of the cavity, we performed simulations in a variety of cavities with different lifetimes, corresponding to both the weak coupling regime ($\tau_c<1/\Omega_\text{R}=21$~fs) and the strong coupling regime ($\tau_c>21$~fs)~\cite{Torma2015}. Because in addition to the deactivation of the $\vert\phi_{S_n,0}\rangle$ states a decay channel described by the terms $-i\hbar\gamma_c$ and $-2i\hbar\gamma_c$ exists (Equation~\ref{eq:Hamiltonian_matrix}), we cannot associate the GS population with EEA only. Instead, because the efficiency of EEA depends on the population of the $\vert\phi_{S_n,0}\rangle$ states, we plot this population in Figure~\ref{fig:figure4} for systems with three different values of the exciton coupling strength, namely $\langle J\rangle=5$~meV, $30$~meV, and $100$~meV. 

For the lowest intermolecular coupling strength ($\langle J\rangle=5$~meV), the population of the $S_n$-excitonic states under strong light-matter coupling exceeds the population outside the cavity (blue and red solid lines in Figure~\ref{fig:figure4}\textbf{a}), which points to an enhancement of EEA as for the ideal cavity (Figures~\ref{fig:figure2}~and~\ref{fig:figure3}). In contrast, under weak light-matter coupling conditions, the population of the $S_n$-excitonic states is reduced  compared to bare molecules (blue dashed lines in Figure~\ref{fig:figure4}\textbf{a}), implying suppression of EEA. As the intermolecular exciton coupling increases to $\langle J\rangle=30$~meV, the effect of light-matter hybridisation on EEA diminishes, and the population of the $\vert\phi_{S_n,0}\rangle$ states is lower inside the cavity than outside the cavity for cavities with $\tau_c\le100$~fs (Figure~\ref{fig:figure4}\textbf{b}). Only in high Q-factor cavities with $\tau_c>100$~fs does the enhancement of the EEA rate remain possible. Upon increasing $\langle J\rangle$ even further, the outside-cavity population of states with one $S_n$-exciton reaches the population in the ideal cavity with $\tau_c\rightarrow\infty$ (red and black lines in Figure~\ref{fig:figure4}\textbf{c}). Because a finite cavity lifetime reduces the EEA efficiency, we conclude that placing materials with high intrinsic exciton mobility inside the cavity will suppress rather than enhance EEA.

Thus, the results of our simulations suggest that the overall effect of strong light-matter coupling on the rate of exciton annihilation depends on the relationship between material properties, in particular the exciton coupling strength $J$ and the disorder strength $\sigma_E$, and the cavity lifetime, $\tau_c$. In contrast, under weak coupling, the EEA rate appears to always be reduced due to cavity decay, at least in the case of a small number of molecules of $N\le100$ as considered in this work.

Recently, Philipp \textit{et al.} utilised Fermi's Golden rule to estimate the EEA rate in the strong coupling regime~\cite{Mitric2026}. The authors proposed that the cavity provides an additional source of exciton annihilation due to a resonant interaction between the $S_1\rightarrow S_n$ molecular transition and the cavity photons. To explore the relevance of this so-called exciton-photon annihilation in our system with molecular parameters characteristic of R6G, we incorporated coupling terms $\tilde{g}$ between the $|\phi_{S_1,1}\rangle$ and $|\phi_{S_n,0}\rangle$ states in the system's Hamiltonian (Equation~\ref{eq:Ham_Rabi}) and repeated the simulation of $N=50$ TLMs with exciton coupling strength $\langle J\rangle=50$~meV under strong coupling without cavity decay ($\gamma_c=0$). Because the TDM $|\boldsymbol{\mu}_{1\rightarrow n}|$ corresponding to the $S_1\rightarrow S_n$ transition is 28 times smaller than the TDM $|\boldsymbol{\mu}_{0\rightarrow1}|$ corresponding to the $S_0\rightarrow S_1$ transition in the optimised geometry of the R6G monomer, and six times smaller in the optimised geometry of the R6G H-dimer at the Time-Dependent Density Functional level of theory~\cite{Hohenberg1964,Runge1984} (Section~4.8 in the SI), the coupling strength responsible for exciton-photon annihilation was set to $\tilde{g}=g/28$ and $\tilde{g}=g/6$. In both situations, the resulting ground state population was found to be the same as in the absence of exciton-photon annihilation (Figure~S11), pointing to a negligible contribution of this effect to the total EEA rate in our system. We note, however, that in molecules with comparable oscillator strengths for the $S_0\rightarrow S_1$ and $S_1\rightarrow S_n$ transitions, \textit{i.e.} when $\tilde{g}\approx g$, exciton-photon annihilation is no longer negligible and should be taken into account when estimating the total rate of EEA (Figure~S11). 

\begin{figure}[!htb]
\centering
\includegraphics[width=0.6\textwidth ]{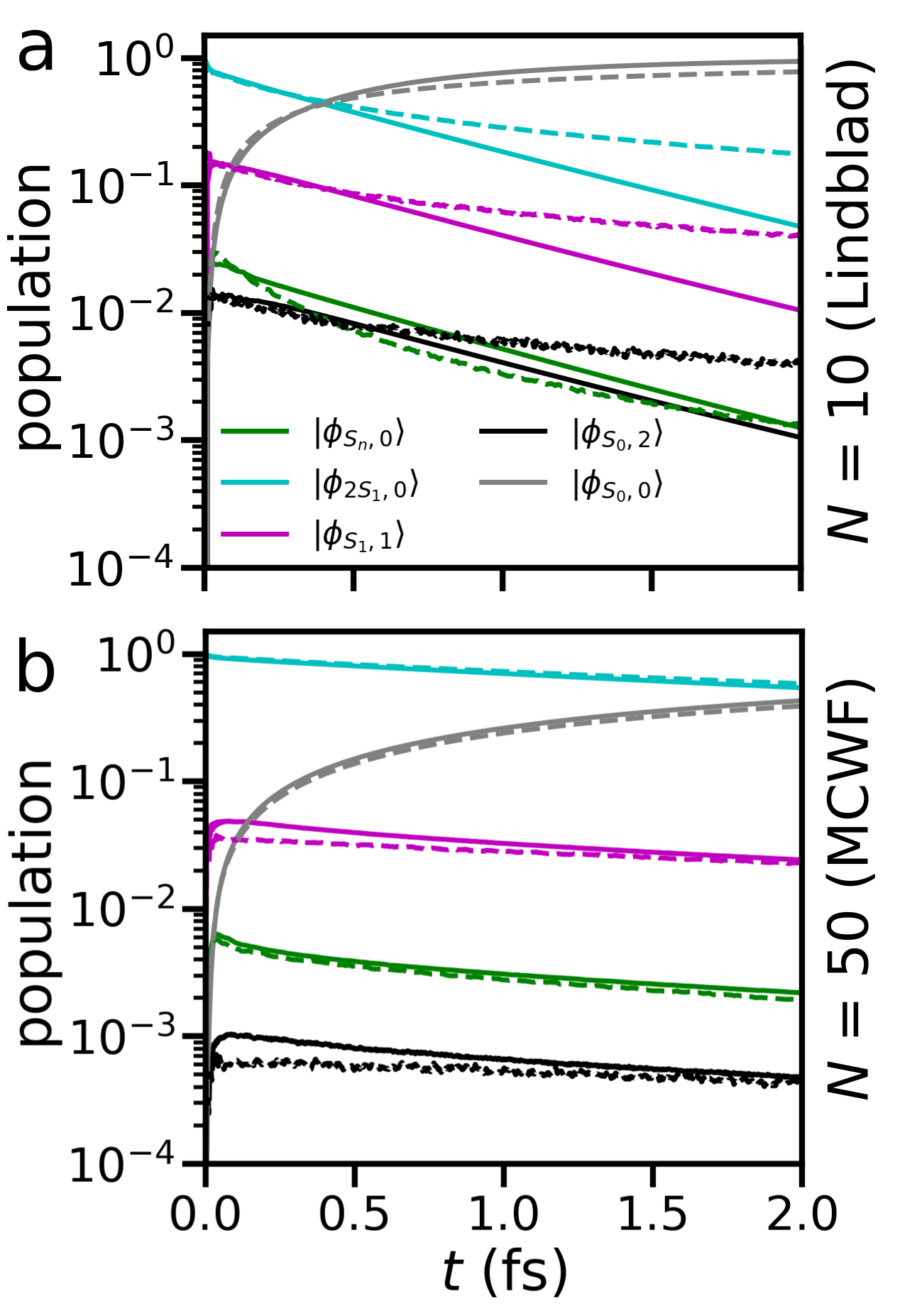}
  \caption{Total populations of all $\vert\phi_{S_n,0}\rangle$ states (green), $\vert\phi_{2S_1,0}\rangle$ states (cyan), $\vert\phi_{S_1,1}\rangle$ states (purple), as well as of the $\vert\phi_{S_0,2}\rangle$ state (black) and the ground state, $\vert\phi_{S_0,0}\rangle$ (grey), in simulations of polariton dynamics in a system of $N=10$ (panel~\textbf{a}) and $N=50$ (panel~\textbf{b}) molecules with an exciton coupling strength of $\langle J\rangle=50$~meV and a collective light-matter coupling strength of $g\sqrt{N}=175$~meV. The dashed lines correspond to the numerical integration of the Schr{\"o}dinger equation without dephasing (Equation~\ref{eq:SE_diabatic}), whereas the solid lines correspond to simulations with dephasing ($\tau_\text{deph}=100$~fs), performed by solving the Lindblad master equation (Equation~\ref{eq:Lindblad_standard}) in the case of $N=10$ and by the Monte-Carlo wave function propagation in the case of $N=50$. The lines corresponding to numerical integration of the Schr{\"o}dinger equation are averages over $1000$ realisations of the disordered Hamiltonian (Equation~\ref{eq:Hamiltonian_matrix}), whereas the lines corresponding to the Lindblad and MCWF method are averages of $100$ and $10000$ realisations, respectively.}
  \label{fig:figure5}
\end{figure}

Finally, we estimate the influence of pure molecular dephasing on the EEA rate arising from the weak interaction of molecules with the environment. Figure~\ref{fig:figure5} shows the populations of product states in the strong light-matter coupling regime in simulation of $N=10$ TLMs obtained by directly solving the Lindblad master equation (Equation~\ref{eq:Lindblad_standard}, panel~\textbf{a}), and $N=50$ TLMs obtained with the MCWF method (panel~\textbf{b}), which provides reasonable agreement with the former method (Figure~S13). Compared to the isolated system, the initially excited states with two $S_1$-excitons, \textit{i.e.}, $|\phi_{2S_1,0}\rangle$, exhibit a stronger depopulation towards the ground state in the system with dephasing. However, this effect decreases with increasing number of molecules, resulting in a smaller discrepancy in the total population of all kinds of product states. In particular, the deviation of the GS population after 2~ps of simulation with dephasing from simulation without dephasing reduces from $\approx18\%$ for $N=10$ to $\approx9\%$ for $N=50$. Importantly, this deviation does not change the conclusion that the EEA rate can either be increased or reduced depending on the magnitudes of the disorder strength and exciton coupling strength, as well as the cavity Q-factor. This is illustrated in Figure~S14, which shows a gradual transition from suppression to enhancement of EEA with increasing cavity lifetime for a system with $\langle J\rangle=30$~meV, in line with Figure~\ref{fig:figure4}\textbf{b}.

\clearpage

\section*{Conclusions and Outlook}

\bigskip

In this work, we performed numerical simulations of exciton-exciton annihilation in the regime of strong light-matter coupling. Although the presence of the cavity does not directly influence exciton-exciton interactions~\cite{Li2025}, \textit{i.e.} coupling terms $J$ and $V$, the EEA rate is modified by two effects related to the matter and photonic components of the resulting exciton–polaritons. These effects are, respectively, (i) \textit{cavity-enhanced exciton delocalisation}, which allows for better connectivity between excitons with poor and moderate mobilities via interaction with the common cavity light modes, and (ii) \textit{decay into the ground state} through the photonic component of the polaritons. Whereas the first effect increases the EEA rate, the second effect competes with exciton annihilation and hence reduces its efficiency. Thus, whether the strong light-matter coupling enhances or suppresses EEA depends on both the matter and cavity components. Matter properties that influence this trade-off are exciton coupling strength and the strength of excitation energy  disorder, while for the cavity the Q-factor is the most important.

As was discussed in the Introduction, experiments reported both suppression and enhancement of EEA in organic and inorganic materials under strong coupling. Based on the results of our simulations, we suggest that these discrepancies might stem from different parameter regimes characterising the materials and cavities in the respective experiments, \textit{i.e.} $\langle J\rangle, \sigma_E$ and $\tau_c$. For example, in the experiment of Qureshi \textit{et al.} in R6G~\cite{Qureshi2025}, which showed suppression of EEA under strong coupling in contrast to the experiment conducted by Wu \textit{et al.} with the same material~\cite{Pullerits2025}, the formation of J- and H-aggregates was reported. This could have led to better exciton mobility, resulting in a diminished effect of strong coupling on exciton annihilation (Figure~\ref{fig:figure3}) and, consequently,  on the overall suppression of EEA due to cavity decay despite a higher Q-factor cavity compared to the experiments of Wu \textit{et al}.

In addition, it was suggested that a high effective group velocity could bring some exciton-polaritons away from the excitation spot and locally depopulate the exciton density. While, in principle, such polariton-mediated transport can be described with our model~\cite{Sokolovskii2023}, doing this for multiple excitons in the second excitation subspace becomes computationally prohibitive: the total number of states increases from $N_\text{st}=5151$ in the case of a single mode and 100~TLMs to $N_\text{st}=20100$ in the case of 100 cavity modes and 100 TLMs. Nevertheless, in the future we will aim to explore this effect of in-plane polariton transport on EEA with a more efficient algorithm for the wave function's propagation. 

For experimental validation of our finding that EEA can be both suppressed and enhanced depending on the relationship between the $\langle J\rangle/\sigma_E$ ratio and the cavity lifetime, $\tau_c$, a good platform could be a material with moderate exciton mobility. By placing this material in cavities with increasing Q-factors, a gradual transition from suppression to enhancement of the EEA rate would confirm the results of our simulations with $\langle J\rangle=30$~meV (Figure~\ref{fig:figure4}\textbf{b}).

In contrast to strong coupling, in the weak coupling regime we observed a suppression of the EEA rate regardless of the exciton mobility, which is due to the competing cavity decay channel. This result appears to contradict the experiment of Wu \textit{et al.}, who observed that the lifetime of transient transmission deviated from a constant value at lower pump intensities for both strong and weak coupling compared to bare films of LH2 complexes and R6G dye~\cite{Pullerits2025}. Although the authors explained this result by enhanced EEA, an alternative explanation based on an increase in the intensity of absorbed light at weak coupling is also possible. This necessitates further experimental research to elucidate whether the enhancement of the EEA rate is possible in the weak coupling regime.

Regarding achieving BEC and lasing of polaritons, the best option would be to choose a material that does not undergo EEA at all yet can strongly couple to cavity light modes~\cite{Akselrod2010}. However, given that polariton lasers require relatively high external pumping, at which EEA occurs in most molecules, it could be very difficult to find such a material. Nevertheless, by carefully designing organic semiconductors with favourable molecular packing~\cite{Khairutdinov1997,Pandya2021} or quantum phase relationship between exciton wave functions~\cite{Tempelaar2017,Huang2023}, or by choosing materials with a mismatch between the energies of the $S_1$ and $S_n$ states, the effect of EEA can be mitigated. Moreover, such materials with low exciton annihilation propensity can be combined with cavity structures that allow for polariton lifetimes to be longer than the rates of polariton-phonon relaxation~\cite{Sfeir2024,He2026}. Eventually, this may lead to efficient polariton relaxation to the bottom of the LP branch, as is required for the operation of polariton lasers.

\bigskip




\bibliographystyle{unsrtnat}
\bibliography{rsc}

@article{Green_NonMarkovianity_Lineshape,
  author = {Green, Dale and Humphries, Ben S. and Dijkstra, Arend G. and Jones, Garth A.},
  title = {Quantifying non-Markovianity in underdamped versus overdamped environments and its effect on spectral lineshape},
  journal = {J. Chem. Phys.},
  year = {2021},
  volume = {155},
  number = {8},
  pages = {084701},
  doi = {10.1063/5.0057352}
}

@article{Green_SpectralFiltering_2DS,
  author = {Green, Dale and Camargo, Franco V. A. and Heisler, Ismael A. and Dijkstra, Arend G. and Jones, Garth A.},
  title = {Spectral Filtering as a Tool for Two-Dimensional Spectroscopy: A Theoretical Model},
  journal = {J. Phys. Chem. Lett.},
  year = {2018},
  volume = {9},
  number = {9},
  pages = {2286--2291},
  doi = {10.1021/acs.jpclett.8b00732}
}

@ARTICLE{Pullerits2025,
  author = {Fan Wu and Tu C. Nguyen-Phan and Richard Cogdell and T{\"o}nu Pullerits},
  title = {Efficient cavity-mediated energy transfer between photosynthetic light harvesting complexes from strong to weak coupling regime},
  journal = {Nat. Commun.},
  year = {2025},
  volume = {16},
  pages = {5300},
}

@article{Qureshi2025,
author = {Qureshi, Hassan A. and Papachatzakis, Michael A. and Abdelmagid, Ahmed Gaber and Salomäki, Mikko and Mäkilä, Ermei and Tuomi, Oskar and Siltanen, Olli and Daskalakis, Konstantinos S.},
title = {Giant Rabi Splitting and Polariton Photoluminescence in an all Solution-Deposited Dielectric Microcavity},
journal = {Adv. Opt. Mater.},
volume = {13},
number = {16},
pages = {2500155},
year = {2025}
}

@article{Tempelaar2017,
author = {Tempelaar, Roel and Jansen, Thomas L. C. and Knoester, Jasper},
title = {Exciton–Exciton Annihilation Is Coherently Suppressed in H-Aggregates, but Not in J-Aggregates},
journal = {J. Phys. Chem. Lett.},
volume = {8},
number = {24},
pages = {6113-6117},
year = {2017},
}

@article{Doveiko2025,
author = {Doveiko, Daniel and Kubiak-Ossowska, Karina and Chen, Yu},
title = {Binding Energy Calculations of Anthracene and Rhodamine 6G H-Type Dimers: A Comparative Study of DFT and SMD Methods},
journal = {J. Phys. Chem. A},
volume = {129},
number = {12},
pages = {2946-2957},
year = {2025},
}

@article{Kowalewski2025,
author = {Borges, Lucas and Schnappinger, Thomas and Kowalewski, Markus},
title = {Impact of Dark Polariton States on Collective Strong Light–Matter Coupling in Molecules},
journal = {J. Phys. Chem. Lett.},
volume = {16},
number = {31},
pages = {7807-7815},
year = {2025},
}

@article{Rider2021,
author = {M. S. Rider and W. L. Barnes},
title = {Something from nothing: linking molecules with virtual light},
journal = {Contemp. Phys.},
volume = {62},
number = {4},
pages = {217--232},
year = {2021},
}

@article{Chen2015,
  title = {Generalized rotating-wave approximation for the two-qubit quantum Rabi model},
  author = {Zhang, Yu-Yu and Chen, Qing-Hu},
  journal = {Phys. Rev. A},
  volume = {91},
  issue = {1},
  pages = {013814},
  numpages = {7},
  year = {2015},
}

@article{Tian2012,
author = {Lu, Tian and Chen, Feiwu},
title = {Multiwfn: A multifunctional wavefunction analyzer},
journal = {J. Comput. Chem.},
volume = {33},
number = {5},
pages = {580-592},
year = {2012}
}

@article{Tian2024,
    author = {Lu, Tian},
    title = {A comprehensive electron wavefunction analysis toolbox for chemists, Multiwfn},
    journal = {J. Chem. Phys.},
    volume = {161},
    number = {8},
    pages = {082503},
    year = {2024},
}

@article{Siltanen2025,
author = {Siltanen, Olli and Luoma, Kimmo and Musser, Andrew J. and Daskalakis, Konstantinos S.},
title = {Enhancing the Efficiency of Polariton OLEDs in and Beyond the Single-Excitation Subspace},
journal = {Adv. Opt. Mater.},
volume = {13},
number = {12},
pages = {2403046},
year = {2025}
}

@article{Yuen-Zhou2022,
    author = {Campos-Gonzalez-Angulo, Jorge A. and Yuen-Zhou, Joel},
    title = {Generalization of the Tavis–Cummings model for multi-level anharmonic systems: Insights on the second excitation manifold},
    journal = {J. Chem. Phys.},
    volume = {156},
    number = {19},
    pages = {194308},
    year = {2022},
}

@article{Li2025,
author = {Li, Kai and Kok, Hui Taou and Gui, Manting and Kaul, Nidhi and Wang, Yuanheng and Rand, Barry P. and Scholes, Gregory D.},
title = {Localized Probe of Molecular Interaction under Strong Light-Matter Coupling},
journal = {J. Phys. Chem. Lett.},
volume = {16},
number = {47},
pages = {12209-12215},
year = {2025},
}

@article{Forrest2022,
author = {Bin Liu and Xinjing Huang and Shaocong Hou and Dejiu Fan and Stephen R. Forrest},
journal = {Optica},
number = {9},
pages = {1029--1036},
title = {Photocurrent generation following long-range propagation of organic exciton--polaritons},
volume = {9},
year = {2022},
}

@article{deJong2024,
title = {Enhancement of the internal quantum efficiency in strongly coupled P3HT-C60 organic photovoltaic cells using Fabry–Perot cavities with varied cavity confinement},
author = {Lianne M. A. de Jong and Anton Matthijs Berghuis and Mohamed S. Abdelkhalik and Tom P. A. van der Pol and Martijn M. Wienk and Rene A. J. Janssen and Jaime Gómez Rivas},
pages = {2531--2540},
volume = {13},
number = {14},
journal = {Nanophotonics},
year = {2024},
}

@article{Peruffo2025,
author = {Peruffo, Nicola and Chen, Runwan and Li, Dongyan and Wu, Qinghe and Börjesson, Karl},
title = {Entering the Strong Coupling Regime in Conventional Organic Solar Cells},
journal = {Adv. Funct. Mater},
pages = {e08994},
year = {2026},
volume = {36},
}

@article{Forrest2010,
  author = {K\'ena-Cohen and S. R. Forrest},
  title = {Room-temperature polariton lasing in an organic single-crystal microcavity},
  journal = {Nat. Photon.},
  year = {2010},
  volume = {4},
  pages = {371-375},
}

@article{Kavokin2022,
  author = {Alexey Kavokin and Timothy C. H. Liew and Christian Schneider and Pavlos G. Lagoudakis and Sebastian Klembt and Sven Hoefling},
  title = {Polariton condensates for classical and quantum computing},
  journal = {Naat. Rev. Phys.},
  year = {2022},
  volume = {4},
  pages = {435-451},
}

@article{Ebbesen2021,
author = {Francisco J. Garcia-Vidal  and Cristiano Ciuti  and Thomas W. Ebbesen},
title = {Manipulating matter by strong coupling to vacuum fields},
journal = {Science},
volume = {373},
number = {6551},
pages = {eabd0336},
year = {2021},
}

@article{Maie2023,
  author = {Luca Sortino and Merve Gülmüs and Benjamin Tilmann and Leonardo de S. Menezes and Stefan A. Maie},
  title = {Room-temperature polariton lasing in an organic single-crystal microcavity},
  journal = {Light Sci. Appl.},
  year = {2023},
  volume = {12},
  pages = {202},
}

@article{Kena-Cohen2010,
  author = {S. K\'{e}na-Cohen and S. R. Forrest},
  title = {Room-temperature polariton lasing in an organic single-crystal microcavity},
  year = 2010,
  journal = {Nat. Photonics},
  volume = 4,
  pages = {371--375}
}

@article{Ramezani2017,
author = {Mohammad Ramezani and Alexei Halpin and Antonio I. Fern\'{a}ndez-Dom\'{i}nguez and Johannes Feist and Said Rahimzadeh-Kalaleh Rodriguez and Francisco J. Garcia-Vidal and Jaime G\'{o}mez Rivas},
journal = {Optica},
number = {1},
pages = {31--37},
publisher = {Optica Publishing Group},
title = {Plasmon-exciton-polariton lasing},
volume = {4},
}

@article{Chen2023,
author = {Chen, Xingzhou and Alnatah, Hassan and Mao, Danqun and Xu, Mengyao and Fan, Yuening and Wan, Qiaochu and Beaumariage, Jonathan and Xie, Wei and Xu, Hongxing and Shi, Zhe-Yu and Snoke, David and Sun, Zheng and Wu, Jian},
title = {Bose Condensation of Upper-Branch Exciton-Polaritons in a Transferable Microcavity},
journal = {Nano Letters},
volume = {23},
number = {20},
pages = {9538-9546},
year = {2023},
}

@article{Yamamoto1996,
  title = {Nonequilibrium condensates and lasers without inversion: Exciton-polariton lasers},
  author = {Imamoglu, A. and Ram, R. J. and Pau, S. and Yamamoto, Y.},
  journal = {Phys. Rev. A},
  volume = {53},
  issue = {6},
  pages = {4250--4253},
  numpages = {0},
  year = {1996},
}

@ARTICLE{Perez-Sanchez2025,
  author = {Juan B. P\'erez-S\'anchez and Joel Yuen-Zhou},
  title = {Radiative pumping vs vibrational relaxation of molecular polaritons: a bosonic mapping approach},
  journal = {Nat. Commun.},
  year = {2025},
  volume = {16},
  pages = {3151},
}

@Article{Schwennicke2025,
author ="Schwennicke, Kai and Koner, Arghadip and Pérez-Sánchez, Juan B. and Xiong, Wei and Giebink, Noel C. and Weichman, Marissa L. and Yuen-Zhou, Joel",
title  ="When do molecular polaritons behave like optical filters?",
journal  ="Chem. Soc. Rev.",
year  ="2025",
volume  ="54",
issue  ="13",
pages  ="6482-6504",
}

@article{Keeling2020,
   author = "Keeling, Jonathan and Kéna-Cohen, Stéphane",
   title = "Bose–Einstein Condensation of Exciton-Polaritons in Organic Microcavities", 
   journal= "Annu. Rev. Phys. Chem.",
   year = "2020",
   volume = "71",
   pages = "435-459",
  }

@article{Bhuyan2023,
author = {Bhuyan, Rahul and Mony, J{\"u}rgen and Kotov, Oleg and Castellanos, Gabriel W. and Gómez Rivas, Jaime and Shegai, Timur O. and B{\"o}rjesson, Karl},
title = {The Rise and Current Status of Polaritonic Photochemistry and Photophysics},
journal = {Chem. Rev.},
volume = {123},
number = {18},
pages = {10877-10919},
year = {2023},
}

@Article{Mikhnenko2015,
author ="Mikhnenko, Oleksandr V. and Blom, Paul W. M. and Nguyen, Thuc-Quyen",
title  ="Exciton diffusion in organic semiconductors",
journal  ="Energy Environ. Sci.",
year  ="2015",
volume  ="8",
issue  ="7",
pages  ="1867-1888",
}

@article{Akselrod2013,
author = {Gleb M. Akselrod and Elizabeth R. Young and M. Scott Bradley and Vladimir Bulovi\'{c}},
journal = {Opt. Express},
number = {10},
pages = {12122--12128},
title = {Lasing through a strongly-coupled mode by intra-cavity pumping},
volume = {21},
year = {2013},
}

@article{Grant2016,
author = {Grant, Richard T. and Michetti, Paolo and Musser, Andrew J. and Gregoire, Pascal and Virgili, Tersilla and Vella, Eleonora and Cavazzini, Marco and Georgiou, Kyriacos and Galeotti, Francesco and Clark, Caspar and Clark, Jenny and Silva, Carlos and Lidzey, David G.},
title = {Efficient Radiative Pumping of Polaritons in a Strongly Coupled Microcavity by a Fluorescent Molecular Dye},
journal = {Adv. Opt. Mater.},
volume = {4},
number = {10},
pages = {1615-1623},
year = {2016}
}

@article{Zaumseil2021,
author = {L{\"u}ttgens, Jan M. and Berger, Felix J. and Zaumseil, Jana},
title = {Population of Exciton–Polaritons via Luminescent sp3 Defects in Single-Walled Carbon Nanotubes},
journal = {ACS Photonics},
volume = {8},
number = {1},
pages = {182-193},
year = {2021},
}

@article{Agranovich2003,
  title = {Cavity polaritons in microcavities containing disordered organic semiconductors},
  author = {Agranovich, V. M. and Litinskaia, M. and Lidzey, D. G.},
  journal = {Phys. Rev. B},
  volume = {67},
  issue = {8},
  pages = {085311},
  numpages = {10},
  year = {2003},
}

@article{Litinskaya2004,
title = {Fast polariton relaxation in strongly coupled organic microcavities},
journal = {J. Lumin.},
volume = {110},
number = {4},
pages = {364-372},
year = {2004},
author = {M. Litinskaya and P. Reineker and V.M. Agranovich},
}

@article{Coles2011,
author = {Coles, David M. and Michetti, Paolo and Clark, Caspar and Tsoi, Wing Chung and Adawi, Ali M. and Kim, Ji-Seon and Lidzey, David G.},
title = {Vibrationally Assisted Polariton-Relaxation Processes in Strongly Coupled Organic-Semiconductor Microcavities},
journal = {Adv. Funct. Mater.},
volume = {21},
number = {19},
pages = {3691-3696},
year = {2011}
}

@article{Somaschi2011,
    author = {Somaschi, N. and Mouchliadis, L. and Coles, D. and Perakis, I. E. and Lidzey, D. G. and Lagoudakis, P. G. and Savvidis, P. G.},
    title = {Ultrafast polariton population build-up mediated by molecular phonons in organic microcavities},
    journal = {Appl. Phys. Lett.},
    volume = {99},
    number = {14},
    pages = {143303},
    year = {2011},
}

@article{delPino2015,
    author = {Javier del Pino and Johannes Feist and Francisco J. Garcia-Vidal},
    title = {Quantum theory of collective strong coupling of molecular vibrations with a microcavity mode},
    journal = {New J. Phys.},
    volume = {17},
    pages = {053040},
    year = {2015},
}

@article{Eizner2019,
author = {Elad Eizner and Luis A. Martínez-Martínez and Joel Yuen-Zhou and Stéphane Kéna-Cohen },
title = {Inverting singlet and triplet excited states using strong light-matter coupling},
journal = {Sci. Adv.},
volume = {5},
number = {12},
pages = {eaax4482},
year = {2019},
}

@article{Ulusoy2019,
author = {Ulusoy, Inga S. and Gomez, Johana A. and Vendrell, Oriol},
title = {Modifying the Nonradiative Decay Dynamics through Conical Intersections via Collective Coupling to a Cavity Mode},
journal = {J. Phys. Chem. A.},
volume = {123},
number = {41},
pages = {8832-8844},
year = {2019},
}

@article{Bhuyan2024,
author = {Bhuyan, Rahul and Lednev, Maksim and Feist, Johannes and Börjesson, Karl},
title = {The Effect of the Relative Size of the Exciton Reservoir on Polariton Photophysics},
journal = {Adv. Opt. Mater.},
volume = {12},
number = {2},
pages = {2301383},
year = {2024},
}

@article{Davidsson2023,
    author = {Davidsson, Eric and Kowalewski, Markus},
    title = {The role of dephasing for dark state coupling in a molecular Tavis–Cummings model},
    journal = {J. Chem. Phys.},
    volume = {159},
    number = {4},
    pages = {044306},
    year = {2023},
}

@article{Sokolovskii2024,
    author = {Sokolovskii, Ilia and Morozov, Dmitry and Groenhof, Gerrit},
    title = {One molecule to couple them all: Toward realistic numbers of molecules in multiscale molecular dynamics simulations of exciton-polaritons},
    journal = {J. Chem. Phys.},
    volume = {161},
    pages = {134106},
    year = {2024},
}

@article{Forster1949,
title = {Experimentelle und theoretische Untersuchung des zwischenmolekularen Übergangs von Elektronenanregungsenergie},
author = {Theodor Förster},
pages = {321--327},
volume = {4},
number = {5},
journal = {ZNA},
year = {1949},
lastchecked = {2025-12-27},
}

@article{Dexter1953,
    author = {Dexter, D. L.},
    title = {A Theory of Sensitized Luminescence in Solids},
    journal = {J. Chem. Phys.},
    volume = {21},
    number = {5},
    pages = {836-850},
    year = {1953},
}

@Article{Kasha1950,
author = "Kasha, Michael",
title  = "Characterization of electronic transitions in complex molecules",
journal  = "Discuss. Faraday Soc.",
year  = "1950",
volume  = "9",
pages = "14-19",
}

@article{Swenberg1976,
title = {Bimolecular quenching of excitons and fluorescence in the photosynthetic unit},
journal = {Biophys. J.},
volume = {16},
number = {12},
pages = {1447-1452},
year = {1976},
author = {C.E. Swenberg and N.E. Geacintov and M. Pope},
}

@article{Akselrod2010,
  title = {Exciton-exciton annihilation in organic polariton microcavities},
  author = {Akselrod, G. M. and Tischler, Y. R. and Young, E. R. and Nocera, D. G. and Bulovic, V.},
  journal = {Phys. Rev. B},
  volume = {82},
  issue = {11},
  pages = {113106},
  year = {2010},
}

@article{Kirton2019,
author = {Kirton, Peter and Roses, Mor M. and Keeling, Jonathan and Dalla Torre, Emanuele G.},
title = {Introduction to the Dicke Model: From Equilibrium to Nonequilibrium, and Vice Versa},
journal = {Adv. Quantum Technol.},
volume = {2},
number = {1-2},
pages = {1800043},
year = {2019},
}

@article{Knoester2001,
    author = {Ryzhov, I. V. and Kozlov, G. G. and Malyshev, V. A. and Knoester, J.},
    title = {Low-temperature kinetics of exciton–exciton annihilation of weakly localized one-dimensional Frenkel excitons},
    journal = {J. Chem. Phys.},
    volume = {114},
    number = {12},
    pages = {5322-5329},
    year = {2001},
}

@article{Fan2023,
  title = {Ultrafast multiexciton dynamics in molecular systems: Inclusion of exciton-exciton annihilation},
  author = {Fan, Xuyang and Wei, An and Klamroth, Tillmann and Zhang, Yuan and Gao, Kun and Wang, Luxia},
  journal = {Phys. Rev. B},
  volume = {107},
  issue = {13},
  pages = {134301},
  numpages = {11},
  year = {2023},
}

@article{Beljonne2025,
author = {Cerd\'a, Jes\'us and Giannini, Samuele and Xu, Lai and Wang, Linjun and Beljonne, David},
title = {Tuning Exciton Diffusion in Organic Semiconductors through Hybridization with Charge-Transfer Excitations},
journal = {J. Phys. Chem. Lett.},
volume = {16},
number = {34},
pages = {8673-8682},
year = {2025},
}

@article{Solano2019,
  title = {Ultrastrong coupling regimes of light-matter interaction},
  author = {Forn-D\'{\i}az, P. and Lamata, L. and Rico, E. and Kono, J. and Solano, E.},
  journal = {Rev. Mod. Phys.},
  volume = {91},
  issue = {2},
  pages = {025005},
  numpages = {48},
  year = {2019},
}

@article{Manzano2020,
  author = {D. Manzano},
  title = {A short introduction to the Lindblad master equation},
  journal = {AIP Advances},
  volume = {10},
  year = {2020},
  pages = {025106},
}

@article{Sokolovskii2024b,
    author = {Sokolovskii, I. and Groenhof, G.},
    title = {Non-Hermitian molecular dynamics simulations of exciton–polaritons in lossy cavities},
    journal = {J. Chem. Phys.},
    volume = {160},
    pages = {092501},
    year = {2024},
}

@article{Selwyn1972,
author = {Selwyn, Judith E. and Steinfeld, Jeffrey I.},
title = {Aggregation of equilibriums of xanthene dyes},
journal = {J. Phys. Chem.},
volume = {76},
number = {5},
pages = {762-774},
year = {1972},
}

@article{Lagemaat2024,
author = {Fei, Rao and Hautzinger, Matthew P. and Rose, Aaron H. and Dong, Yifan and Smalyukh, Ivan I. and Beard, Matthew C. and van de Lagemaat, Jao},
title = {Controlling Exciton/Exciton Recombination in 2-D Perovskite Using Exciton–Polariton Coupling},
journal = {J. Phys. Chem. Lett.},
volume = {15},
number = {6},
pages = {1748-1754},
year = {2024},
}

@article{Tichauer2022,
  title={Identifying Vibrations that Control Non-Adiabatic Relaxation of Polaritons in Strongly Coupled Molecule-Cavity Systems},
  author={R. H. Tichauer and D. Morozov and I. Sokolovskii and J. J. Toppari and G. Groenhof},
  journal={J. Phys. Chem. Lett.},
  volume = {13},
  pages = {6259--6267},
  year={2022},
}

@ARTICLE{Sokolovskii2023,
  author = {I. Sokolovskii and R.H. Tichauer and D. Morozov and J. Feist and G. Groenhof},
  title = {Multi-scale molecular dynamics simulations of enhanced energy transfer in organic molecules under strong coupling},
  journal = {Nat. Commun.},
  year = {2023},
  volume = {14},
  pages = {6613},
}

@article{Buttner2025,
    author = {Büttner, Simon and Philipp, Luca Nils and Lüttig, Julian and Rödel, Maximilian and Hensen, Matthias and Pflaum, Jens and Mitric, Roland and Brixner, Tobias},
    title = {Probing plexciton dynamics with higher-order spectroscopy},
    journal = {J. Chem. Phys.},
    volume = {163},
    number = {4},
    pages = {044702},
    year = {2025},
}

@article{Renger2001,
title = {Ultrafast excitation energy transfer dynamics in photosynthetic pigment–protein complexes},
journal = {Phys. Rep.},
volume = {343},
number = {3},
pages = {137-254},
year = {2001},
author = {Thomas Renger and Volkhard May and Oliver Kühn},
}

@book{Birks1973,
  author    = {Birks, John B.},
  title     = {Organic Molecular Photophysics},
  publisher = {John Wiley \& Sons},
  year      = {1973},
  address   = {New York},
  isbn      = {0471064709}
}

@article{ORCA,
author = {Neese,F.},
title = {The ORCA program system},
journal = {WIRES Comput. Molec. Sci.},
volume = {2},
number = {1},
pages = {73-78},
year = {2012},
}

@article{NAMD,
    author = {Phillips, James C. and Hardy, David J. and Maia, Julio D. C. and Stone, John E. and Ribeiro, João V. and Bernardi, Rafael C. and Buch, Ronak and Fiorin, Giacomo and Hénin, Jérôme and Jiang, Wei and McGreevy, Ryan and Melo, Marcelo C. R. and Radak, Brian K. and Skeel, Robert D. and Singharoy, Abhishek and Wang, Yi and Roux, Benoît and Aksimentiev, Aleksei and Luthey-Schulten, Zaida and Kalé, Laxmikant V. and Schulten, Klaus and Chipot, Christophe and Tajkhorshid, Emad},
    title = {Scalable molecular dynamics on CPU and GPU architectures with NAMD},
    journal = {J. Chem. Phys.},
    volume = {153},
    number = {4},
    pages = {044130},
    year = {2020},
}

@article{Noginov2006,
    author = {Gavrilenko, V. I. and Noginov, M. A.},
    title = {Ab initio study of optical properties of rhodamine 6G molecular dimers},
    journal = {J. Chem. Phys.},
    volume = {124},
    number = {4},
    pages = {044301},
    year = {2006},
}

@article{Valdes1989,
author = {Valdes-Aguilera, Oscar and Neckers, D. C.},
title = {Aggregation phenomena in xanthene dyes},
journal = {Acc. Chem. Res.},
volume = {22},
number = {5},
pages = {171-177},
year = {1989},
}

@article{Antonov1999,
title = {UV–Vis spectroscopic and chemometric study on the aggregation of ionic dyes in water},
journal = {Talanta},
volume = {49},
number = {1},
pages = {99-106},
year = {1999},
author = {L. Antonov and G. Gergov and V. Petrov and M. Kubista and J. Nygren},
}

@article{Hohenberg1964,
  title = {Inhomogeneous Electron Gas},
  author = {Hohenberg, P. and Kohn, W.},
  journal = {Phys. Rev.},
  volume = {136},
  issue = {3B},
  pages = {B864-B871},
  numpages = {0},
  year = {1964},
}

@article{Runge1984,
  title = {Density-Functional Theory for Time-Dependent Systems},
  author = {Runge, Erich and Gross, E. K. U.},
  journal = {Phys. Rev. Lett.},
  volume = {52},
  issue = {12},
  pages = {997--1000},
  numpages = {0},
  year = {1984},
}

@article{Chai2008,
    author = {Chai, Jeng-Da and Head-Gordon, Martin},
    title = {Systematic optimization of long-range corrected hybrid density functionals},
    journal = {J. Chem. Phys.},
    volume = {128},
    number = {8},
    pages = {084106},
    year = {2008},
}

@article{Caldeweyher2019,
    author = {Caldeweyher, Eike and Ehlert, Sebastian and Hansen, Andreas and Neugebauer, Hagen and Spicher, Sebastian and Bannwarth, Christoph and Grimme, Stefan},
    title = {A generally applicable atomic-charge dependent London dispersion correction},
    journal = {J. Chem. Phys.},
    volume = {150},
    number = {15},
    pages = {154122},
    year = {2019},
}

@Article{Weigend2005,
author ="Weigend, Florian and Ahlrichs, Reinhart",
title  ="Balanced basis sets of split valence{,} triple zeta valence and quadruple zeta valence quality for H to Rn: Design and assessment of accuracy",
journal  ="Phys. Chem. Chem. Phys.",
year  ="2005",
volume  ="7",
issue  ="18",
pages  ="3297-3305",
}

@article{Madjet2006,
author = {Madjet, M. E. and Abdurahman, A. and Renger, T.},
title = {Intermolecular Coulomb Couplings from Ab Initio Electrostatic Potentials: Application to Optical Transitions of Strongly Coupled Pigments in Photosynthetic Antennae and Reaction Centers},
journal = {J. Phys. Chem. B.},
volume = {110},
number = {34},
pages = {17268-17281},
year = {2006},
}

@article{Michaud2011,
author = {Michaud-Agrawal, Naveen and Denning, Elizabeth J. and Woolf, Thomas B. and Beckstein, Oliver},
title = {MDAnalysis: A toolkit for the analysis of molecular dynamics simulations},
journal = {J. Comput. Chem.},
volume = {32},
number = {10},
pages = {2319-2327},
year = {2011},
}

@article{Gowers2016,
  author = {Gowers, Richard J. and Linke, Max and Barnoud, Jonathan and Reddy, Tyler J. E. and Melo, Manuel N. and Seyler, Sean L. and Domański, Jan and Dotson, David L. and Buchoux, Sébastien and Kenney, Ian M. and Beckstein, Oliver},
  title = {MDAnalysis: A Python Package for the Rapid Analysis of Molecular Dynamics Simulations},
  journal = {SciPy 2016},
  year = {2016},
}

@article{Huang2013,
author = {Huang, Jing and MacKerell Jr, Alexander D.},
title = {CHARMM36 all-atom additive protein force field: Validation based on comparison to NMR data},
journal = {J. Comput. Chem.},
volume = {34},
number = {25},
pages = {2135-2145},
year = {2013}
}

@article{Giannini2022,
  title = {Exciton transport in molecular organic semiconductors boosted by transient quantum delocalization},
  author = {Samuele Giannini and Wei-Tao Peng and Lorenzo Cupellini and Daniele Padula and Antoine Carof and Jochen Blumberger},
  journal = {Nat. Commun.},
  volume = {13},
  pages = {2755},
  year = {2022},
}

@article{Torma2015,
year = {2014},
publisher = {IOP Publishing},
volume = {78},
number = {1},
pages = {013901},
author = {Törmä, P and Barnes, W L},
title = {Strong coupling between surface plasmon polaritons and emitters: a review},
journal = {Rep. Prog. Phys.},
}

@article{Mitric2026,
    title={Exciton-Exciton and Exciton-Photon Annihilation in Polaritonic Systems}, 
    author={Luca Nils Philipp and Julian Lüttig and Roland Mitric},
    year={2026},
    journal = {J. Phys. Chem. C},
    volume = {XXXX},
    number = {XXX},
    pages = {XXX-XXX},
}

@article{Stojanovic2024,
author = {Stojanovic, Ljiljana and Giannini, Samuele and Blumberger, Jochen},
title = {Exciton Transport in the Nonfullerene Acceptor O-IDTBR from Nonadiabatic Molecular Dynamics},
journal = {J. Chem. Theory Comput.},
volume = {20},
number = {14},
pages = {6241-6252},
year = {2024},
}

@ARTICLE{Ivanovich2025,
  author = {Filip Ivanovi\'{c} and Samuele Giannini and Wei-Tao Peng and Jochen Blumberger},
  title = {Transiently delocalised hybrid quantum states are gateways for efficient exciton dissociation at organic donor-acceptor interfaces},
  journal = {Nat. Commun.},
  year = {2025},
  volume = {16},
  pages = {11560},
}

@article{Rao2022,
author = {Sneyd, Alexander J. and Beljonne, David and Rao, Akshay},
title = {A New Frontier in Exciton Transport: Transient Delocalization},
journal = {J. Phys. Chem. Lett.},
volume = {13},
number = {29},
pages = {6820-6830},
year = {2022},
}

@article{Malyshev1999,
title = {Exciton–exciton annihilation in linear molecular aggregates at low temperature},
journal = {Chem. Phys. Lett.},
volume = {305},
number = {1},
pages = {117-122},
year = {1999},
author = {V.A. Malyshev and H. Glaeske and K.-H. Feller},
}

@book{Pope1999,
  author    = {Martin Pope and Charles E. Swenberg},
  title     = {Electronic Processes in Organic Crystals and Polymers},
  publisher = {Oxford University Press},
  year      = {1999},
  address   = {Oxford},
  isbn      = {9780195129632}
}

@book{Brutting2005,
  author    = {Wolfgang Br{\"u}tting},
  title     = {Physics of Organic Semiconductors},
  publisher = {Wiley VCH},
  year      = {2005},
  isbn      = {352740550X}
}

@article{Plint2025,
title = {Direct Measurement of Density for Evaporated Thin Films},
journal = {Small methods},
volume = {9},
number = {11},
pages = {e01438},
year = {2025},
author = {Trevor Plint and Halynne R. Lamontagne and Joseph Manion and Beno\^{i}t H. Lessard},
}

@article{Aroeira2024,
title = {Coherent transient exciton transport in disordered polaritonic wires},
author = {Gustavo J. R. Aroeira and Kyle T. Kairys and Raphael F. Ribeiro},
pages = {2553--2564},
volume = {13},
number = {14},
journal = {Nanophotonics},
year = {2024},
}

@article{Khairutdinov1997,
author = {Khairutdinov, R. F. and Serpone, N.},
title = {Photophysics of Cyanine Dyes: Subnanosecond Relaxation Dynamics in Monomers, Dimers, and H- and J-Aggregates in Solution},
journal = {J. Phys. Chem. B},
volume = {101},
number = {14},
pages = {2602-2610},
year = {1997},
}

@article{Pandya2021,
author = {Pandya, Raj and Alvertis, Antonios M. and Gu, Qifei and Sung, Jooyoung and Legrand, Laurent and Kr{\'e}her, David and Barisien, Thierry and Chin, Alex W. and Schnedermann, Christoph and Rao, Akshay},
title = {Exciton Diffusion in Highly-Ordered One Dimensional Conjugated Polymers: Effects of Back-Bone Torsion, Electronic Symmetry, Phonons and Annihilation},
journal = {J. Phys. Chem. Lett.},
volume = {12},
number = {14},
pages = {3669-3678},
year = {2021},
}

@ARTICLE{Huang2023,
  author = {Sarath Kumar and Ian S. Dunn and Shibin Deng and Tong Zhu and Qiuchen Zhao and Olivia F. Williams and Roel Tempelaar and Libai Huang},
  title = {Exciton annihilation in molecular aggregates suppressed through quantum interference},
  journal = {Nat. Chem.},
  year = {2023},
  volume = {15},
  pages = {1118-1126},
}

@article{Sfeir2024,
author = {Michail, Evripidis and Rashidi, Kamyar and Liu, Bin and He, Guiying and Menon, Vinod M. and Sfeir, Matthew Y.},
title = {Addressing the Dark State Problem in Strongly Coupled Organic Exciton-Polariton Systems},
journal = {Nano Lett.},
volume = {24},
number = {2},
pages = {557-565},
year = {2024},
}

@article{He2026,
author = {Zhang, Nie and Liu, Yanzhao and Huang, Huihao and Zhai, Xiaokun and Xiao, Fupeng and Wang, Fuqing and Gao, Tingge and Wang, Sheng and Li, Yan and Zhang, Zhiyong and He, Xiaowei},
title = {Exciton-Polariton Relaxation and Emission in Carbon Nanotube Microcavities with Varied Quality Factors},
journal = {ACS Nano},
volume = {20},
number = {6},
pages = {5189-5198},
year = {2026},
}

@article{Herrera2022,
    author = {Triana, Johan F. and Herrera, Felipe},
    title = {Open quantum dynamics of strongly coupled oscillators with multi-configuration time-dependent Hartree propagation and Markovian quantum jumps},
    journal = {J. Chem. Phys.},
    volume = {157},
    number = {19},
    pages = {194104},
    year = {2022},
}

@article{Dalibard1992,
  title = {Wave-function approach to dissipative processes in quantum optics},
  author = {Dalibard, Jean and Castin, Yvan and M\o{}lmer, Klaus},
  journal = {Phys. Rev. Lett.},
  volume = {68},
  issue = {5},
  pages = {580-583},
  numpages = {0},
  year = {1992},
}

@article{Molmer1993,
author = {Klaus M{\o}lmer and Yvan Castin and Jean Dalibard},
journal = {J. Opt. Soc. Am. B},
number = {3},
pages = {524-538},
title = {Monte Carlo wave-function method in quantum optics},
volume = {10},
year = {1993},
}

@article{Tremblay2022,
    author = {Mandal, Souvik and Gatti, Fabien and Bindech, Oussama and Marquardt, Roberto and Tremblay, Jean Christophe},
    title = {Stochastic multi-configuration time-dependent Hartree for dissipative quantum dynamics with strong intramolecular coupling},
    journal = {J. Chem. Phys.},
    volume = {157},
    number = {14},
    pages = {144105},
    year = {2022},
}

@article{Scala2007,
  title = {Microscopic derivation of the Jaynes-Cummings model with cavity losses},
  author = {Scala, M. and Militello, B. and Messina, A. and Piilo, J. and Maniscalco, S.},
  journal = {Phys. Rev. A},
  volume = {75},
  issue = {1},
  pages = {013811},
  numpages = {8},
  year = {2007},
}

@Article{Jaynes1963,
  author =       "E. T. Jaynes and F. W. Cummings",
  title =        "Comparison of quantum and semiclassical radiation theories with application to the beam maser",
  journal =      "Proc. IEEE",
  year =         1963,
  volume =       51,
  pages =        "89--109",
}

@article{Briegel1993,
  title = {Quantum optical master equations: The use of damping bases},
  author = {Briegel, Hans-J\"urgen and Englert, Berthold-Georg},
  journal = {Phys. Rev. A},
  volume = {47},
  issue = {4},
  pages = {3311-3329},
  year = {1993},
}

@article{Sokolovskii2025,
    author = {Sokolovskii, Ilia and Blumberger, Jochen},
    title = {Strong intermolecular coupling protects delocalisation and transport of organic exciton-polaritons against static excitation energy disorder},
    journal = {J. Chem. Phys.},
    volume = {163},
    pages = {234127},
    year = {2025},
}

@article{Ivanovich2026,
    author = {Filip Ivanovi\'{c} and Ilia Sokolovskii and Samuele Giannini and Wei-Tao Peng and Jochen Blumberger},
    title = {Exciton dissociation from non-adiabatic molecular dynamics: The role of initial conditions, dimensionality and disorder},
    journal = {J. Chem. Phys.},
    volume = {164},
    pages = {154112},
    year = {2026},
}

\end{document}


\setlength{\marginparwidth}{3cm}
\newpage
\tableofcontents
\newpage

\section{Modified Dicke Hamiltonian in matrix form}

Equation~\ref{eq:Hamiltonian} represents the matrix form of the modified Dicke Hamiltonian, introduced in Equation~1 in the main text, for the case of $N=3$ three-level molecules (TLMs). Here, the diagonal elements highlighted in blue, represent the molecular and/or photonic excitations in the zeroth (ground state with energy $E_g=0$), first (excitation of either a $S_1$-exciton with energy $E_{S_1,i}$ or a cavity photon with energy $E_c$), and second excitation subspace (excitation of either a $S_n$-exciton with energy $E_{S_n,i}$, or two $S_1$-excitons with energies $E_{S_1,i}$ and $E_{S_1,j}$ simultaneously, or one $S_1$-exciton and one cavity photon simultaneously, or two cavity photons with total energy $2E_c$). The red elements, $J$, describe transfers of $S_1$-excitons between different TLMs, whereas the green elements, $V$, are responsible for exciton-exciton annihilation (EEA). Finally, the elements highlighted in cyan, describe light-matter couplings within the first excitation subspace ($\hbar g$), within the second excitation subspace ($\hbar g$ and $\sqrt{2}\hbar g$), as well as between the zeroth and the second excitation subspaces ($\hbar g$). Within the Rotating wave approximation, the counter-rotating terms (\textit{i.e.} $\hbar g$ terms in the first row and the first column) that simultaneously create or destroy an exciton and a photon, are neglected, and the block of the Hamiltonian corresponding to the second excitation subspace becomes decoupled from the block corresponding to the zeroth excitation subspace. Therefore, in the current work, we reduce the Hamiltonian's size and consider only the block corresponding to the second excitation subspace.

\begin{equation}
\begin{turn}{90}
{$\scriptsize\hat{\text{H}} = \left(\begin{array}{ccccccccccccccc} 
   \color{blue}{E_g} & 0 & 0 & 0 & 0 & 0 & 0 & 0 & 0 & 0 & 0 & {\color{cyan} \hbar g} & {\color{cyan} \hbar g} & {\color{cyan} \hbar g} & 0 \\
   0 &  \color{blue}{E_{S_1,1}} & {\color{red} J} & {\color{red} J} & {\color{cyan} \hbar g} & 0 & 0 & 0 & 0 & 0 & 0 & 0 & 0 & 0 & 0 \\
   0 & {\color{red} J} &  \color{blue}{E_{S_1,2}} & {\color{red} J} & {\color{cyan} \hbar g} & 0 & 0 & 0 & 0 & 0 & 0 & 0 & 0 & 0 & 0 \\
   0 & {\color{red} J} & {\color{red} J} &  \color{blue}{E_{S_1,3}} & {\color{cyan} \hbar g} & 0 & 0 & 0 & 0 & 0 & 0 & 0 & 0 & 0 & 0 \\
   0 & {\color{cyan} \hbar g} & {\color{cyan} \hbar g} & {\color{cyan} \hbar g} &  \color{blue}{E_c} & 0 & 0 & 0 & 0 & 0 & 0 & 0 & 0 & 0 & 0 \\
   0 & 0 & 0 & 0 & 0 &  \color{blue}{E_{S_n,1}} & 0 & 0 & {\color{green} V} & {\color{green} V} & 0 & 0 & 0 & 0 & 0 \\
   0 & 0 & 0 & 0 & 0 & 0 &  \color{blue}{E_{S_n,2}} & 0 & {\color{green} V} & 0 & {\color{green} V} & 0 & 0 & 0 & 0 \\
  0 & 0 & 0 & 0 & 0 & 0 & 0 &  \color{blue}{E_{S_n,3}} & 0 & {\color{green} V} & {\color{green} V} & 0 & 0 & 0 & 0 \\
  0 & 0 & 0 & 0 & 0 & {\color{green} V} & {\color{green} V} & 0 &  \color{blue}{E_{S_1,1}}+\color{blue}{E_{S_1,2}} & {\color{red} J} & {\color{red} J} & {\color{cyan} \hbar g} & {\color{cyan} \hbar g} & 0 & 0 \\
  0 & 0 & 0 & 0 & 0 & {\color{green} V} & 0 & {\color{green} V} & {\color{red} J} &  \color{blue}{E_{S_1,1}}+\color{blue}{E_{S_1,3}} & {\color{red} J} & {\color{cyan} \hbar g} & 0 & {\color{cyan} \hbar g} & 0 \\
  0 & 0 & 0 & 0 & 0 & 0 & {\color{green} V} & {\color{green} V} & {\color{red} J} & {\color{red} J} &   \color{blue}{E_{S_1,2}}+\color{blue}{E_{S_1,3}} & 0 & {\color{cyan} \hbar g} & {\color{cyan} \hbar g} & 0\\
 {\color{cyan} \hbar g} & 0 & 0 & 0 & 0 & 0 & 0 & 0 & {\color{cyan} \hbar g} & {\color{cyan} \hbar g} & 0 & \color{blue}{E_{S_1,1}}+\color{blue}{E_c} & {\color{red} J} & {\color{red} J} & {\color{cyan} \sqrt{2}\hbar g} \\
  {\color{cyan} \hbar g} & 0 & 0 & 0 & 0 & 0 & 0 & 0 & {\color{cyan} \hbar g} & 0 & {\color{cyan} \hbar g} & {\color{red} J} & \color{blue}{E_{S_1,2}}+\color{blue}{E_c} & {\color{red} J} &  {\color{cyan} \sqrt{2}\hbar g} \\
  {\color{cyan} \hbar g} & 0 & 0 & 0 & 0 & 0 & 0 & 0 & 0 & {\color{cyan} \hbar g} & {\color{cyan} \hbar g} & {\color{red} J} & {\color{red} J} &  \color{blue}{E_{S_1,3}}+\color{blue}{E_c} & {\color{cyan} \sqrt{2}\hbar g} \\
  0 & 0 & 0 & 0 & 0 & 0 & 0 & 0 & 0 & 0 & 0 & {\color{cyan} \sqrt{2}\hbar g} & {\color{cyan} \sqrt{2}\hbar g} & {\color{cyan} \sqrt{2}\hbar g} & \color{blue}{2E_c}
\end{array}\right)$}
\label{eq:Hamiltonian}
\end{turn}
\end{equation}

\clearpage

\section{Rabi splitting in the second excitation subspace}

In the absence of the excitonic and EEA coupling, the Hamiltonian in the second-excitation subspace can be written as
\begin{equation}
\scriptsize{\hat{\text{H}} = \left(\begin{array}{cccccccccc} 
   E_{S_n,1} & 0 & 0 & 0 & 0 & 0 & 0 & 0 & 0 & 0 \\
   0 & E_{S_n,2} & 0 & 0 & 0 & 0 & 0 & 0 & 0 & 0 \\
  0 & 0 & E_{S_n,3} & 0 & 0 & 0 & 0 & 0 & 0 & 0 \\
  0 & 0 & 0 & {\color{blue} E_{S_1,1}+E_{S_1,2}} & {\color{blue} 0} & {\color{blue} 0} & {\color{blue}\hbar g} & {\color{blue}\hbar g} & {\color{blue} 0} & {\color{blue} 0} \\
  0 & 0 & 0 & {\color{blue} 0} & {\color{blue} E_{S_1,1}+E_{S_1,3}} & {\color{blue} 0} & {\color{blue}\hbar g} & {\color{blue} 0} & {\color{blue}\hbar g} & {\color{blue} 0} \\
  0 & 0 & 0 & {\color{blue} 0} & {\color{blue} 0} &  {\color{blue} E_{S_1,2}+E_{S_1,3}} & {\color{blue} 0} & {\color{blue}\hbar g} & {\color{blue}\hbar g} & {\color{blue} 0}\\
  0 & 0 & 0 & {\color{blue}\hbar g} & {\color{blue}\hbar g} & {\color{blue} 0} & {\color{red} E_{S_1,1}+E_c} & {\color{red} 0} & {\color{red} 0} & {\color{red} \sqrt{2}\hbar g} \\
  0 & 0 & 0 & {\color{blue}\hbar g} & {\color{blue} 0} & {\color{blue}\hbar g} & {\color{red} 0} & {\color{red} E_{S_1,2}+E_c} & {\color{red} 0} &  {\color{red} \sqrt{2}\hbar g} \\
  0 & 0 & 0 & {\color{blue} 0} & {\color{blue}\hbar g} & {\color{blue}\hbar g} & {\color{red} 0} & {\color{red} 0} & {\color{red}E_{S_1,3}+E_c} & {\color{red} \sqrt{2}\hbar g} \\
  0 & 0 & 0 & {\color{blue} 0} & {\color{blue} 0} & {\color{blue} 0} & {\color{red} \sqrt{2}\hbar g} & {\color{red} \sqrt{2}\hbar g} & {\color{red} \sqrt{2}\hbar g} & {\color{red} 2E_c}
\end{array}\right)}
\label{eq:Hamiltonian_second}
\end{equation}
where, for the sake of space, we assumed that the number of TLMs was equal to three ($N=3$).

The red block of the Hamiltonian in Equation~\ref{eq:Hamiltonian_second} is an arrowhead matrix, representing the coupling between the states with the excitation of a molecule into its first excited state and of a cavity Fock state ($\vert\phi_{S_1,1}\rangle$), and the state with two excitations of the cavity photon ($\vert\phi_{S_0,2}\rangle$), hence the off-diagonal elements are multiplited by $\langle2|\langle S_0|\hat{\sigma}^-\hat{a}^\dagger|S_1\rangle|1\rangle=\sqrt{2}$.\cite{Chen2015} In the absence of disorder and under the resonant condition (\textit{i.e.} $E_{S_1,j}=E_0~\forall j$ and $E_c=E_0$), the diagonalisation of this matrix yields two polaritonic states with energies $2E_0\pm \hbar g\sqrt{2N}$ 
and $N-1$ dark states with energy $2E_0$. Because the dark states lack a contribution from the cavity photons, the Hamiltonian in Equation~\ref{eq:Hamiltonian_second} can be represented in terms of the coupling between the states with two molecules excited into their first excited state ($\vert\phi_{2S_1,0}\rangle$; the blue block of the Hamiltonian) and the two polaritonic states:

\begin{equation}
\footnotesize{\hat{\text{H}} = \left(\begin{array}{cccccccc} 
  E_{S_n,1} & 0 & 0 & 0 & 0 & 0 & 0 & 0 \\
  0 & E_{S_n,2} & 0 & 0 & 0 & 0 & 0 & 0 \\
  0 & 0 & E_{S_n,3} & 0 & 0 & 0 & 0 & 0 \\
  0 & 0 & 0 & {\color{blue} E_{S_1,1}+E_{S_1,2}} & {\color{blue} 0} & {\color{blue} 0}  & {\color{blue} \sqrt{2/N}\hbar g} & {\color{blue} \sqrt{2/N}\hbar g} \\
  0 & 0 & 0 & {\color{blue} 0} & {\color{blue} E_{S_1,1}+E_{S_1,3}} & {\color{blue} 0} & {\color{blue} \sqrt{2/N}\hbar g} & {\color{blue} \sqrt{2/N}\hbar g}\\
  0 & 0 & 0 & {\color{blue} 0} & {\color{blue} 0} &  {\color{blue} E_{S_1,2}+E_{S_1,3}} & {\color{blue} \sqrt{2/N}\hbar g} & {\color{blue} \sqrt{2/N}\hbar g} \\
  0 & 0 & 0 & {\color{blue} \sqrt{2/N}\hbar g} & {\color{blue} \sqrt{2/N}\hbar g} & {\color{blue} \sqrt{2/N}\hbar g} & {\color{red} 2E_c-g\sqrt{2N}} & {\color{red} 0} \\
  0 & 0 & 0 & {\color{blue} \sqrt{2/N}\hbar g} & {\color{blue} \sqrt{2/N}\hbar g} & {\color{blue} \sqrt{2/N}\hbar g} & {\color{red} 0} & {\color{red} 2E_c+g\sqrt{2N}}
\end{array}\right)}
\label{eq:Hamiltonian_second_modified}
\end{equation}
where the coupling elements are multiplied by $\sqrt{2/N}$ to account for the fact that each $\vert\phi_{2S_1,0}\rangle$ state interacts with only two $\vert\phi_{S_1,1}\rangle$ states out of $N$ such states.

The splitting between the highest and lowest energy states of the Hamiltonian in Equation~\ref{eq:Hamiltonian_second_modified} can be defined in a similar way to the general expression for the Rabi splitting in the first-excitation subspace,\cite{Rider2021} $\Omega_\text{R}^{(1)}=2\sqrt{g^2N+(\omega_0-\omega_c)^2/4}$ with $\omega_0=E_0/\hbar$ and $\omega_c=E_c/\hbar$, resulting in the following expression:
\begin{equation}
\begin{array}{ccl}
\Omega_\text{R}^{(2)}&=&2\sqrt{2\left(\sqrt{\frac{2}{N}}g\right)^2N_\text{comb}+(2\Delta E/\hbar)^2/4}=\\
\\
&&2\sqrt{4\frac{g^2}{N}\frac{(N-1)N}{2}+(2g\sqrt{2N})^2/4}= \\
\\
&&2g\sqrt{2\left(N-1\right)+2N}=2g\sqrt{4N-2}
\label{eq:Rabi_2}
\end{array}
\end{equation}
with 
$2\Delta E$ being the sum of the energy shifts between the $\vert\phi_{2S_1,0}\rangle$ states of energy $2E_0$ and the polaritonic states of energy $2E_0\pm \hbar g\sqrt{2N}$, \textit{i.e.} $2\Delta E=2\hbar g\sqrt{2N}$. In addition, $N_\text{comb}=C_N^2$ is the number of combinations of two different molecules excited into their first excited state, \textit{i.e.} the number of diagonal elements in the blue-coloured block of the Hamiltonian in Equation~\ref{eq:Hamiltonian_second_modified}. Finally, the first term under the square root is multiplied by two to account for the interaction with two polaritonic states.

\clearpage

\begin{figure*}[!htb]
\centering
\includegraphics[width=1\textwidth]{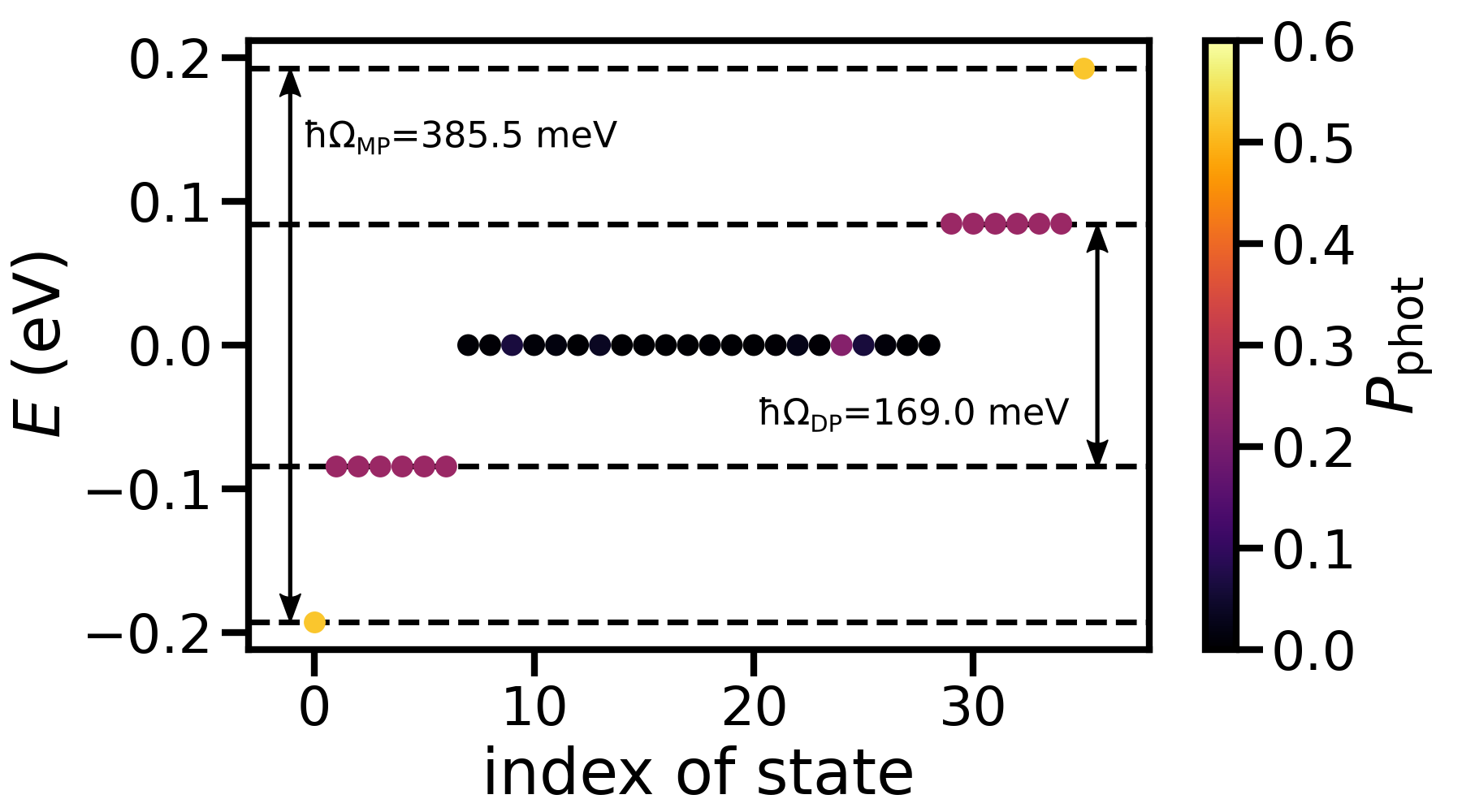}
  \caption{Eigenvalues of the light-matter Hamiltonian in Equation~\ref{eq:Hamiltonian_second} for $N=7$ molecules and $g\sqrt{N}=100$~meV.}\label{fig:figureS1}
\end{figure*}

As an example, Figure~\ref{fig:figureS1} depicts the eigenvalues of the Hamiltonian in Equation~\ref{eq:Hamiltonian_second} for a system of $N=7$ isoenergetic molecules ($E_0$) resonantly coupled to a cavity mode with the collective coupling strength $g\sqrt{N}=100$~meV in the absence of the coupling terms responsible for exciton hopping and EEA, \textit{i.e.} $J=V=0$~meV. There, one can distinguish three types of polaritonic states with different contributions from the cavity, $P_\text{phot}$: \textit{i)} multi-polariton states (yellow) with the photonic content $P_\text{phot}\approx0.5$ and the energy splitting $\hbar\Omega_\text{MP}=2\hbar g\sqrt{4N-2}=385.5$~meV; \textit{ii)} dark polariton states (magenta) with the photonic content $P_\text{phot}\approx0.25$, energies $E_\text{DP}=E_0\pm\hbar g\sqrt{N-2}$ and $E_0$ and the highest splitting $\hbar\Omega_\text{DP}=2\hbar g\sqrt{N-2}=169.0$~meV; and \textit{iii)} dark states with the photonic content $P_\text{phot}\approx0$ and energy $E_\text{DS}=E_0$.

\clearpage

\section{Molecular Dynamics simulations of Rhodamine 6G H-dimer}

\begin{figure*}[!tb]
\centering
\includegraphics[width=0.8\textwidth]{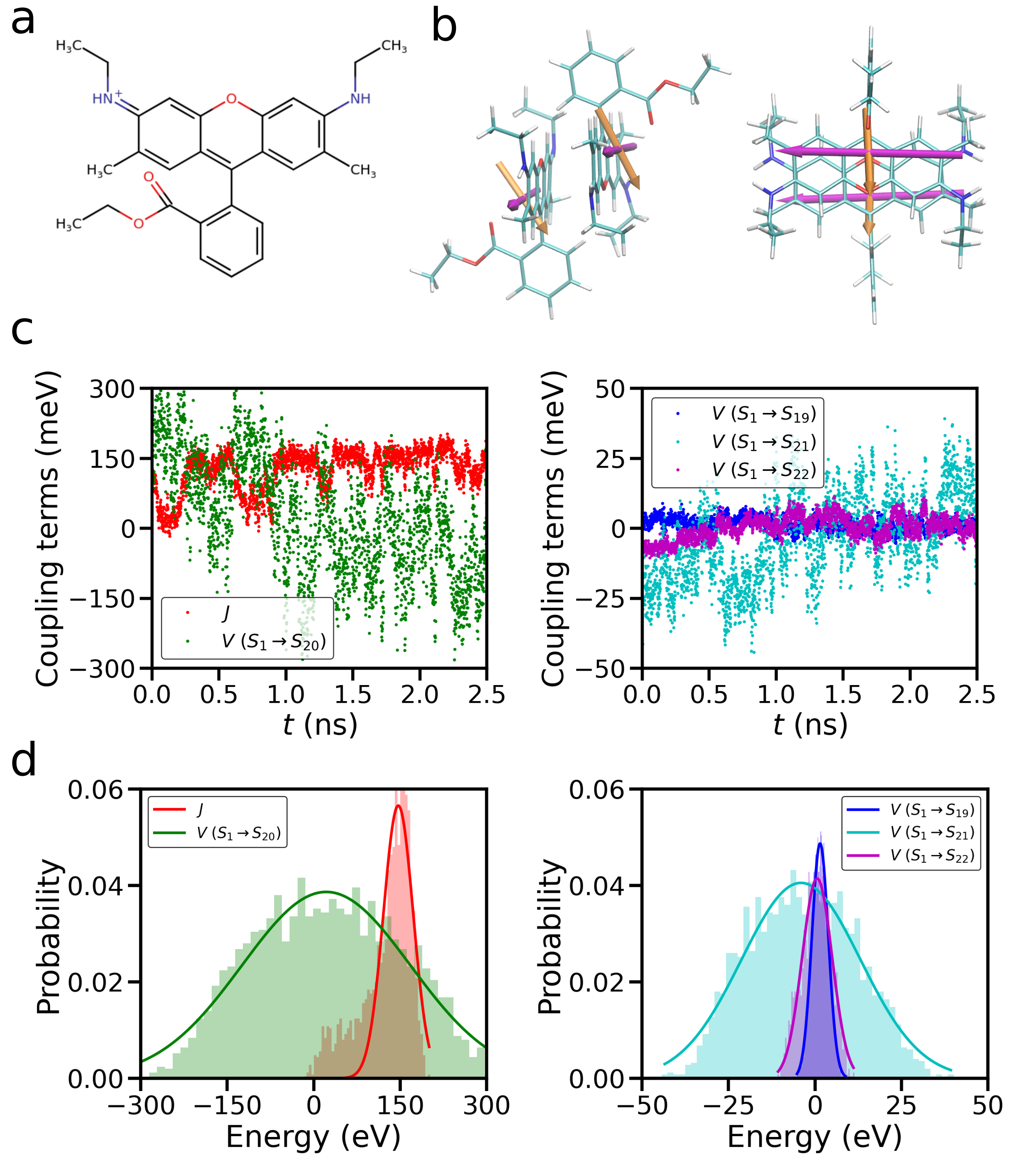}
  \caption{Panel~\textbf{a}: Chemical structure of Rhodamine 6G (R6G). Panel~\textbf{b}: Structure of the H-dimer of R6G after geometry optimisation at the TD-DFT/$\omega$B97X-D4//def-TZVPP level of theory. Colors indicate the following atoms: green - carbon, white - hydrogen, red - oxygen, blue - nitrogen. The purple arrows are transition dipole moment (TDM) vectors corresponding to the $S_0\rightarrow S_1$ excitation of the monomers ($\boldsymbol{\mu}_{0\rightarrow1}$), and the orange arrows are TDM vectors corresponding to the $S_1\rightarrow S_n$ excitation of the monomers ($\boldsymbol{\mu}_{1\rightarrow n}$) with energy of the $S_n$-state given by $E_{S_n}\approx2E_{S_1}$. Panel~\textbf{c}: Fluctuation of the exciton coupling $J$ and the dipole-dipole coupling $V$ between $\boldsymbol{\mu}_{0\rightarrow1}$ and $\boldsymbol{\mu}_{1\rightarrow n}$ with $n=19,20,21,22$, extracted from molecular dynamics simulations. Panel~\textbf{d}: Corresponding histograms fitted with a Gaussian function.}\label{fig:figureS2}  
\end{figure*}

In two experiments on cavity-modified exciton-exciton annihilation (EEA), Rhodamine 6G (R6G, Figure~\ref{fig:figureS2}\textbf{a}) fluorophore was strongly coupled to the cavity.\cite{Pullerits2025,Qureshi2025} It is known that this molecule tends to aggregation at high concentrations with the formation of both J- and H-dimers, as well as trimers.\cite{Valdes1989,Antonov1999} Therefore, to estimate the maximum possible coupling strength between excitons in R6G, we considered the R6G H-dimer, since its formation is energetically favourable over the J-dimer formation.\cite{Noginov2006}

First, we performed geometry optimisation of the H-dimer in vacuum at the Density Functional level of Theory\cite{Hohenberg1964} (DFT) using the range-separated hybrid $\omega$B97X functional\cite{Chai2008} with the D4 dispersion correction\cite{Caldeweyher2019} and the def2-TZVPP basis set.\cite{Weigend2005} Figure~\ref{fig:figureS2}\textbf{b} shows the optimised structure of the dimer, in which the magenta arrows denote the transition dipole moment (TDM) vectors of the monomers for excitation from the ground state (GS), $S_0$, to the first excited state, $S_1$ ($\boldsymbol{\mu}_{0\rightarrow1}$), calculated with the Time-Dependent DFT.\cite{Runge1984} The vectors are collinear, which is characteristic of H-dimers. The first excited state of the dimer has a negligible oscillator strength of $f=0.0007$, while the second excited state is optically bright with oscillator strength $f=1.7666$, which also indicates that the dimer is of the H-type. The dipole-dipole interaction between the $S_1$-excitons allows for exciton hops between adjacent molecules. The corresponding coupling term can be calculated using the transition charge approach:\cite{Madjet2006}
\begin{equation}   
    J_{i,k}^\text{TrESP} = \sum_{A\in i}\sum_{B\in k}\frac{q^{0\rightarrow1}_{T,A}q^{0\rightarrow1}_{T,B}}{|\textbf{r}_A-\textbf{r}_B|},
\label{eq:J_TrESP}
\end{equation}
where $A$ and $B$ are atoms belonging to molecules $i$ and $k$, respectively; $q^{0\rightarrow1}_{T,A}$ and $q^{0\rightarrow1}_{T,B}$ are transition electrostatic potential (TrESP) charges for the $S_0\rightarrow S_1$ excitation, and $\textbf{r}_{A/B}$ are atomic positions. For the optimised H-dimer, the transition charge approach gives a coupling strength of $J_\text{opt}=210$~meV.

The orange arrows in Figure~\ref{fig:figureS2}\textbf{b} denote the TDM vectors of the monomers for excitation from the $S_1$-state to the $S_{20}$-state ($\boldsymbol{\mu}_{1\rightarrow20}$). The energy of state $S_{20}$ is nearly twice as high as the energy of the $S_1$-state, and the $S_1\rightarrow S_{20}$ transition has a much greater oscillator strength compared to the transitions to other states with energies close to $2E_{S_1}$ (Table~\ref{tab:tab1}). Therefore, the $S_{20}$-state is expected to make the largest contribution to EEA. The coupling term responsible for EEA is calculated using the transition charge approach as well:

\begin{equation}   
    V_{i,k}^\text{TrESP} = \frac{1}{2}\left[\sum_{A\in i}\sum_{B\in k}\frac{q^{0\rightarrow1}_{T,A}q^{1\rightarrow n}_{T,B}}{|\textbf{r}_A-\textbf{r}_B|}+\sum_{A\in i}\sum_{B\in k}\frac{q^{1\rightarrow n}_{T,A}q^{0\rightarrow1}_{T,B}}{|\textbf{r}_A-\textbf{r}_B|}\right],
\label{eq:V_TrESP}
\end{equation}
where $q^{1\rightarrow n}_{T,A/B}$ is the TrESP charge for excitation from the first excited state to the $n$-th excited state with energy $E_{S_n}\approx2E_{S_1}$. The calculated values of the coupling term $V$ for $n=19,20,21,22$ in the optimised geometry of the R6G H-dimer are presented in Table~\ref{tab:tab2}.

Additionally, we optimised the geometry of the R6G monomer at the same level of theory (DFT/$\omega$B97X//def2-TZVPP) and analysed the excited states. The oscillator strength of the transition to the first excited state with energy $E_{S_1}=3.394$~eV ($f=1.108$) was estimated to be $\approx4$ and $\approx96$ times higher than the oscillator strength of the transition to the $18^\text{th}$  and $19^\text{th}$ excited states with energies $E_{S_{18}}=6.580$~meV ($f=0.311$) and $E_{S_{19}}=6.735$~meV ($f=0.012$), respectively. Therefore, we assumed that the excitation of the latter states was unable to enter strong coupling with the second cavity mode and hence did not include the corresponding coupling term in the system's Hamiltonian (Equation~1 in the main text), nor did we consider the exciton coupling between the $S_n$-excitons.

\begin{table}[b]
\footnotesize
\caption{Energies and oscillator strengths of transitions between different states of the monomers in the geometry optimised H-dimer of Rhodamine 6G.}
\begin{tabular}{c|c|c|c|c|c}
Transition & $S_0\rightarrow S_1$  & $S_1\rightarrow S_{19}$ & $S_1\rightarrow S_{20}$ & $S_1\rightarrow S_{21}$ & $S_1\rightarrow S_{22}$ \\ \hline 
\multicolumn{6}{c}{Monomer 1} \\ \hline 
Energy, eV & 3.428 & 3.391 & 3.494 & 3.583 & 3.626 \\
Osc. strength & 1.123 & 0.003 & 0.046 & 0.001 & 0.022 \\ \hline 
\multicolumn{6}{c}{Monomer 2} \\ \hline 
Energy, eV & 3.427 & 3.399 & 3.504 & 3.579 & 3.619 \\
Osc. strength & 1.128 & 0.006 & 0.200 & 0.018 & 0.0004 \\
\end{tabular}\label{tab:tab1}
\end{table}

\begin{table}[b]
\footnotesize
\caption{Exciton coupling $J$ and couplings $V$ for the dipole interaction between the $S_0\rightarrow S_1$ transition and the $S_1\rightarrow S_n$ transition with $n=19,20,21,$ and $22$. The reported values are the coupling terms in the optimised geometry of the Rhodamine 6G H-dimer (second row), coupling terms averaged over a 2.5 ns long molecular dynamics trajectory (third row), as well as their standard deviations (fourth row). All values are in meV.}
\begin{tabular}{c|c|c|c|c|c}
Coupling term & $J$  & $V_{S_1\rightarrow S_{19}}$ & $V_{S_1\rightarrow S_{20}}$ & $V_{S_1\rightarrow S_{21}}$ & $V_{S_1\rightarrow S_{22}}$ \\ \hline 
In optimised geometry & 210 & 1.5 & 9.3 & 0.3 & 1.3 \\
Average from MD & 147 & 1.5 & 22 & 4.0 & 0.6 \\
Standard deviation & 26 & 2.4 & 146 & 17 & 4.3 \\
\end{tabular}\label{tab:tab2}
\end{table}

To explore how the coupling terms $J$ and $V$ are affected by thermal molecular motions, we performed molecular dynamics (MD) simulations of the R6G H-dimer in water. The parameters of the model and the simulation protocol were taken from the article by Doveiko \textit{et al.}\cite{Doveiko2025} Briefly, the simulation box with 27442 atoms was initially minimised with fixed atoms of R6G and then equilibrated for 10 ps with a time step 1 fs. Temperature and pressure were maintained at 300~K and 1~bar with a Langevin barostat and thermostat in the NPT ensemble. The system was then minimised again, but without imposing constraints on the R6G atoms, with the subsequent heating to 300~K for 30~ps and equilibration for 270~ps. Finally, the MD production was conducted for 2.5~ns in the NVT ensemble using a Langevin thermostat with a damping constant of 5~ps$^{-1}$. In all simulations, the CHARMM36 force field was used to model the interactions between R6G molecules and water molecules.\cite{Huang2013}  

Figure~\ref{fig:figureS2}\textbf{c} shows the time-evolution of the coupling terms $J$ and $V$ during the MD production, which were calculated using Equations~\ref{eq:J_TrESP}~and~\ref{eq:V_TrESP} with frozen TrESP charges corresponding to the geometry optimised H-dimer. It was previously shown that using frozen transition charges is a decent approximation, which provides a satisfactory estimation of excitonic couplings.\cite{Giannini2022} Figure~\ref{fig:figureS2}\textbf{d} depicts histograms of the coupling terms for 2500 frames extracted from the 2.5~ns trajectory. These histograms were fitted with Gaussian functions $A\exp\left[-\frac{\left(J(t)-\langle J\rangle\right)^2}{2\sigma_J^2}\right]$ and $B\exp\left[-\frac{\left(V(t)-\langle V\rangle\right)^2}{2\sigma_V^2}\right]$, from which the average couplings, $\langle J\rangle$ and $\langle V\rangle$, and their fluctuations, $\sigma_J$ and $\sigma_V$, were extracted and presented in Table~\ref{tab:tab2}.

Simulations of polariton dynamics were performed over a wide range of exciton couplings $\langle J\rangle$, and the value $\langle J\rangle_\text{MD}\approx150$~meV obtained from the MD simulation, was considered as a high-concentration limit, at which R6G molecules undergo dimerisation. The average EEA coupling terms, $\langle V\rangle_\text{MD}$, extracted from MD simulations are close to zero for all transitions except for the $S_1\rightarrow S_{20}$ transition, for which $\langle V\rangle_\text{MD}^{S_1\rightarrow S_{20}}\approx20$~meV (Table~\ref{tab:tab2}). This value was used in our simulations of polariton dynamics. We also chose the standard deviations of the coupling terms $\sigma_J$ and $\sigma_V$ both to be 10~meV.

\begin{figure*}[!tb]
\centering
\includegraphics[width=1\textwidth]{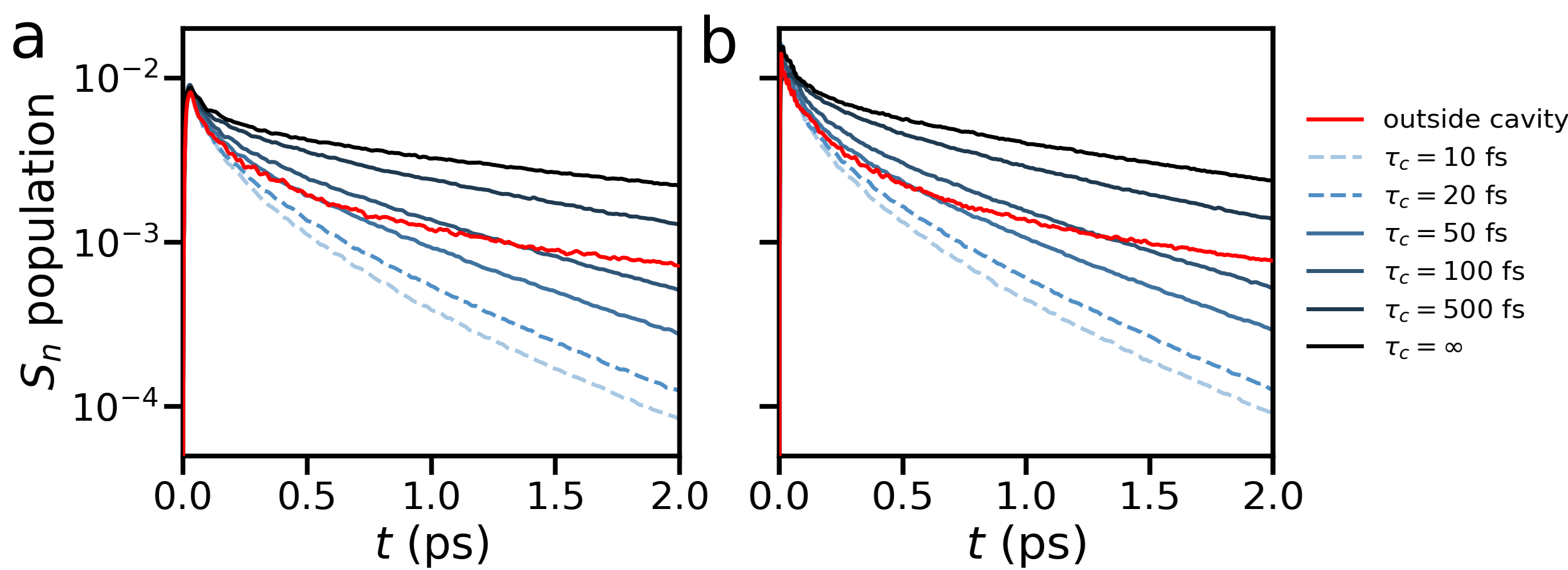}
  \caption{Total population of the $\vert\phi_{S_n,0}\rangle$ states in simulations of polariton dynamics in a system of $N=50$ three-level molecules with an exciton coupling strength of $\langle J\rangle=30$~meV and two different values of the EEA coupling strength and its standard deviation, namely $\langle V\rangle=50$~meV and $\sigma_V=50$~meV (panel~\textbf{a}), and $\langle V\rangle=20$~meV and $\sigma_V=10$~meV (panel~\textbf{b}). The simulations were performed with different cavity lifetimes ranging between $\tau_c=10$~fs and $\tau_c=500$~fs, as well as for the ideal cavity with no decay ($\gamma_c=1/\tau_c=0$) and outside the cavity. For both average values of the EEA coupling strength and its standard deviation, the dynamics of the total population of the $\vert\phi_{S_n,0}\rangle$ states are very similar and suggest a gradual transition from enahcement to suppression of the EEA rate with decreasing $\tau_c$.}\label{fig:figureS3}
\end{figure*}

Because coupling terms $V$ can fluctuate significantly due to thermal motions of the molecules, especially in the case of the transition with the highest oscillator strength (green histogram in Figure~\ref{fig:figureS2}\textbf{d}), we performed additional simulations of polariton dynamics with enhanced EEA coupling and its fluctuation, namely $\langle V\rangle=50$~meV and $\sigma_V=50$~meV. The results of this simulation show a gradual change from suppression to enhancement of the EEA rate with an increase in the cavity lifetime (Figure~\ref{fig:figureS3}) as in the case when $\langle V\rangle=20$~meV and $\sigma_V=10$~meV, which indicates that the conclusions made in the work do not depend on the value of the EEA coupling strength.

\clearpage

\section{Additional results}

\subsection{Convergence of dynamics with respect to time step}

\begin{figure*}[!htb]
\centering
\includegraphics[width=0.7\textwidth]{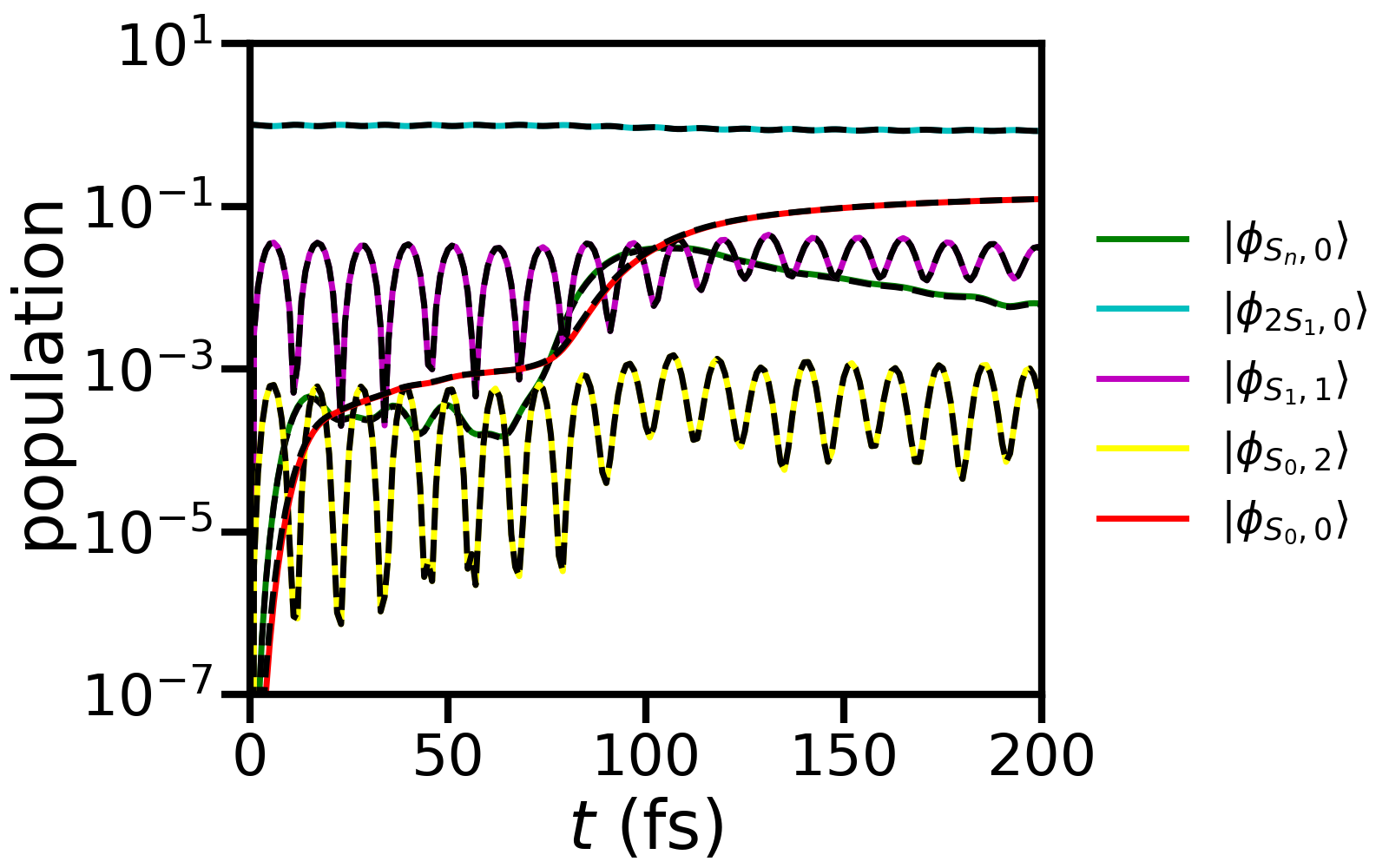}
  \caption{Total population of all $|\phi_{S_n,0}\rangle$ states (green), $|\phi_{2S_1,0}\rangle$ states (cyan), $|\phi_{S_1,1}\rangle$ states (purple), $|\phi_{S_0,2}\rangle$ state (yellow), and the ground state $|\phi_{S_0,0}\rangle$ (red), in simulation of $N=50$ isoenergetic three-level molecules with an exciton coupling strength of $\langle J\rangle=50$~meV, a collective light-matter coupling strength of $g\sqrt{N}=175$~meV, and a time step of $\Delta t=1$~fs. The black dashed lines show the same populations but in simulation with a time step of $\Delta t=0.1$~fs.}\label{fig:figureS4}
\end{figure*}

When time-propagating an electronic wave function, the time step is usually chosen in the range from 0.01~fs to 0.1~fs. In particular, it is essential that the time step is smaller than the shortest time scale of the electronic dynamics. In the TLM-cavity system considered in this work, such a time scale is the period of Rabi oscillations, which is equal to $T_\text{R}=11$~fs for a Rabi splitting of $\hbar\Omega_\text{R}=375$~meV. We therefore selected a time step of $\Delta t=1$~fs in numerical simulations, which is small enough to capture all the relevant effects during the dynamics and large enough to save computational resources. We also ensured that the resulting dynamics were consistent with those at a smaller time step of $\Delta t=0.1$~fs, as shown in Figure~\ref{fig:figureS4} for the case of $N=50$ TLMs without disorder, interacting with the coupling strength $\langle J\rangle=50$~meV with each other.

\clearpage

\subsection{Short-time exciton dynamics}

\begin{figure*}[!htb]
\centering
\includegraphics[width=0.8\textwidth]{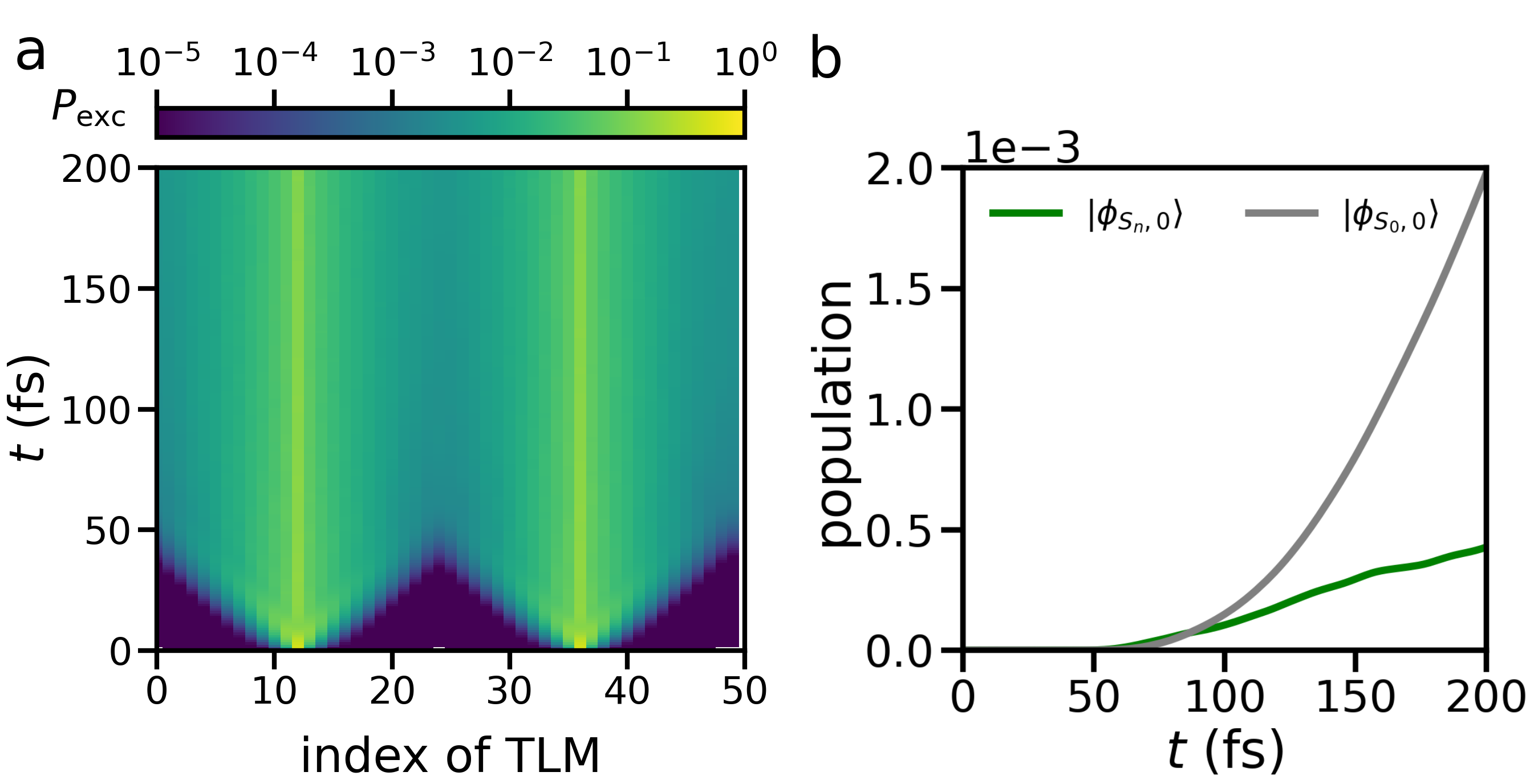}
  \caption{Panel~\textbf{a}: Excitonic probability density, $P_\text{exc}$, as a function of time and the positions of three-level molecules, in simulations of exciton-polariton dynamics in a system of $N=50$ molecules with an exciton coupling strength of $\langle J\rangle=70$~meV outside the cavity, \textit{i.e.} $g\sqrt{N}=0$~meV. Panel~\textbf{b}: Total populations of all $\vert\phi_{S_n,0}\rangle$ states (green) and of the ground state (grey).
  }\label{fig:figureS5}
\end{figure*}

Figure~\ref{fig:figureS5}\textbf{a} demonstrates the exciton probability density, $P_\text{exc}$, as a function of the TLMs' positions and time. The probability density at position $j$ was calculated as $P_\text{exc}(t)=\frac{1}{2}\sum_{i=N+1}^{N(N-1)/2}|d_{i\in j}(t)|^2+\frac{1}{2}\sum_{i=N(N-1)/2+1}^{N_\text{st}-1}|d_{i\in j}(t)|^2$, where the first term sums up the expansion coefficients of the total polariton wave function (Equation~3 in the main text) that correspond to $\vert\phi_{2S_1,0}\rangle$ states, while the second term sums up the expansion coefficients that correspond to $\vert\phi_{S_1,1}\rangle$ states. In the absence of light-matter coupling, only the first term remains. The subscript $i\in j$ implies that only those states are taken into account, which describe the excitation of molecule $j$ into the $S_1$-state. Both terms are multiplied by $1/2$ because in both the $\vert\phi_{2S_1,0}\rangle$ states and the $\vert\phi_{S_1,1}\rangle$ states, only half the excitation falls on this molecule. The exciton probability density is normalised at each time step to give the total probability of finding an exciton in the chain of TLMs equal to one.

Figure~\ref{fig:figureS5}\textbf{b} shows the population of the $\vert\phi_{S_n,0}\rangle$ states (green) and the GS population (grey). The build up of both populations starts after approximately 50~fs, which is the time required for the probability densities of two excitons to approach each other, allowing for population transfer from the states with two $S_1$-excitons to the states with one $S_n$-exciton.

\clearpage

\subsection{Contribution of Rabi oscillations to the ground state population}

\begin{figure*}[!htb]
\centering
\includegraphics[width=1\textwidth]{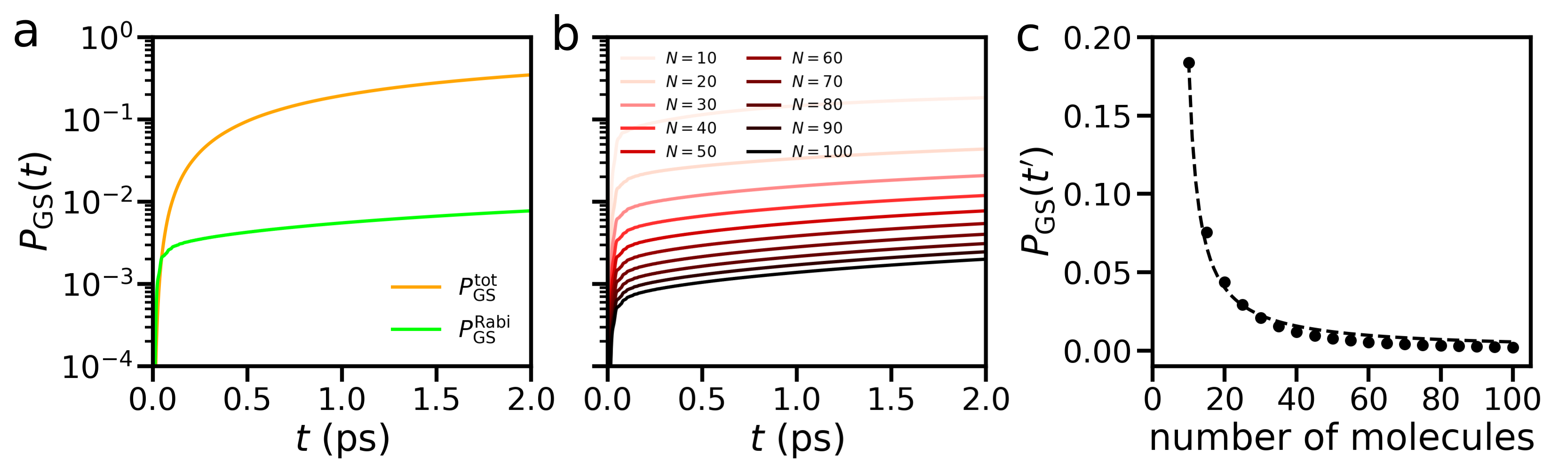}
  \caption{Panel~\textbf{a}: Ground state (GS) population as a function of time in simulations of polariton dynamics in a system of $N=50$ three-level molecules (TLMs) inside the cavity ($g\sqrt{N}=175$~meV) with an exciton coupling strength of $\langle J\rangle=70$~meV (orange) and $J=0$~meV (green). Panel~\textbf{b}: GS population in a system with $\langle J\rangle=0$~meV with different numbers of TLMs. Panel~\textbf{c}: GS population at time $t^\prime=2$~ps as a function of the number of molecules. The dashed line is a fit to function $\frac{a}{b+N}$ with $a=0.513$ and $b=-7.229$.
  }\label{fig:figureS6}
\end{figure*}

To estimate the contribution of Rabi oscillations due to light-matter coupling to the total GS population, we performed simulation of polariton dynamics in the absence of exciton coupling, \textit{i.e.} $J=0$~meV. In this case, excitons are localised at the initial positions and hence unable to interact other than via Rabi oscillations. Figure~\ref{fig:figureS6}\textbf{a} demonstrates that the GS population due to Rabi oscillations is very small compared to the GS population in the case when exciton interactions are included, \textit{i.e.} $J>0$. Moreover, the contribution of Rabi oscillations rapidly decreases with the number of molecules as $1/N$ (Figure~\ref{fig:figureS6}\textbf{b},\textbf{c}). Thus, enhancement of the EEA rate observed in a molecular system with low exciton mobility cannot be explained based on a simple Rabi oscillations argument. Instead, our simulations suggest that the cavity helps excitons to partially overcome disorder, leading to an increase in the likelihood of exciton interaction.

\clearpage

\subsection{Comparison between molecule-cavity systems with and without disorder}

\begin{figure*}[!htb]
\centering
\includegraphics[width=1\textwidth]{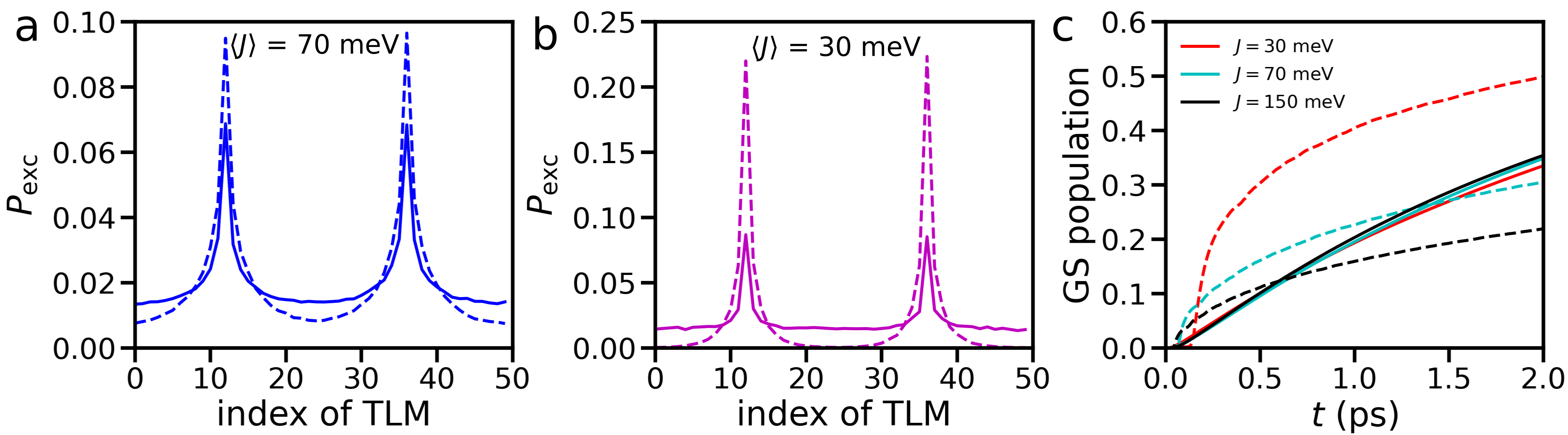}
  \caption{Panels~\textbf{a},\textbf{b}: Excitonic probability density, $P_\text{exc}$, as a function of the positions of three-level molecules (TLMs), in simulations of exciton-polariton dynamics in a system of $N=50$ molecules with an exciton coupling strength of $\langle J\rangle=70$~meV (panel~\textbf{a}) and $\langle J\rangle=30$~meV (panel~\textbf{a}), outside the cavity ($g\sqrt{N}=0$~meV, dashed lines) and inside the cavity ($g\sqrt{N}=175$~meV, solid lines). The exciton population was averaged between $t=1$~ps and $t=2$~ps. Panel~\textbf{c}: The ground state population as a function of time in simulations of exciton-polariton dynamics in a system of $N=50$ TLMs with disorder ($\sigma_E=100$~meV, solid lines) and without disorder ($\sigma_E=0$~meV, dashed lines) under strong light-matter coupling. The exciton coupling strength was set to $\langle J\rangle=30$~meV (red), $\langle J\rangle=70$~meV (cyan), and $\langle J\rangle=150$~meV (black).}\label{fig:figureS7}
\end{figure*}

Figure~\ref{fig:figureS7}\textbf{a},\textbf{b} depicts the distribution of the exciton probability along the TLMs chain with $\langle J\rangle=70$~meV (panel~\textbf{a}) and $\langle J\rangle=30$~meV (panel~\textbf{b}). Whereas outside the cavity (dashed lines), the exciton propagation is suppressed, and the probability density is mostly localised at the initial spot, especially in the case of low exciton coupling, strong light-matter coupling ensures exciton delocalisation inside the cavity, resulting in a non-zero uniform distribution of the probability density outside the initial spot (solid lines). Due to exciton delocalisation, EEA is almost as efficient at low exciton coupling as at high exciton coupling when molecules are placed inside the cavity in the presence of disorder ($\sigma_E=100$~meV, solid lines in Figure~\ref{fig:figureS7}\textbf{c}). Interestingly, in an ideal system without disorder ($\sigma_E=0$~meV), the GS population, and hence the EEA rate, decreases with increasing exciton coupling strength. We attribute this effect to a decrease in the time that excitons spend near each other when $\langle J\rangle$ becomes higher, leading to less intense exciton annihilation.

\clearpage

\subsection{Exciton propagation in systems with different exciton mobility}

\begin{figure*}[!htb]
\centering
\includegraphics[width=1\textwidth]{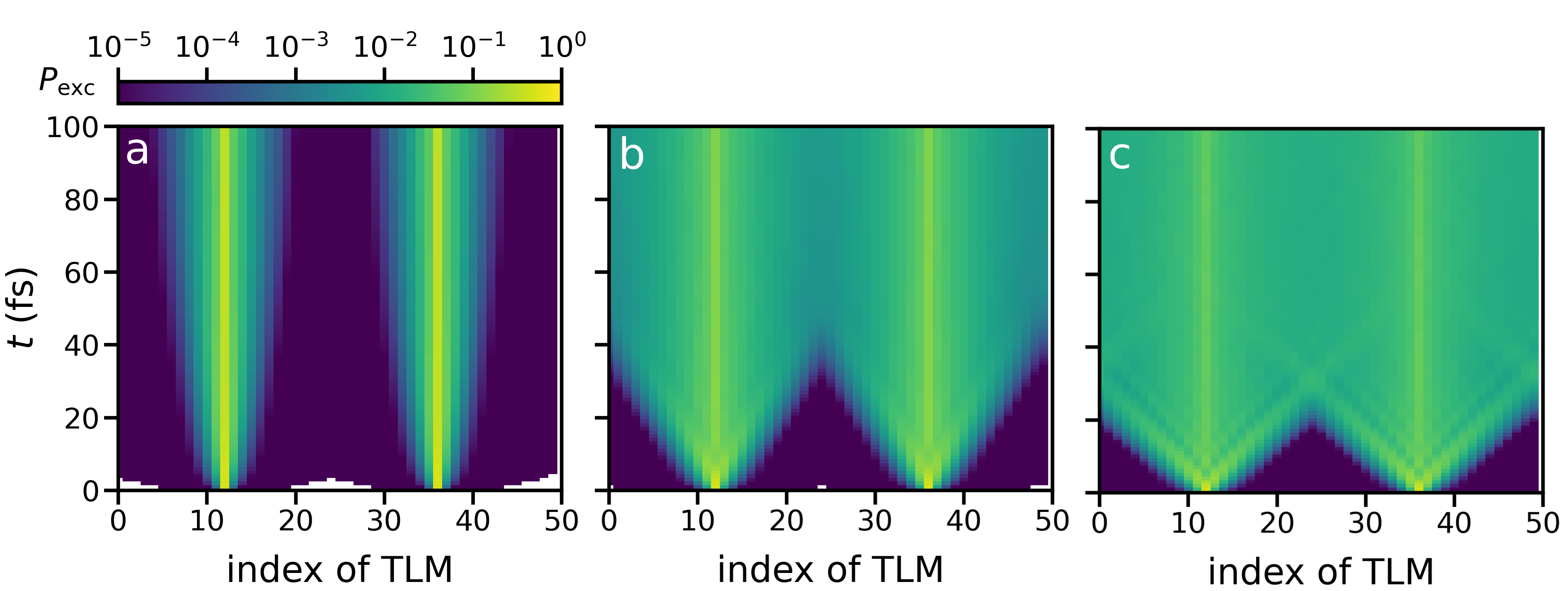}
  \caption{Excitonic probability density, $P_\text{exc}$, as a function of time and the positions of three-level molecules in simulations with $\langle J\rangle=20$~meV (panel~\textbf{a}), $\langle J\rangle=80$~meV (panel~\textbf{b}), and $\langle J\rangle=140$~meV (panel~\textbf{c}), under strong light-matter coupling. The excitation energy disorder strength is $\sigma_E=100$~meV.}\label{fig:figureS8}
\end{figure*}

Figure~\ref{fig:figureS8} shows the excitonic probability amplidute, $P_\text{exc}$, as a function of the TLMs' positions and time in simulation of $N=50$ TLMs under strong coupling with three different exciton coupling strengths, namely $\langle J\rangle=20$~meV, $80$~meV, and $140$~meV. In the former case, exciton propagation is suppressed by disorder ($\sigma_E=100$~meV), which is manifested as Anderson localisation of the exciton density. As $\langle J\rangle$ increases, exciton propagation becomes possible and the degree of the excitonic wave packet's propagation increases with the value of the exciton coupling strength. As a result, the EEA rate also increases (Figure~3\textbf{a} in the main text).

\clearpage

\subsection{Exciton escape probability}

\begin{figure*}[!htb]
\centering
\includegraphics[width=0.65\textwidth]{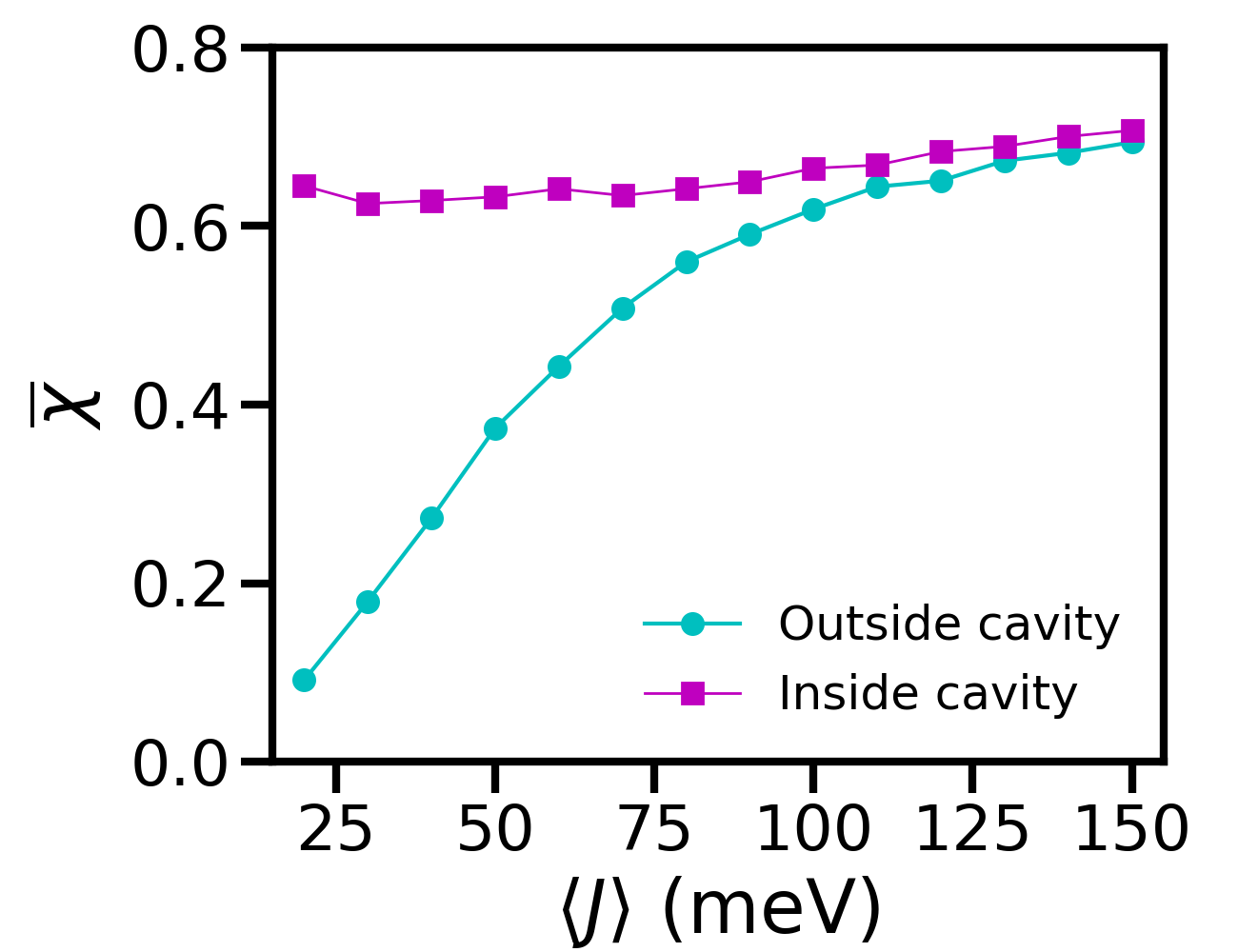}
  \caption{Exciton escape probability, $\overline{\chi}$, averaged between $t=1$~ps and $t=2$~ps, in simulation of $N=50$ three-level molecules inside (purple squares) and outside the cavity (cyan circles) with different values of exciton coupling strength $\langle J\rangle$.
  }\label{fig:figureS9}
\end{figure*}

To quantify how much of the exciton population extends beyond the initial excitation point, we computed the exciton escape probability using Equation~7 in the main text in a system of $N=50$ TLMs. As shown in Figure~\ref{fig:figureS9}, the average escape probability, $\overline{\chi}$, is rather small at low exciton coupling strength ($\overline{\chi}=0.09$ at $\langle J\rangle=20$~meV) outside the cavity due to Anderson localisation of excitons caused by disorder (Figure~\ref{fig:figureS8}\textbf{a}). As $\langle J\rangle$ increases, exciton transport is less impeded by disorder, resulting in a gradual increase in $\overline{\chi}$, which reaches a value of $\overline{\chi}=0.69$ at the highest exciton coupling strength of $\langle J\rangle=150$~meV. In contrast, the escape probability is high ($\overline{\chi}>0.60$) in the whole range of exciton coupling strengths between $\langle J\rangle=20$~meV and $\langle J\rangle=150$~meV inside the cavity and only slightly increases with $\langle J\rangle$. This is because a strong light-matter coupling extends the exciton probability density over the whole molecular chain (Figure~\ref{fig:figureS7}\textbf{a},\textbf{b}) regardless of the values of $\langle J\rangle$. This little dependence of the escape probability on exciton mobility correlates well with the little dependence of the EEA rate on $\langle J\rangle$ as shown in Figure~3\textbf{b} in the main text. 

\clearpage

\subsection{Estimation of the threshold density for the onset of EEA}

\begin{figure*}[!htb]
\centering
\includegraphics[width=0.5\textwidth]{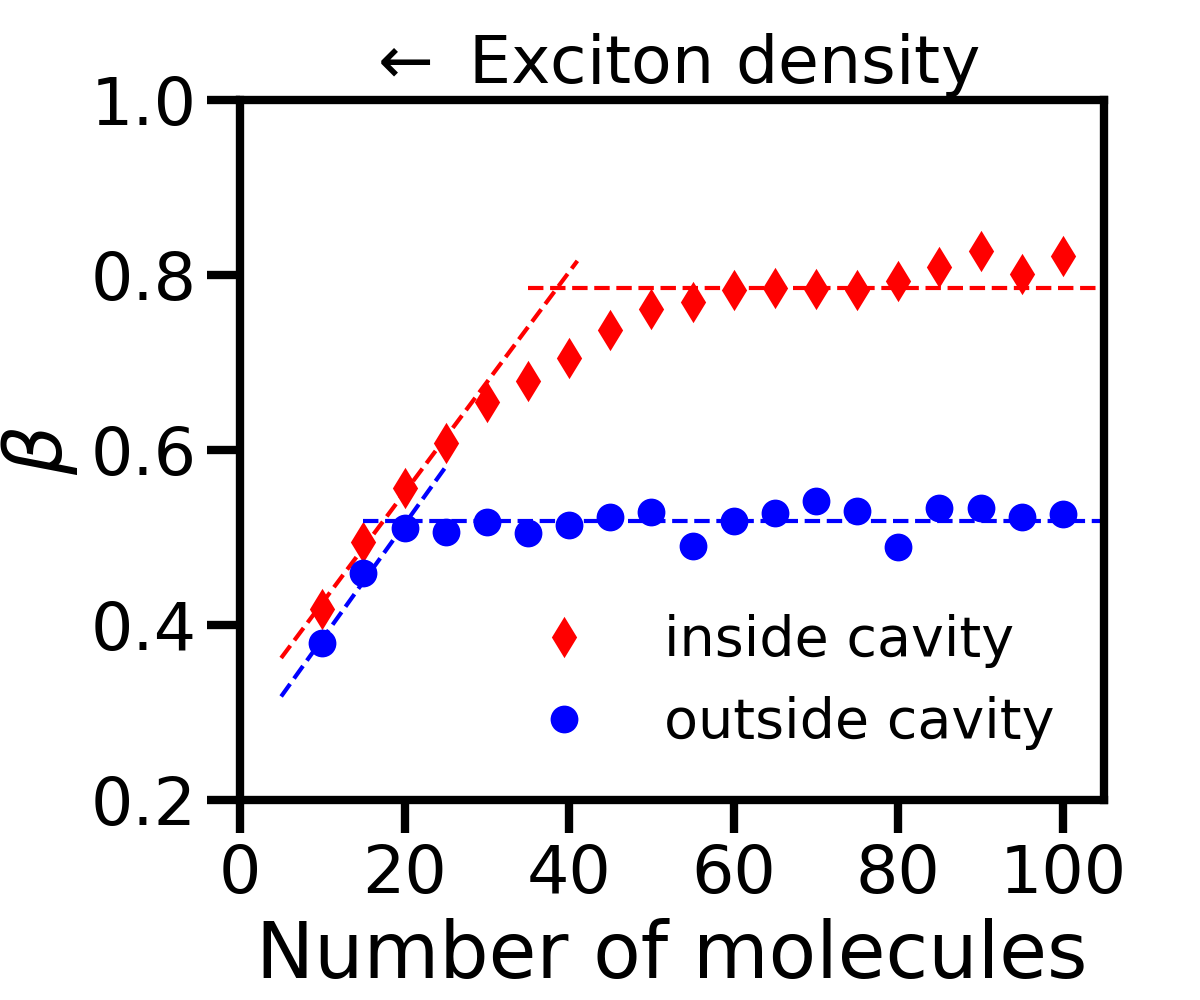}
  \caption{Dependence of $\beta$ on the number of molecules, $N$, inside (red diamonds) vs. outside the cavity (blue circles), with $\beta$ extracted from the fit of the GS population at different values of $N$ in an ideal cavity ($\gamma_c=0$) to the function $P_\text{GS}(t)=\alpha t^\beta$. The dashed lines are fits to a linear function. The intersection of the lines indicates the onset of EEA, which occurs at a larger $N$ (smaller exciton density) when molecules are strongly coupled to the cavity.}\label{fig:figureS10}
\end{figure*}

In the main text, the threshold number of molecules, and hence the threshold exciton density, was estimated from the change in the slope of the $\alpha$ coefficient as a function of $N$, where $\alpha$ was extracted from fitting the GS populations in systems with different numbers of molecules to the power-law function, $P_\text{GS}(t)=\alpha t^\beta$. Similarly, we can estimate the threshold density from the change in the slope of the $\beta$ coefficient, as shown in Figure~\ref{fig:figureS10}. There, we can distinguish two regimes: a constant $\beta$ at large $N$ (low exciton density) and a linearly changing $\beta$ at small $N$ (large exciton density). The latter implies a faster initial growth of the GS population with increasing exciton density, as expected for the power-law function $at^b$ with $0<b<1$, indicating an increase in the EEA rate after the exciton density exceeds a certain threshold. This threshold was estimated from the intersection of the lines corresponding to the two regimes of $\beta$ (dashed lines in Figure~\ref{fig:figureS10}). The resulting threshold density of $\rho=5\%$ (or a threshold number of molecules of $N_\text{th}=40$) for the system under strong light-matter coupling was smaller than the threshold density outside the cavity ($\rho=10\%$, or $N_\text{th}=20$), further supporting the conclusion that strong coupling can enhance exciton annihilation.

\clearpage

\subsection{Contribution of exciton-photon annihilation to the total EEA rate}

Recently, it was proposed that in addition to exciton annihilation caused by the dipole-dipole interaction between a pair of molecules (\textit{i.e.}, EEA), the cavity provides another source of exciton annihilation.\cite{Mitric2026} This so-called exciton-photon annihilation results from a resonant interaction between the $S_1\rightarrow S_n$ molecular transition and a cavity photon, characterised by coupling strength $\tilde{g}$. To investigate the relevance of this effect in our system with the parameters of TLMs characteristic of R6G, we incorporated coupling terms $\tilde{g}$ in the system's Hamiltonian (Equation~2 in the main text) within the Rotating-wave approximation as follows:
\begin{equation}
\begin{array}{ccl}
{\hat{H}} &=&\sum_i^{N} E_{S_1,i}\hat{\sigma}^+_{S_1,i}\hat{\sigma}^-_{S_1,i}+E_{\text{c}}\hat{a}^\dagger\hat{a}-\sum_i^Ng\left(\hat{\sigma}_{S_1,i}^+\hat{a}+\hat{\sigma}_{S_1,i}^-\hat{a}^\dagger\right)- \\
\\
&&{\color{magenta}\sum_i^N\tilde{g}\left(\hat{\sigma}_{S_1,i}^-\hat{\sigma}_{S_n,i}^+\hat{a}+\hat{\sigma}_{S_1,i}^+\hat{\sigma}_{S_n,i}^-\hat{a}^\dagger\right)+} \\
\\
&&\sum_{i,k}^NJ_{i,k}\left(\hat{\sigma}_{S_1,i}^{+}\hat{\sigma}_{S_1,k}^{-}+\text{h.c.}\right) + \\
\\
&&\sum_j^NE_{S_n,j}\hat{\sigma}^+_{S_n,j}\hat{\sigma}^-_{S_n,j} + \sum_{i,k}^NV_{i,k}\left[\left(\hat{\sigma}^+_{S_n,i}+\hat{\sigma}^+_{S_n,k}\right)\hat{\sigma}^-_{S_1,i}\hat{\sigma}^-_{S_1,k} +\text{h.c.}\right],
\label{eq:Ham_Rabi_EPA}
\end{array}
\end{equation}
where the coupling terms responsible for exciton-photon annihilation are highlighted in pink.

\begin{figure*}[!htb]
\centering
\includegraphics[width=0.7\textwidth]{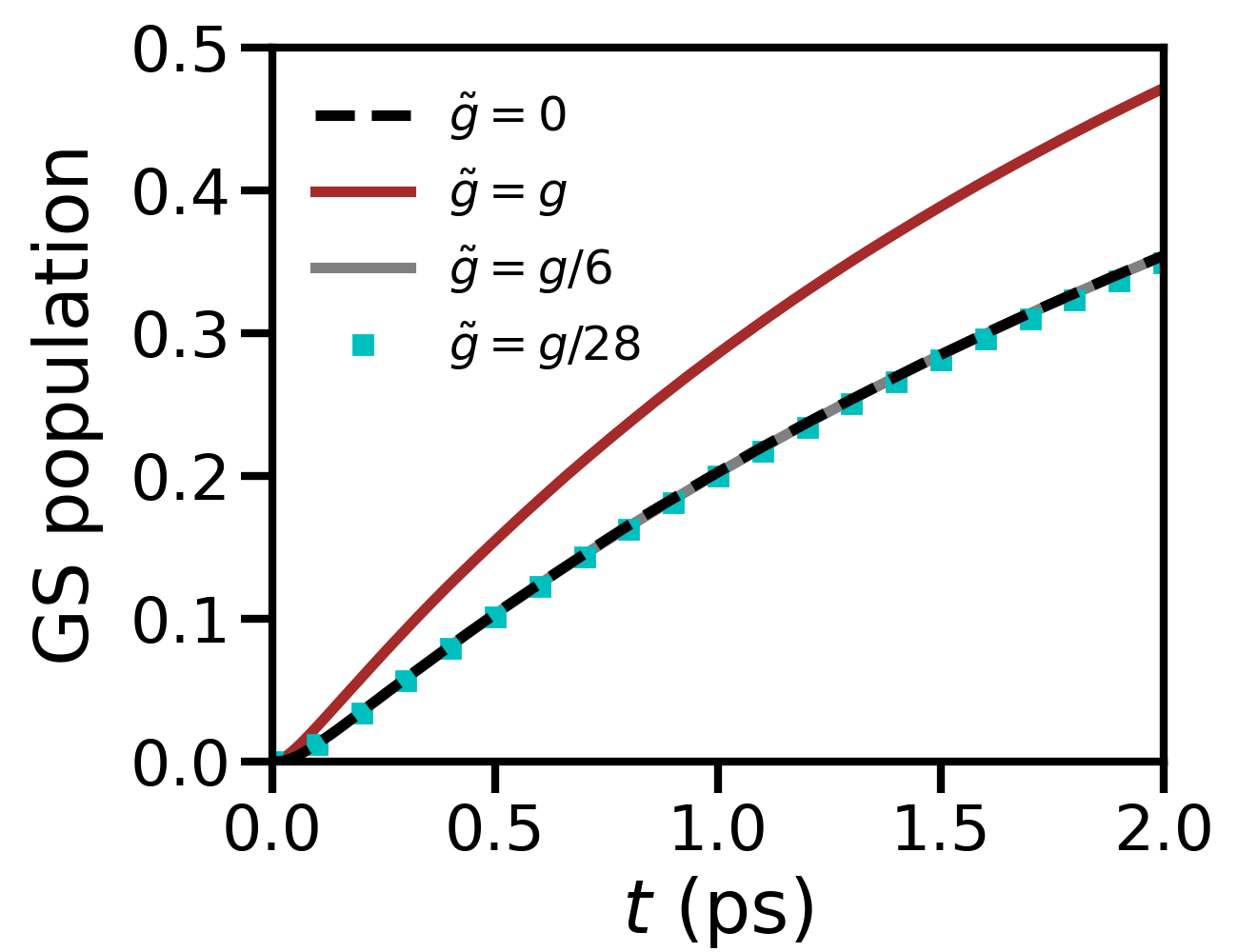}
  \caption{Ground state (GS) population as a function of time in simulations of polariton dynamics in a system of $N=50$ three-level molecules (TLMs) inside the cavity ($g\sqrt{N}=175$~meV) with an exciton coupling strength of $\langle J\rangle=50$~meV. Additionally, coupling terms $\tilde{g}$ between the $|\phi_{S_n,0}\rangle$ and $|\phi_{S_1,1}\rangle$ states were included in the system's Hamiltonian (Equation~\ref{eq:Ham_Rabi_EPA}) to account for exciton-photon annihilation. The following values of $\tilde{g}$ were chosen: $\tilde{g}=0$ (dashed black line), $\tilde{g}=g$ (solid brown line), $\tilde{g}=g/6$ (solid gray line), and $\tilde{g}=g/28$ (cyan squares).
  }\label{fig:figureS11}
\end{figure*}

With the obtained Hamiltonian, we repeated simulations of $N=50$ TLMs with exciton coupling strength $\langle J\rangle=50$~meV under strong light-matter coupling without cavity decay ($\gamma_c=0$). Because the strongest oscillator strength for the $S_1\rightarrow S_n$ transition ($f=0.040$) with the energy difference close to resonance with the $S_0\rightarrow S_1$ transition ($E_{S_n}-E_{S_1}=3.341$~eV with $n=19$) in the optimised geometry of the R6G monomer at the TD-DFT/$\omega$B97X//def2-TZVPP level of theory is $\approx28$ times smaller than the oscillator strength for the $S_0\rightarrow S_1$ transition ($f=1.108$, $E_{S_1}-E_{S_0}=3.394$~eV), we set $\tilde{g}=g/28$ in our simulations. With such a small coupling strength, exciton-photon annihilation has no contribution to the total EEA rate, as can be seen in the time-dependence of the GS population shown in Figure~\ref{fig:figureS11} (cyan squares and black dashed line).

In contrast to the R6G monomer, the ratio between the oscillator strengths corresponding to the $S_0\rightarrow S_1$ and $S_1\rightarrow S_n$ transitions is lower in the optimised geometry of the R6G H-dimer. Indeed, with oscillator strength $f=1.128$ for the $S_0\rightarrow S_1$ transition and oscillator strength $f=0.200$ for the  $S_1\rightarrow S_{20}$ transition in one of the two monomers of the H-dimer (Table~\ref{tab:tab1}), the ratio is $\approx6$. Repeating exciton-polariton dynamics simulations with $\tilde{g}=g/6$ showed no difference in the GS population compared to the simulation with $\tilde{g}=0$ (solid gray and black dashed lines in Figure~\ref{fig:figureS11}), which indicated a negligible role of exciton-photon annihilation in the total EEA rate.

We note, however, that in molecular systems with comparable oscillator strengths for the $S_0\rightarrow S_1$ and $S_1\rightarrow S_n$ transitions, exciton-photon annihilation can become a significant source of exciton annihilation, as was confirmed by simulating TLMs under strong coupling with $\tilde{g}=g$ (brown solid line in Figure~\ref{fig:figureS11}).

\section{Effect of molecular dephasing on the EEA rate}

The time evolution of a closed quantum system can be described using the time-dependent Schr{\"o}dinger equation. In a system with electronic and photonic degrees of freedom, as is the case in the current study, the loss of population associated with the leakage of the electromagnetic field through the cavity mirrors can be treated implicitly by introducing an imaginary component to the corresponding diagonal elements of the system's Hamiltonian (Equation~2 in the main text). While such a method allows us to appreciate the effect of cavity decay on the system's dynamics, it does not treat this process explicitly, nor does it account for other processes resulting from interaction with the environment, such as molecular relaxation or decoherence. One of the methods to explicitly describe the evolution of an open quantum system in the limit of weak interaction with the environment is the Lindblad master equation. In standard form, this equation is defined as follows:\cite{Manzano2020}
\begin{equation}
    \dot{\hat{\rho}} = -\frac{i}{\hbar}\left[\hat{H},\hat{\rho}\right] + \sum_n\left(\hat{L}_n\hat{\rho}\hat{L}_n^\dagger - \frac{1}{2}\{\hat{L}_n^\dagger\hat{L}_n,\hat{\rho}\}\right)
\label{eq:Lindblad_standard}
\end{equation}
with $\hat{\rho}$ being the density matrix, $\hat{L}_n$ being a Lindblad jump operator associated with the $n^{\text{th}}$ interaction channel with the environment, and the sum spanning all interaction channels. 

Without population relaxation ($\gamma_c=\gamma_\text{mol}=0$) and in the presence of pure dephasing, $\hat{L}_\text{deph}=\sqrt{\gamma_\text{deph}}\hat{\sigma}_z$, the Lindblad master equation takes the following form:
\begin{equation}
    \dot{\hat{\rho}} = -\frac{i}{\hbar}\left[\hat{H},\hat{\rho}\right] + \gamma_\text{deph}\hat{\sigma}_z\hat{\rho}\hat{\sigma}_z^\dagger - \frac{\gamma_\text{deph}}{2}\{\hat{\sigma}_z^\dagger\hat{\sigma}_z,\hat{\rho}\}.
\end{equation}

Because $\hat{\sigma}_z^\dagger\hat{\sigma}_z=\hat{\mathbb{I}}$ and $\hat{\sigma}_z=\hat{\sigma}_z^\dagger$, the master equation is reduced to
\begin{equation}
    \dot{\hat{\rho}} = -\frac{i}{\hbar}\left[\hat{H},\hat{\rho}\right] + \gamma_\text{deph}\left(\hat{\sigma}_z\hat{\rho}\hat{\sigma}_z - \hat{\rho}\right).
\end{equation}

With inclusion of the Lindblad operator responsible for cavity decay, $\hat{L}_c=\sqrt{\gamma_c}\hat{a}$, the master equation is transformed to
\begin{equation}
    \dot{\hat{\rho}} = -\frac{i}{\hbar}\left[\hat{H},\hat{\rho}\right] + \gamma_\text{deph}\left(\hat{\sigma}_z\hat{\rho}\hat{\sigma}_z - \hat{\rho}\right) + \gamma_c\hat{a}\hat{\rho}\hat{a}^\dagger - \frac{\gamma_c}{2}\{\hat{a}^\dagger\hat{a},\hat{\rho}\}.
\label{eq:Lindblad_deph_plus_decay}
\end{equation}

\subsection{Jaynes-Cummings model}

Consider a system of one molecule strongly interacting with one cavity photon. Under the rotating-wave approximation, this system can be described by the Jaynes-Cummings Hamiltonian:\cite{Jaynes1963}
\begin{equation}
    \hat{H}_\text{JC} = \hbar\omega_m\hat{\sigma}^+\hat{\sigma}^- + \hbar\omega_c\hat{a}^\dagger\hat{a} + \hbar g\left(\hat{\sigma}^+\hat{a}+\hat{\sigma}^-\hat{a}^\dagger\right).
\label{eq:JC_Hamiltonian}
\end{equation}

An analytical solution can be obtained for the master equation with the Jaynes-Cummings Hamiltonian in the presence of dephasing and cavity decay. In the case $\gamma_\text{deph},\gamma_c\ll\Omega_\text{R}$ and initial excitation into the molecule, the solution looks as follows:\cite{Briegel1993,Scala2007}
\begin{equation}
    P_e(t) \approx \frac{1}{2}\left[1+e^{-\gamma_\text{deph}t/2}\cos\left(\Omega_\text{R}t\right)\right]e^{-\gamma_ct/2}.
\label{eq:analytical_deph_plus_decay}
\end{equation}

\begin{figure*}[!htb]
\centering
\includegraphics[width=1\textwidth]{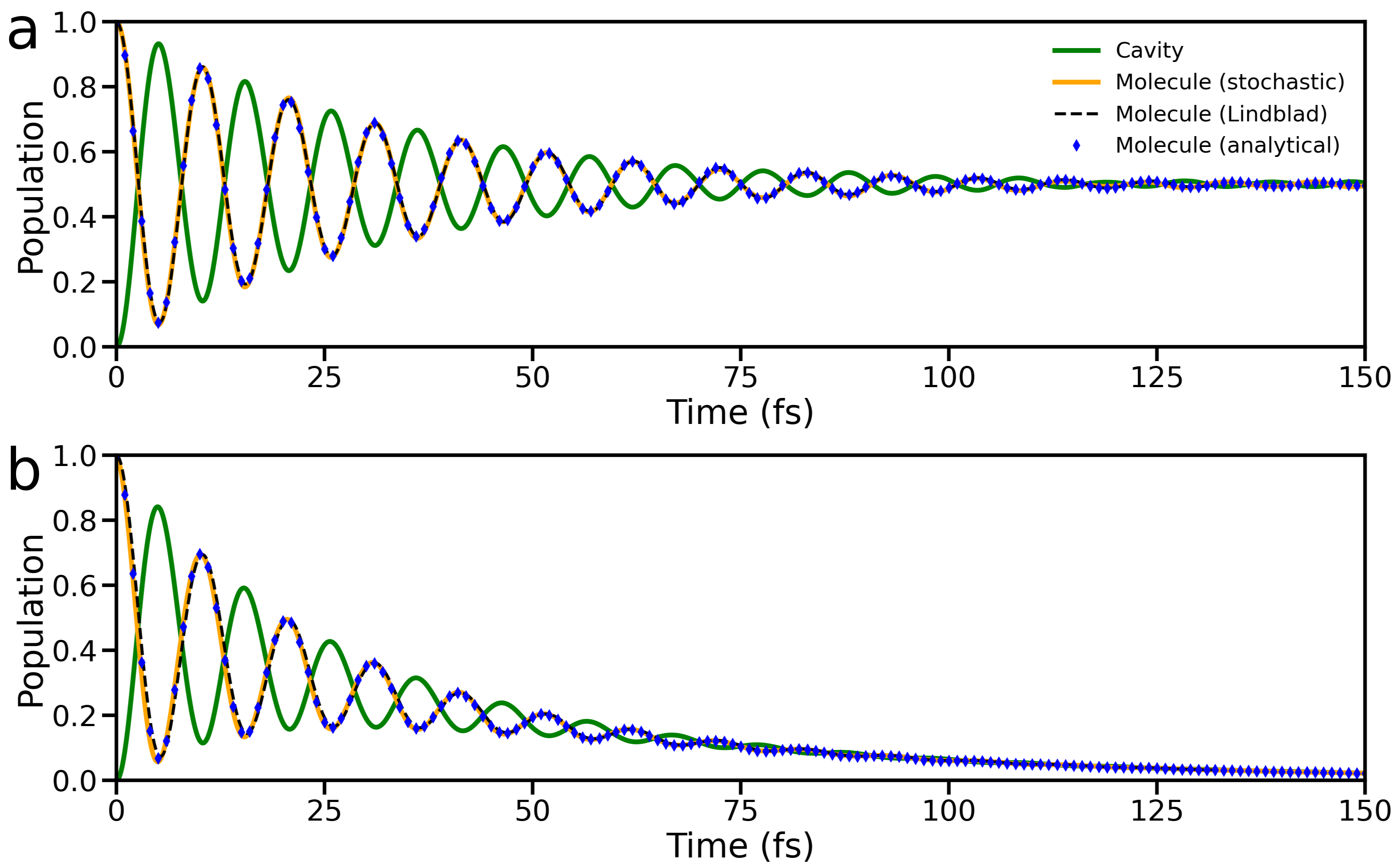}
  \caption{Population of the cavity mode (solid green lines) and of the molecular excitation in the Jaynes-Cummings model under the rotating-wave approximation (Equation~\ref{eq:JC_Hamiltonian}) in the presence of (\textbf{a}) dephasing ($\tau_\text{deph}=100$~fs) and (\textbf{b}) dephasing and cavity decay ($\tau_c=150$~fs). Population of the molecular excitation is shown for the case of the direct solution of the Lindblad master equation (Equation~\ref{eq:Lindblad_standard}, dashed black line) and the Monte Carlo wave function (MCWF) propagation (solid orange lines) and compared with the analytical solution (Equation~\ref{eq:analytical_deph_plus_decay} with $\gamma_c=0$ (\textbf{a}) and $\gamma_c=1/\tau_c$ (\textbf{b}), blue diamonds). The populations obtained with the MCWF propagation were averaged over 5000 runs.
  }\label{fig:figureS12}
\end{figure*}

Figure~\ref{fig:figureS12} shows the time-evolution of the molecular and photonic populations in the Jaynes-Cummings model with molecular dephasing (panel~\textbf{a}) and with both dephasing and cavity decay (panel~\textbf{b}). The population of the molecular component obtained by solving the Lindblad master equation (Equation~\ref{eq:Lindblad_deph_plus_decay}; dashed black lines) is identical to the analytical solution (blue diamonds), demonstrating the ability of the Lindblad master equation to correctly capture the effect of cavity decay and molecular dephasing.

\subsection{Simulation of EEA with the Lindblad master equation}

\subsubsection{Computational scaling of direct Lindblad simulations}

To assess the feasibility of simulating EEA using direct integration of the Lindblad master equation, we analyse the computational scaling of a bespoke solver (which is part of the Jones Group SpecSuite~\cite{Green_SpectralFiltering_2DS,Green_NonMarkovianity_Lineshape}) with system size. In this framework, the dominant numerical cost arises from dense linear algebra operations acting on the system density matrix. The computational cost is therefore governed primarily by the Hilbert space dimension $d$.

Benchmark simulations were performed for systems of $N=10$ and $N=20$ TLMs under identical numerical conditions (identical propagation window and solver parameters). The corresponding Hilbert space dimensions are $d_{10}=67$ and $d_{20}=232$, with measured runtimes for a single propagation chunk ($0 \rightarrow 0.001$ ps, corresponding to 20 HEOMsolve steps) of 18 seconds and 9.2 minutes, respectively. These values imply a power-law scaling of the form
\begin{equation}
    t_{\text{chunk}} \sim d^{p},
\end{equation}
with an exponent $p \approx 4.9$.
This indicates a very rapid increase in computational cost with system size. 

We extrapolate up to a single $0.2$ ps propagation chunk, and a full $2 \, \mathrm{ps}$ run for the following analysis. Using the measured runtime for $N=20$ as a reference, we would require approximately 52 days for $N=30$, 2 years for $N=40$, and over 17 years for $N=50$. Extending to the full simulation time of 2 ps (corresponding to 10 of the $0.2 \, \mathrm{ps}$ chunks) results in total runtimes ranging from several days ($N=20$) to many years ($N=50$).

\subsubsection{Implications for EEA simulations}

The steep scaling of direct Lindblad simulations imposes severe limitations on the accessible system sizes. While systems up to $N \sim 20$ can be treated using direct density matrix propagation, larger ensembles rapidly become computationally prohibitive. Importantly, this limitation arises from the growth of the Hilbert space rather than memory usage, which remains relatively modest for the system sizes considered.

\subsubsection{Projected Runtimes for Larger Systems}

Using the measured $N=20$ chunk runtime as a reference,
\begin{equation}
t^{(\mathrm{chunk})}(N) = t_{20}^{(\mathrm{chunk})}
\left(\frac{d(N)}{d_{20}}\right)^{4.9},
\end{equation}
with
$t_{20}^{(\mathrm{chunk})} \approx 1.3~\mathrm{days}$,
and taking the Hilbert space dimensions as
\begin{align*}
d_{20} &= 232, \, \, 
d_{30} = 496, \\
d_{40} &= 861, \, \, 
d_{50} = 1326,
\end{align*}
we obtain the per-chunk runtimes for a single $0.200~\mathrm{ps}$ propagation as displayed in Table~\ref{tab:tableS3}.

\begin{table}[h]
\centering
\begin{tabular}{c c c c}
\hline
$N$ & $d(N)$ &  $t^{(\mathrm{0.001 \, chunk})}$ (mixed) & $t^{(\mathrm{0.2 \,chunk})}$ (mixed) \\
\hline
20 & 232 & 9.2 minutes  & 1.3 days \\
30 & 496 & 6.3 hours   & 52.5 days \\
40 & 861  & 3.9 days   & 2.1 years \\
50 & 1326 & 32 days  & 17.6 years \\
\hline
\end{tabular}
\caption{Estimated per-chunk runtimes.}
\label{tab:tableS3}
\end{table}

\subsubsection{Full $2.0~\mathrm{ps}$ Runtime Estimates}

Since the target simulation length is $2.0~\mathrm{ps}$ and we have measured chunks of $0.2~\mathrm{ps}$, the full simulation requires approximately
\begin{equation}
N_{\mathrm{chunks}} = \frac{2.0}{0.2} = 10,
\end{equation}
and hence
\begin{equation}
t^{(\mathrm{full})}(N) \approx 10 \, t^{(\mathrm{chunk})}(N).
\end{equation}

Table~\ref{tab:tableS4} gives the approximate full runtimes.

\begin{table}[h]
\centering
\begin{tabular}{c c c c}
\hline
$N$ & $t^{(\mathrm{full})}$ (hours) & $t^{(\mathrm{full})}$ (days) & $t^{(\mathrm{full})}$ (mixed units) \\
\hline
10 & 10 & 0.42 & $10~\mathrm{hours}$ \\
20 & 720 & 30 & $1~\mathrm{month}$ \\
30 & $2.9\times10^4$ & 1210 & $3.3~\mathrm{years}$ \\
40 & $4.2\times10^5$ & $1.75\times10^4$ & $48~\mathrm{years}$ \\
50 & $3.5\times10^6$ & $1.46\times10^5$ & $400~\mathrm{years}$ \\
\hline
\end{tabular}
\caption{Estimated full runtimes for a $2.0~\mathrm{ps}$ simulation ($10$ $0.2 \, \mathrm{ps}$ chunks).}
\label{tab:tableS4}
\end{table}

These considerations strongly motivate the use of stochastic wave function approaches, such as the Monte-Carlo wave function (MCWF) method described below. By propagating pure states instead of density matrices, the MCWF method avoids the quadratic scaling associated with the density matrix dimension and enables simulations of significantly larger molecular ensembles. This allows us to access the regime of $N=50$ molecules considered in the main text, while still incorporating the effects of dephasing and cavity decay.

In practice, the MCWF approach introduces an additional statistical averaging step over stochastic trajectories. However, as demonstrated in Figure \ref{fig:figureS12}, convergence to the Lindblad master equation results is achieved with a manageable number of trajectories, making this approach a computationally efficient alternative for large-scale EEA simulations.

\subsubsection{Memory requirement of direct Lindblad solution}

In addition to computational cost, practical limitations arise from memory usage and data transfer. In the standard representation, the density matrix is stored as a full complex matrix of size $d \times d$, where $d$ is the Hilbert space dimension. A single density matrix therefore requires storage that scales as $\mathcal{O}(d^2)$.

For illustration, a system with $d=20$ corresponds to a $20 \times 20$ complex matrix, i.e.\ 400 complex numbers, which is negligible in terms of memory. However, for system sizes relevant to this work ($d \sim 10^2$), the storage becomes significantly larger. For example, $d=200$ corresponds to $4\times10^4$ complex numbers per density matrix. When storing a full trajectory over $N_{\mathrm{save}} \sim 10^3$ time points, this results in $\sim 4\times10^7$ complex numbers, corresponding to several hundred megabytes per simulation when using double-precision complex arithmetic.

In the present implementation, the dynamics are propagated in the Fortran solver using a fixed internal time step of 50 as up to the end of each propagation chunk. However, the density matrix is not written to disk at every internal step, but only at a smaller set of output times chosen for analysis and plotting. As a result, the computational cost and storage cost are controlled by different time scales.

If $N_{\mathrm{step}}$ denotes the number of internal propagation steps and $N_{\mathrm{save}}$ the number of saved time points, then the total storage required for a full trajectory scales as
\begin{equation}
    \mathcal{O}(N_{\mathrm{save}} d^2),
\end{equation}
whereas the dominant computational effort scales as
\begin{equation}
    \mathcal{O}(N_{\mathrm{step}} d^3),
\end{equation}
since each internal propagation step requires dense matrix operations such as commutators and dissipative terms. Because $N_{\mathrm{step}} \gg N_{\mathrm{save}}$, the runtime is determined by the fine internal propagation grid, while the data volume is determined by the coarser saved output grid.

Even when the number of saved time points is moderate, the dense representation leads to large output files for increasing system size. In practice, this makes it difficult to transfer full simulation results between local machines and ARCHER2 (a supercomputer used in the current work), particularly when multiple runs or parameter sweeps were required. This limitation arises primarily from the cumulative size of dense trajectory data and associated input/output overhead.

To address these issues, we implemented a bespoke sparse matrix formulation of the Lindblad dynamics. In this approach, the Hamiltonian and Lindblad operators are stored in sparse format, retaining only the non-zero matrix elements and their indices. This is justified by the structure of the exciton--photon model, in which couplings are restricted to specific manifolds (e.g.\ nearest-neighbour exciton hopping, light--matter coupling, and ladder-type photon transitions), resulting in a large fraction of zero entries in the full matrices.

The sparse implementation therefore reduces memory usage from $\mathcal{O}(d^2)$ to $\mathcal{O}(\mathrm{nnz})$, where $\mathrm{nnz}$ is the number of non-zero elements, which scales much more favourably with system size. In addition, storing trajectories in sparse format significantly reduces file sizes, making it feasible to transfer and analyse simulation outputs.

Overall, while dense representations are straightforward to implement, their quadratic scaling in both memory and storage makes them impractical for large-scale simulations. The adoption of a sparse representation provides a practical route to overcoming these limitations in exciton--polariton dynamics. However, overall, the good agreement between the MCWF formulation and the full Lindblad (at least to this level of approximation of physical processes) makes it highly preferable for the chosen system. 

\subsection{Simulation of EEA with the Monte-Carlo wave function propagation}

Due to the unfortunate scaling of the simulation time with the system size, we were unable to propagate the density matrix for molecule-cavity systems with $N>20$ with the Lindblad master equation. Therefore, to study EEA in such large molecular ensembles, we instead propagated the full exciton-polariton wave function (Equation~3 in the main text) together with stochastic Markovian jumps associated with the Lindblad jump operators, responsible for dephasing and cavity decay. In the MCWF method,\cite{Dalibard1992,Molmer1993} the expansion coefficients of the wave function are numerically propagated according to the Schr\"{o}dinger equation (Equation~4 in the main text). At each time step of a simulation, the norm of the wave function is reduced according to 
\begin{equation}
    |\Psi(t+\Delta t)|^2=1-\delta p,
\end{equation}
where $\Delta t$ is the time step size, and $\delta p=\sum_n\delta p_n$ with $\delta p_n$ being the contribution of the $n^\text{th}$ jump operator to the total reduction of the wave function's norm, defined as $\delta p_n=\langle\Psi(t)|\hat{L}_n^\dagger\hat{L}_n|\Psi(t)\rangle$.\cite{Herrera2022,Tremblay2022} Then, a number $\xi$ is randomly drawn from the uniform distribution between zero and one, $0<\xi<1$, and is compared to $\delta p$. 

In most cases, $\delta p$ is smaller than $\xi$, and we assume that no quantum jump has occurred. In this case, the wave function is simply renormalised before the next time step: 
\begin{equation}
    |\Psi_\text{renorm}(t+\Delta t)\rangle=\frac{|\Psi(t+\Delta t)\rangle}{\sqrt{1-\delta p}}.
\end{equation}

Occasionally, however, the condition $\delta p>\xi$ is fulfilled. In this situation, we assume that a quantum jump has occured, and a Lindblad operator then acts on the wave function. The choice of a specific operator to act on the wave function is made randomly with weights given by $\delta p_n/\delta p$.

When the selected operator is associated with pure dephasing, $\hat{L}_n=\sqrt{\gamma_\text{deph}}\hat{\sigma}_z$, the amplitude of the expansion coefficient of the state, on which the dephasing operator acts, receives a random phase shift,
\begin{equation}
    |d_n(t+\Delta t)|\rightarrow|d_n(t+\Delta t)|\exp\left(i\zeta\right),
\end{equation}
with $\zeta$ being a random phase number in the range $[0;2\pi]$.

When the selected operator is associated with cavity decay, $\hat{L}_n=\sqrt{\gamma_c}\hat{a}$, the system collapses into the total ground state. Because this state is not explicitly included in the system's Hamiltonian, we set all expansion coefficients to zero until the end of the simulation, \textit{i.e.} $d_j=0$~$\forall j$, resulting in the norm of the wave function being zero at any time after the jump.

In this work, we consider four types of operators acting on different exciton-photon product states, namely
\begin{enumerate}
    \item pure dephasing operator $\hat{L}_\text{deph}=\sqrt{\gamma_\text{deph}}\hat{\sigma}_z$, acting on the states, in which only one TLM is excited, \textit{i.e.} $|\phi_{S_n,0}\rangle$ and $|\phi_{S_1,1}\rangle$;
    \item pure dephasing operator $\hat{L}_\text{deph}^\prime=\sqrt{2\gamma_\text{deph}}\hat{\sigma}_z$, acting on the states, in which two TLMs are simultaneously excited, \textit{i.e.} $|\phi_{2S_1,0}\rangle$;
    \item cavity decay operator $\hat{L}_c=\sqrt{\gamma_c}\hat{a}$, acting on the states, in which only one cavity photon is present, \textit{i.e.} $|\phi_{S_1,1}\rangle$;
    \item cavity decay operator $\hat{L}_c^\prime=\sqrt{2\gamma_c}\hat{a}$, acting on the state, in which two cavity photons are present, \textit{i.e.} $|\phi_{S_0,2}\rangle$.
\end{enumerate}

We first validated the MCFW approach against directly solving the Lindblad master equation for the Jaynes-Cummings model. As shown in Figure~\ref{fig:figureS12}, the populations of the molecular excited state obtained with the two approaches (solid orange line and dashed black line) agree very well when \textit{i)} only dephasing is present and \textit{ii)} both dephasing and cavity decay are included, as long as the results obtained with the MCFW method are averaged over a sufficiently large number of runs (in Figure~\ref{fig:figureS12}, the populations are shown for $N_\text{runs}=5000$).

\begin{figure*}[!htb]
\centering
\includegraphics[width=1\textwidth]{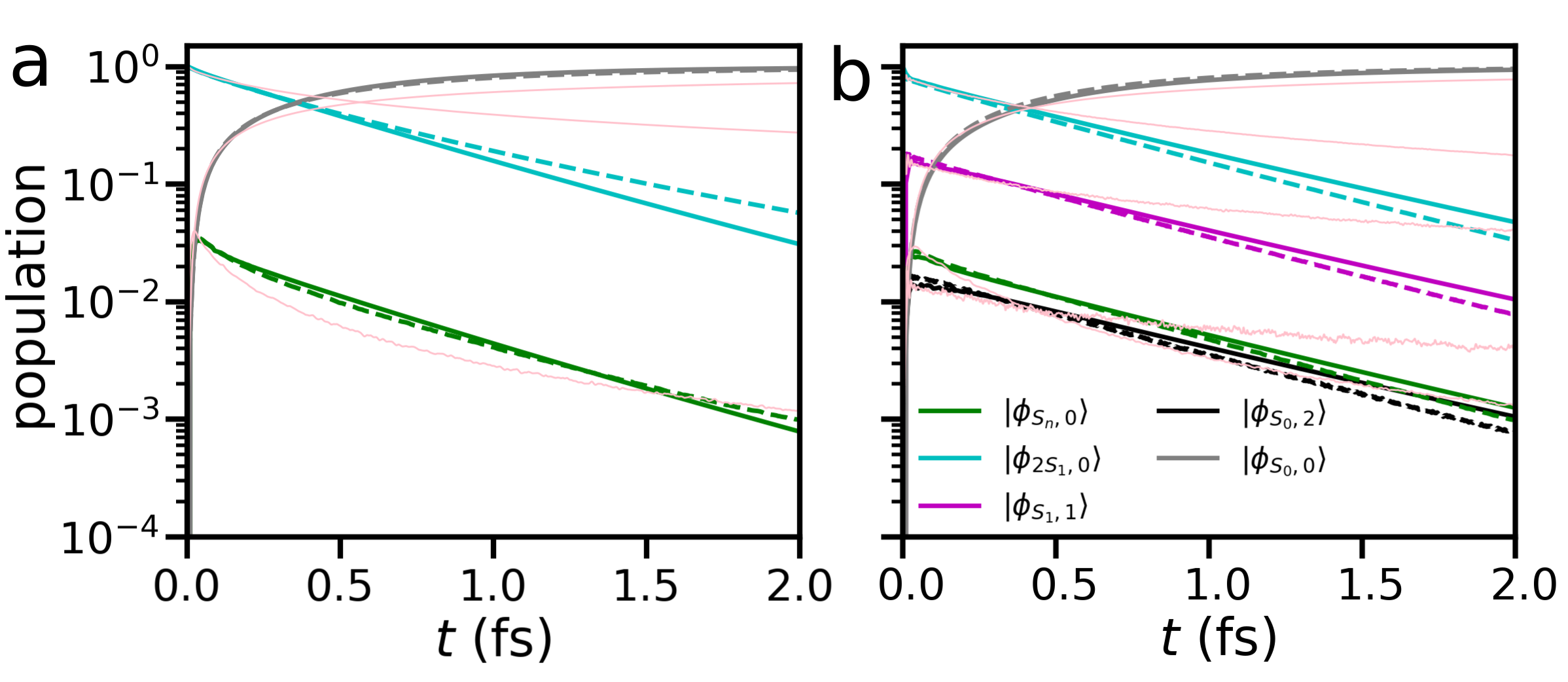}
  \caption{Total population of all $|\phi_{S_n,0}\rangle$ states (green), $|\phi_{2S_1,0}\rangle$ states (cyan), $|\phi_{S_1,1}\rangle$ states (purple), $|\phi_{S_0,2}\rangle$ state (black), and the ground state $|\phi_{S_0,0}\rangle$ (grey), in simulation of $N=10$ three-level molecules (TLMs) with an exciton coupling strength of $\langle J\rangle=50$~meV and a collective light-matter coupling strength of $g\sqrt{N}=0$~meV (panel~\textbf{a}) and $g\sqrt{N}=175$~meV (panel~\textbf{b}). In both panels, the solid lines correspond to direct solution of the Lindblad master equation, and the dashed lines correspond to the Monte Carlo wave function (MCWF) propagation with the inclusion of dephasing ($\tau_\text{deph}=100$~fs). The pink lines show the populations of product states in simulations without dephasing. The populations obtained with the Lindblad master equation and with the MCWF propagation were averaged over 100 and 10000 runs, respectively.
  }\label{fig:figureS13}
\end{figure*}

Next, we compared the two methods for the TLM-cavity system considered in the current work. Figure~\ref{fig:figureS13} shows the populations of the product states for a system of $N=10$ 
TLMs in the absence and presence of light-matter coupling. Although the agreement between the methods is not perfect, the populations obtained with the MCFW are much closer to those obtained by directly solving the Lindblad master equation than to the populations obtained by numerical propagation of the Schr{\"o}dinger equation in the absence of dephasing (pink lines in Figure~\ref{fig:figureS13}). Therefore, we reckon it reasonable to use the MCFW for studying the effect of dephasing on EEA in large molecular ensembles with $N>20$, which are not tractable with the full Lindblad equation.   

\begin{figure*}[!htb]
\centering
\includegraphics[width=0.6\textwidth]{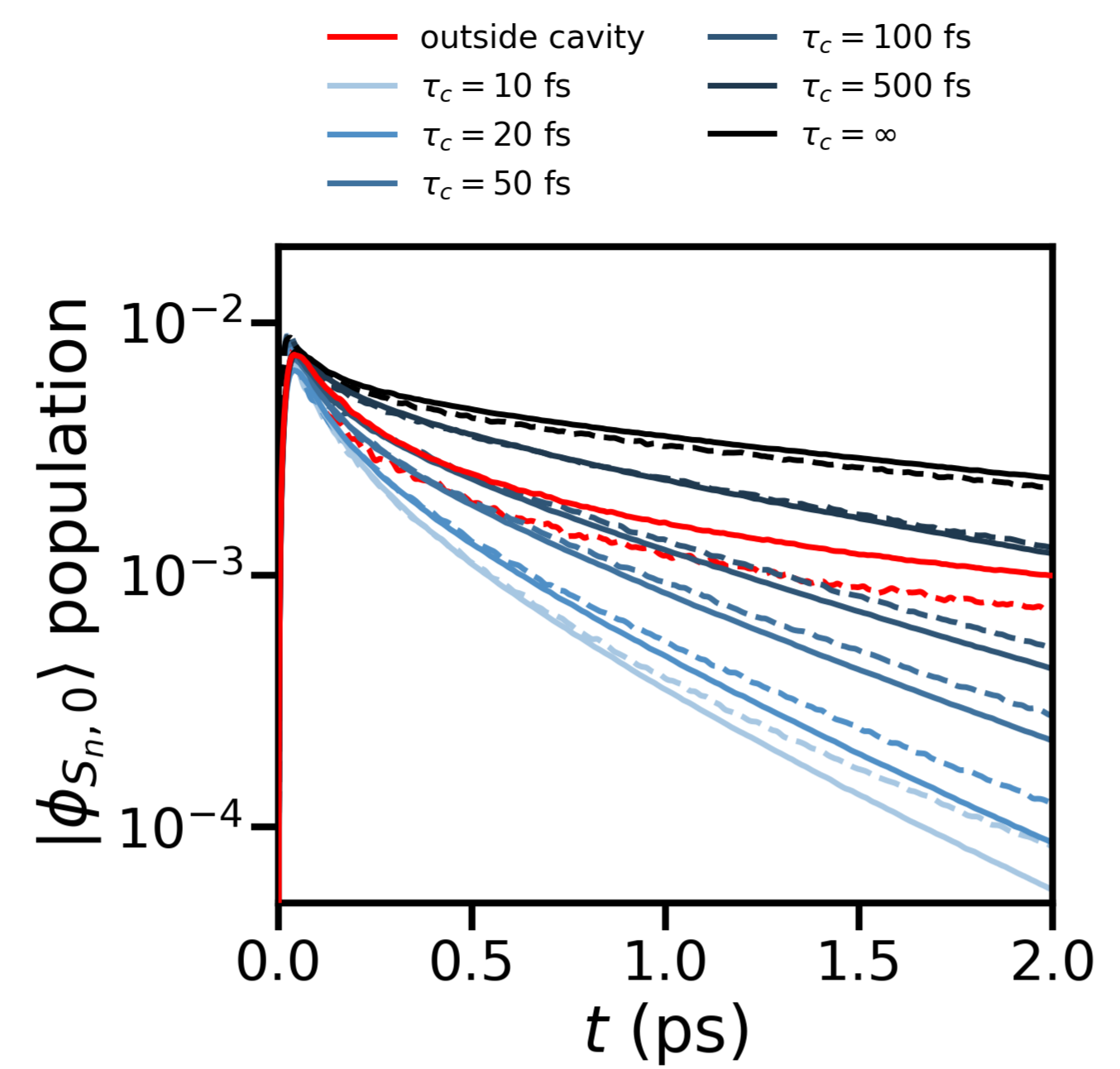}
  \caption{
  Total population of the $\vert\phi_{S_n,0}\rangle$ states in simulations of $N=50$ TLMs with an exciton coupling strength of $\langle J\rangle=30$~meV and a collective light-matter coupling strength of $g\sqrt{N}=100$~meV. The simulations were performed with different cavity lifetimes ranging between $\tau_c=10$~fs and $\tau_c=500$~fs, as well as for the ideal cavity with no decay ($\gamma_c=1/\tau_c=0$) and outside the cavity. In both panels, the solid lines correspond to the Monte-Carlo wave function (MCWF) propagation with the inclusion of dephasing ($\tau_\text{deph}=100$~fs), and the dashed lines correspond to the numerical integration of the Schr{\"o}dinger equation without dephasing. The populations obtained with the MCWF propagation were averaged over 10000 runs.
  }\label{fig:figureS14}
\end{figure*}

After validating the MCFW approach, we performed exciton-polariton dynamics simulations with $N=50$, which is larger than the number of molecules that can be handled using the Lindblad master equation with our available computational resources. As Figure~5 in the main text depicts, pure dephasing results in a more rapid depopulation of the initially excited states with two $S_1$-excitons, \textit{i.e.} $|\phi_{2S_1,0}\rangle$, and hence a more rapid build-up of the population of the remaining states, including the ground state $|\phi_{S_0,0}\rangle$ (grey lines). In Figure~\ref{fig:figureS14}, we also show the GS population in simulations with different cavity lifetimes, $\tau_c=1/\gamma_c$. Although dephasing leads to a quantitative change in the GS population for all lifetimes, the trends remain similar to those in the system without dephasing. Importantly, EEA undergoes a transition from suppression at short lifetimes to enhancement at long lifetimes, in line with simulations without dephasing.

\clearpage

\bibliography{rsc}